\documentclass[article]{JHEP3}
\let\ifpdf\ULOS
\usepackage{amssymb,amsfonts,bm}
\usepackage{epsfig}
\usepackage{amsmath}
\usepackage{amssymb,amsfonts}
\usepackage{ifpdf}

\renewcommand{\theequation}{\arabic{section}.\arabic{equation}}
 \def\II{\relax{\rm I\kern-.18em I}}
\def\be{\begin{equation}}
\def\ee{\end{equation}}
 \def\beq{\begin{equation}}
 \def\eeq{\end{equation}}
 \newcommand{\bea}{\begin{eqnarray}}
 \newcommand{\eea}{\end{eqnarray}}
 \def\nn{\nonumber}

 \def\l{\lambda}
 \def\m{\mu}
 \def\n{\nu}

 \def\e{\epsilon}
 \def\d{\partial}
 \def\p{\partial}

\newcommand{\f}{{\kappa}} 

 \def \lta {\mathrel{\vcenter
      {\hbox{$<$}\nointerlineskip\hbox{$\sim$}}}}
 \def \gta {\mathrel{\vcenter
      {\hbox{$>$}\nointerlineskip\hbox{$\sim$}}}}

 \def\hre#1#2{\href{http://arxiv.org/abs/#1/#2}{[arXiv:#1/#2]}}

 \def\hri#1#2{\href{http://arxiv.org/abs/#1}{[arXiv:#1#2]}}

\newcommand{\rmi}[1]{{\mbox{\scriptsize #1}}}
\newcommand{\fr}[2]{{\frac{#1}{#2}\,}}
\newcommand{\fra}[2]{{\textstyle{\frac{#1}{#2}\,}}}  
\newcommand{\la}[1]{\label{#1}}
\def\CL{{\cal L}}
\def\CO{{\cal O}}

\newcommand{\ba}{\begin{eqnarray}}
\newcommand{\ea}{\end{eqnarray}}
\newcommand{\bi}{\begin{itemize}}
\newcommand{\ei}{\end{itemize}}
\newcommand{\nr}[1]{(\ref{#1})}

\title{On finite-temperature holographic QCD in the Veneziano limit}
\author{T. Alho$^a$, M. J\"arvinen$^b$,  K. Kajantie$^c$, E. Kiritsis$^{b,d}$,
K. Tuominen$^{a,c}$\\
~\\
$^a$ \href{https://www.jyu.fi/fysiikka/en}
{Department of Physics}, P.O.Box 35, FI-40014 University of Jyv\"askyl\"a, Finland\\
~\\
$^b$ \href{http://hep.physics.uoc.gr}{Crete Center for Theoretical Physics},
Department of Physics, University of Crete, PO Box 2208, 71003 Heraklion, Greece\\
~\\
$^c$ \href{http://www.hip.fi}
{Helsinki Institute of Physics}, P.O.Box 64, FI-00014 University of Helsinki, Finland\\
~\\
$^d$ \href{http://www.apc.univ-paris7.fr}{APC, Universit\'e Paris 7}, CNRS/IN2P3, CEA/IRFU,
Obs. de Paris, Sorbonne Paris\\ Cit\'e,
B\^atiment Condorcet, F-75205, Paris Cedex 13, France (UMR du CNRS 7164).
}

\preprint{CCTP-2012-17, HIP-2012-16/TH}

\abstract{Holographic models in the $T=0$ universality class of QCD in the limit of
large number $N_c$ of colors  and $N_f$ massless fermion flavors,   but constant ratio
$x_f=N_f/N_c$, are analyzed at finite temperature. The models contain a 5-dimensional metric and two scalars,
a dilaton sourcing ${\rm Tr}F^2$ and a tachyon dual to $\bar qq$. The phase structure
on the $T,x_f$ plane is computed and various 1st order, 2nd order transitions and crossovers
 with their chiral symmetry properties are identified. For each $x_f$,
the temperature dependence of $p/T^4$ and the condensate $\langle\bar qq\rangle$ is
computed.
In the simplest case, we find that for $x_f$ up to the critical $x_c\sim 4$ there is a 1st order
transition on which chiral symmetry is broken and the energy density jumps.
In the conformal window $x_c<x_f<11/2$, there is only a continuous
crossover between two conformal phases.
When approaching $x_c$ from below, $x_f\to x_c$,
temperature scales approach zero as specified by Miransky scaling.
}

\DeclareGraphicsExtensions{.png}

\begin{document}

\section{Introduction}

QCD in the Veneziano limit \cite{v},
\be
N_c\to \infty,\quad N_f\to \infty,\quad {N_f\over N_c}=x_f~~{\rm fixed},\quad
\lambda=g_\rmi{YM}^2N_c~~{\rm fixed},
\label{v1}
\ee
is expected to display a host of interesting and mostly non-perturbative phenomena, including:

\begin{itemize}

\item The ``conformal window'' with a nontrivial infrared (IR) fixed point, which extends from
$x_f={11\over 2}$ to smaller values of $x_f$.
      The region $x_f\to {11\over 2}$ has an IR fixed point while the theory is
still weakly coupled,  as was  analyzed in \cite{bz} (see also \cite{cas}).

\item It is expected that at a critical $x_f=x_c$, the conformal window will end, and
for $x_f<x_c$, the theory will exhibit chiral symmetry breaking in the IR. This
behavior is expected to persist down to $x_f=0$.
For $x_f>x_c$ the IR theory is a
conformal field theory 
at strong coupling,  that progressively
becomes weak as $x_f\to {11\over 2}$.

\item Near and below $x_c$, there is the transition region to conventional QCD
IR behavior. In this region the theory is expected to be ``walking'':
This means that the theory appears to be approaching the IR fixed point as
the coupling evolves very slowly for many e-foldings of energies.  But
chiral symmetry breaking is nevertheless triggered and in the deep infrared
the coupling diverges as in QCD.
The slow evolution of the coupling has been correlated with a nontrivial dimension
for the quark mass operator near two, rather than three (the free field value).
IR observables are expected to obey the Miransky scaling \cite{bkt,miransky}
as $x_f \to x_c$ from below.

\item New phenomena are expected to appear at finite density driven by strong
coupling and the presence of quarks. These include color superconductivity
\cite{cs,cfl}. In this case, however, gauge invariant
vevs are effectively double trace operators
and the phase structure is determined at the next to leading order in $1/N_c$.

\end{itemize}

The existence of the ``walking" region makes the theory extremely
interesting for
applications in dynamical electroweak symmetry breaking (technicolor). This has also
motivated an intensive lattice Monte Carlo work
during recent years \cite{lat2012,teper,fodor_nf12}. The bulk of this work has been
done at zero temperature; recently there appeared the first attempts to go
to finite $T$ for QCD with $N_c=3$, $N_f$ up to 8
\cite{jinmawhinney_nf8,miuralombardo,fodor_nf12} and for non-QCD-like theories \cite{kogutsinclair}.
Chiral effective theories have also been applied
\cite{pisarskilargeNQCD,pisarskidofs,appelquistVQCD,sannino_xc,rischke_chiralmodel,
fiveloop_pisarski_wilczek,karchobannon,sunsun_chiralmodelEW}.

The aim of the present work is to study  a
class of holographic bottom-up models
(V-QCD) 
that belong to the universality class of QCD
with massless quarks in
the Veneziano limit \cite{jk} at finite temperature and zero chemical potential.
 We will calculate the temperature dependence
of the free energy density (= $-$pressure
= $-p(T)$) and of the quark condensate ($\langle\bar qq\rangle(T)$).
The former acts as an effective
order parameter for deconfinement (at large $N_c$), for which there is no true order
parameter associated with a symmetry.\footnote{A related one, used commonly in
lattice work, is the expectation value of the Polyakov loop.} The quark condensate  is
a true order parameter for chiral symmetry if quarks are massless.
The calculation is carried out for the full range of $x_f$, $0<x_f<11/2$.

Discontinuities or rapid variations in pressure (or energy density) and
quark condensate
can be used to define phase boundaries associated with
deconfinement and chiral symmetry
restoration temperatures $T_d(x_f)$ and $T_\chi(x_f)$.
We will use the usual nomenclature: If the $n$th derivative of $p(T)$
is discontinuous, the transition is of $n$th order. We also consider  continuous
 {\em crossovers} which are identified
by using the
  scaled quantity  $(\epsilon-3p)/T^4$.
Its maximum
  defines the crossover temperature $T_\rmi{crossover}(x_f)$.
The phase diagram is defined as a plot of all phase boundaries
on the $(x_f,T)$ plane. The phase diagrams we present will also contain a rich
structure of metastable states, namely local (but globally subleading) minima of the free energy.

In the holographic approach the thermal transitions will be transitions between
various 5-dimensional black hole and ``thermal gas" metrics and the
nomenclature of transitions, explained later in great detail, will be correspondingly
different.
The holographic approach is constrained but not fully constrained and we cannot
give a precise prediction of the phase diagram of hot V-QCD. We can state the
 most plausible behavior but we can also mention a few other alternatives.
We will always find the analogues of $T_d$ and $T_\chi$, but we will also find
transitions with no obvious QCD interpretation. Whether these
reflect real physics of hot QCD in the Veneziano limit or whether they are
artifacts of the holographic approach will be an interesting problem for further
study.

The usual expectation is that there is a 1st order line at
$T_d=T_\chi$; in the large $N_c$ limit one can actually
prove that $T_\chi\ge T_d$ \cite{pisarskilargeNQCD,pisarskidofs}. The main class
of our predictions reflect these properties: for smaller $x_f$ we find that
deconfinement and chiral symmetry restoration coincide, but for $x_f$
approaching $x_c$ the deconfining
and chiral transitions
can become 
separate so that $T_\chi>T_d$ (see, for example,
Fig.~\ref{figTTransitionsSB} below). The chiral transition is then of 2nd order
(and mean field type).
Furthermore, for smaller $x_f$ the
separate 2nd order
chiral transition is in the metastable region
so that it can be reached if the system is supercooled
\cite{narayananneuberger}. One might here add that $T_\chi<T_d$ for stable
phases may be reached at large chemical potential \cite{quarkyonic,Kahara:2010wh}.

The starting point of our finite temperature analysis is the $T=0$
holographic model introduced in \cite{jk}, based on previous theoretical ideas in
\cite{YM1,YM2,YM3,ckp,ikp}. Moving to finite $T$ implies
studying black hole solutions of the action in \cite{jk}. A defining characteristic of this
class of models
is that they  contain full
backreaction between the duals of the color
and flavor degrees of freedom. Earlier work
\cite{kajantieFT,gubser,kajantieIRFP,kajantie} on thermodynamics in such
bottom-up models
imposed quasiconformality directly on the beta function of the theory.
One should note that walking behavior
and the related ``conformal transition'' at $x_f=x_c$ have 
also been studied
in top-down models \cite{nunez,pare,angel,kutasov1}, 
as well as in simpler bottom-up models \cite{walkingbu,kutasov2} 
which do not attempt to model the backreaction. 
See also the review \cite{unqrev} on introducing backreacted flavor in the top-down models.

In this  introduction we will first describe the special properties of V-QCD from
\cite{jk} and then discuss general properties of its black hole
solutions. Section 2 will contain a detailed discussion of the action of the
model and 
of the two characteristic classes of scalar potentials.
Section 3 presents the Einstein equations of the model, describes how they
are numerically solved and, finally, how thermodynamics is computed from
the numerical bulk fields. A particularly delicate issue here is the fixing of
the quark mass $m$ to zero. We also briefly comment on fixing $m$ to some nonzero value.
An extensive list of numerical results is given
in Section 4. From these, the types of phase transition
lines the models predict are determined. In Section 5,  techniques for computing the condensate
are described and several numerical results are given. One should note that this,
as well as many other numerical issues in the model, are technically very demanding.
Finally, Section 6 contains a discussion of what are the effects of making the quark mass
nonzero. Several detailed considerations are collected in Appendices.

\subsection{V-QCD at zero temperature}
In \cite{jk} a class of bottom-up holographic models was introduced (named V-QCD)
and shown to be in the universality class of QCD in the Veneziano limit
at zero temperature and density.
These were 5-dimensional models of two scalars coupled to gravity.
One of the scalars, the ``dilaton" $\l$,  is dual to ${\rm Tr}[F^2]$ (the QCD gauge
coupling constant, or more precisely the 't Hooft coupling).
The other scalar, the ``tachyon" $\tau$, is dual to the quark mass
operator $\bar q q$.
The potentials and interactions were modeled along successful bottom-up models for YM,
namely Improved Holographic QCD (IHQCD) \cite{YM1,YM2,YM3}
and the idea that string theory tachyon condensation describes chiral symmetry breaking
\cite{ckp, ikp, sstachyon}.

The bulk action considered was
 \be
S=S_g+S_f,\quad  S_{g}=M^3N_c^2\int d^5x\sqrt{-g}\left[R-{4\over 3}
{(\p \l)^2\over \l^2}+V_g(\l)\right],
\label{i1} \ee
 with $\l$ the 't Hooft coupling (exponential of the dilaton $\phi$)
  and the tachyon\footnote{We have taken the tachyon to be real and diagonal
in flavor space.}
  action\footnote{To find the vacuum (saddle point) solution we have set the gauge fields
 $A_{\m}^{L,R}$ dual to the QCD currents to zero, as they are not expected
to have vacuum expectation values at zero density.
We have also suppressed the Wess-Zumino
terms as they also do not contribute to the vacuum solution.}
  \be
  S_{f}=-x_fM^3N_c^2\int d^5x~ V_f(\l,\tau)\sqrt{\det(g_{\m\n}
  +\f(\l)\partial_{\m}\tau\partial_{\nu}\tau )}.
\label{i2}  \ee

The pure glue potential $V_g$ has been determined from previous studies \cite{YM2}.
The tachyon potential $V_f(\l,\tau)$ must satisfy some basic properties
determined by the dual theory or by general properties of tachyons
 in string theory: (a) To provide the proper dimension for the dual
 operator near the boundary (b) To exponentially vanish for $\tau\to \infty$.
The function $\f(\l)$ captures, among other things, the transformation from the string frame to the
Einstein frame in five dimensions.
The class of potentials that were investigated in \cite{jk} are of the form
\be
V_f(\l,\tau)=V_{0f}(\l)e^{-a(\l)\tau^2}.
\label{i3}\ee
In the Veneziano limit, the back-reaction of the flavor sector on the glue sector
is fully  included.

As with IHQCD, it was arranged that the theory is
asymptotically AdS in the UV
up to logarithmic corrections in the bulk coordinate.
The function $V_{0f}(\l)$ is such that the potential $V_g(\l)-x_fV_{0f}(\l)$,
 when the tachyon has not condensed ($\tau=0$) has an extremum
 \footnote{The extremum
 may exist for all $0<x_f<{11\over 2}$ or may disappear at some small $x_f$.
  No changes in the phase structure
 at zero temperature for these two cases were found in \cite{jk}.} at a finite value $\l=\l_*$.
 As we approach the Banks-Zaks region \cite{bz}, $x_f\to {11\over 2}$, the value
 of $\l_*$ approaches zero.
 Without the tachyon, $\tau=0$, the equations of motion imply that also $\beta(\l_*)=0$,
 i.e., $\l_*$ is an IR fixed point. When the dynamics of $\tau$ is included, the system
 approaches $\l_*$ but is driven away from it as long as $x_f<x_c$ (see Fig.7 of \cite{jk}).

 The dimension of the chiral condensate was calculated in
the IR fixed point theory from the bulk equations. It was found that it decreases
monotonically with $x_f$ for reasonably chosen potentials.
It crossed the value 2 at $x_f=x_c$
where $x_c$ corresponds to the end of the conformal window as argued in
\cite{Kaplan}.

The lower edge of the conformal window $x_c$ lies in the vicinity of 4.
Requiring the holographic $\beta$-functions to match with QCD in the UV, we find
that
\beq
 3.7 \lesssim x_c \lesssim 4.2,
\eeq
where the bounds are not strict but hold approximately for potentials that have smooth
$\l$-dependence in the UV.

There is also a phenomenological heuristic argument for the value $x_c\approx4$, simply
from counting degrees of freedom. At low $T$ chiral symmetry is broken and
the massless degrees of freedom are $N_f^2$ Goldstone bosons. At large $T$ there
are $2N_c^2+\fra72 N_cN_f$ weakly coupled degrees of freedom. These numbers are
equal for $x_f=4$.
Conformal window and the location of its edge was also discussed within holographic
frameworks related to V-QCD in \cite{cw,kajantieIRFP}.

Apart from $x_f$, there is a single parameter in the theory, namely ${m\over
\Lambda_\rmi{QCD}}$ where $m$ is the UV value of the (flavor independent)
quark mass.
For
each value of $x_f$,  the bulk equations were solved with fixed sources corresponding
to fixed
$m, \Lambda_\rmi{QCD}$.
The vevs were determined
such that the solution is
``regular''
in the IR. The notion of regularity is tricky even in the case of  IHQCD (pure
glue),  as there is a naked singularity in the far IR. For the dilaton this has been
 settled in \cite{YM1,YM2}. For the tachyon the notion of regularity is
different and has been studied in detail  in \cite{ikp}.

The regularity condition was implemented in the IR. After solving the equations
 from the IR to the UV (this
was
done mostly numerically), there is a
single parameter that determines the
 solutions as well as the UV
coupling constants and vevs, and this is a real number $\tau_0$
controlling the value of the tachyon in the IR. This number reflects the single
dimensionless parameter ${m\over
\Lambda_\rmi{QCD}}$ of the theory.

For different values of $x_f$ and $m$ the following qualitatively
different regions were found in \cite{jk}:
\begin{itemize}

 \item When $x_c \leq x_f <11/2$ and $m=0$, the theory flows to an IR fixed
point.
The IR conformal field theory 
is weakly coupled near $x_f={11\over 2}$ and strongly coupled in the
vicinity of $x_c$.
Chiral symmetry is unbroken in this regime (this is known as the conformal
window).

 \item When $x_c \leq x_f <11/2$ and $m\not=0$, the tachyon has a nontrivial
profile,
 and there is a single solution with the given source, which is ``regular'' in
the IR. The IR theory is a theory with a mass gap.

 \item When $0<x_f<x_c$ and $m=0$, there is an infinite number of regular
solutions with nontrivial tachyon profile, and a special solution with an
identically vanishing tachyon and a nontrivial  IR fixed point.
The infinite number of solutions with nontrivial tachyon are classified by their number of zeros.
The solution with the lowest free energy is the one with no zeros.

 \item When $0<x_f<x_c$ and $m\not=0$, the theory has vacua with nontrivial
profile for the tachyon. For every non-zero $m$, there is a finite number of
regular solutions that grows as $m$ approaches zero.

\end{itemize}

In \cite{jk} two large classes of tachyon potentials were
identified.
Potentials in class I,
have $a(\l)$ constant in (\ref{i3}). In this case the tachyon diverges exponentially in the IR
for the regular solution
\be
\tau\underset{r\to \infty}{\sim} \tau_0\exp\left[Cr\right],
\label{tau0_intro}
\ee
where $C$ is a known constant (see Appendix B) and $\tau_0$ is the only integration
constant controlling the solution. It determines the source (mass) in the UV.
Potentials in class II, have $a(\l)\sim \l^{2\over 3}$ as $\l\to\infty$, and a
tachyon that diverges in a milder way in the IR as
\be
\tau\underset{r\to \infty}{\sim} C\sqrt{r-{r_1}},
\ee
where again $C$ is known and $r_1$ is the single integration constant controlling the regular solution.
The qualitative conclusions above and below were valid for both classes of potentials.

\begin{figure}
\centering
\includegraphics[width=0.45\textwidth]{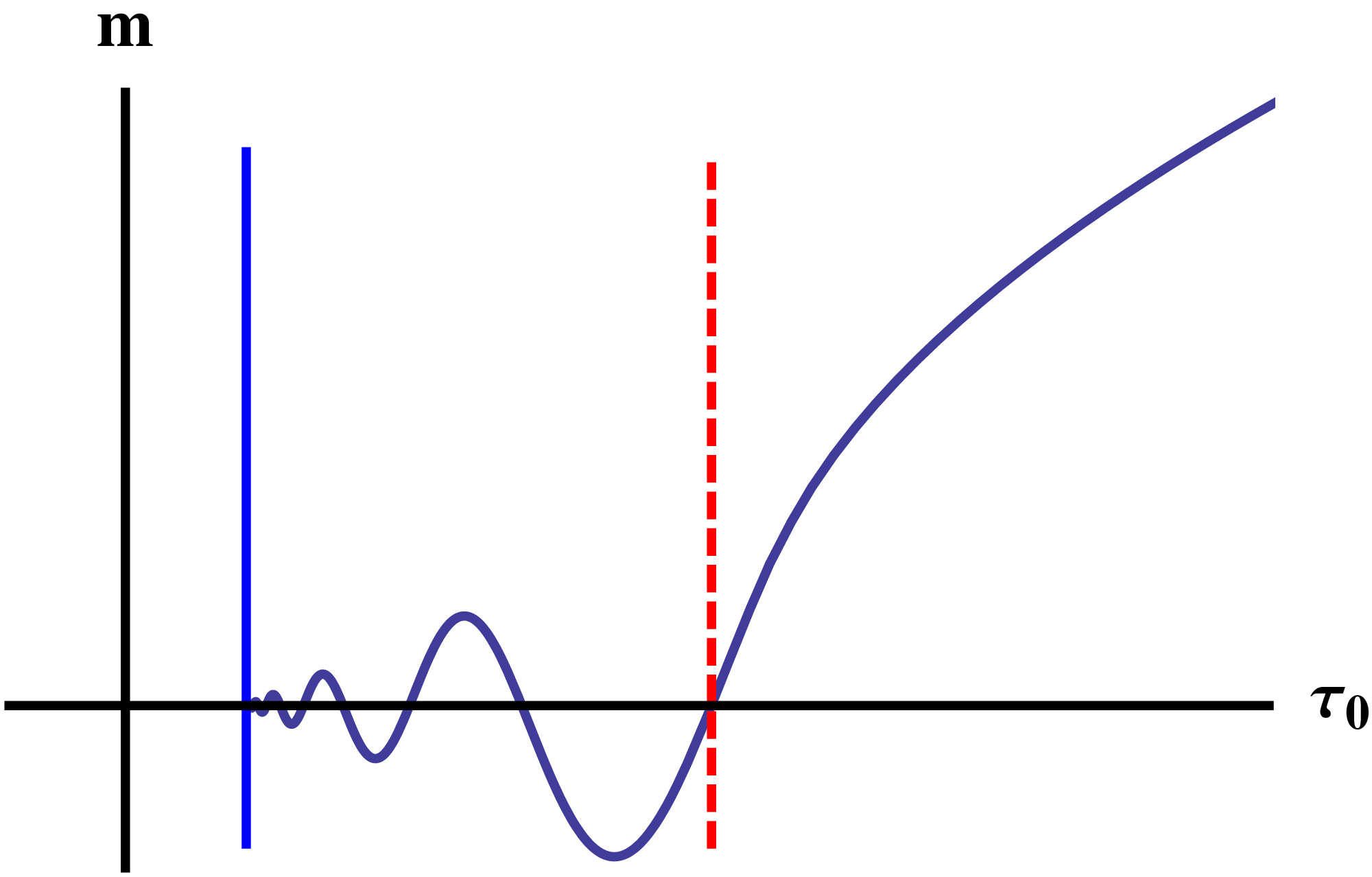}%

\caption{Plot of the UV Mass parameter $m$,  as a function of the
IR
scale
$\tau_0$
in \protect\nr{tau0_intro}, for $x_f<x_c$. The vertical solid blue line marks
the end-point of the existence of regular solutions. The dashed
red line indicates the appearance of more than one regular solution with the same value of $m$.
}
\label{mT0}
\end{figure}

In the region $x_f<x_c$ where several solutions exist, there is a interesting
relation between the IR value $\tau_0$
controlling
 the regular solutions, and the UV parameters, namely $m$.
This is determined numerically, and a relevant plot describing the
 relation between $m$ and $\tau_0$ at fixed $x_f$ is in Fig.~\ref{mT0}.

The solutions are characterized by the number of times $n$ the tachyon field
changes sign
 as it evolves from the UV to the IR. For all values of $m$ there is a single
solution with no tachyon zeroes. In addition, for each positive $n$ there are
two
solutions\footnote{As $m$ and $-m$ are related by a chiral rotation by $\pi$, we can take $m\geq 0$.}
 which exist within a finite  range $0<m<m_n$, where the limiting
value $m_n$ decreases with increasing $n$, and one solution for $m=0$.
In particular, for large enough fixed $m$, we find that only the
solution without tachyon zeroes exists.

For $m\not =0$, out of all regular solutions, the ``first'' one
without tachyon zeroes has the smallest free energy.
The same is true for $m=0$, namely the solution with nontrivial tachyon without
zeroes is energetically favored over the solutions with positive $n$ as well as
over the special solution with identically vanishing tachyon, which appears only
for $m=0$ and would leave chiral symmetry unbroken.
Therefore, chiral symmetry is broken for $x_f<x_c$.

In the region just below $x_c$,  \cite{jk} found Miransky
scaling for the chiral
condensate.
As $x_f \to x_c$,
\be
 \sigma =\langle\bar q q\rangle \sim \Lambda_\rmi{QCD}^3
\exp\left(-\frac{2 \hat K}{\sqrt{x_c-x_f}}\right) .
 \label{i4}\ee
For $x \geq x_c$, let $m_\rmi{IR}(x)$ be the mass of the tachyon at the IR
fixed point and  $\ell_\rmi{IR}(x)$ the IR AdS radius.
The coefficient $\hat K$ is then fixed as
\be
\hat K = \frac{\pi}{\sqrt{\frac{d }{d x}\big[m^2_\rmi{IR}\ell^2_\rmi{IR}
\big]_{x=x_c}}} \ .
 \label{i5}\ee

The behavior at and below the
conformal 
transition at $x_f=x_c$
is to a large extent
independent of the details of the model. In particular, no
information on the nonlinear terms in the tachyon EoM is needed or how the IR boundary
conditions are fixed.
In the same region, ``walking'' of gauge coupling is realized. The YM coupling
flows from small values to values very near $\l_*$, remains approximately constant
for many e-foldings of energy (in this regime the tachyon remains small),
and then runs off to infinity, driven by a large value of the tachyon field in the IR.
The walking is related to a long section of the solution which is similar to the one
studied in earlier bottom-up models for walking \cite{walkingbu}.

%

The finite temperature analysis of V-QCD amounts to studying all black hole
solutions with appropriate boundary conditions.
To start with, any zero temperature solution becomes a candidate saddle point at
finite temperature by compactifying time on a circle of radius $\beta$.
Any other competing black hole solution must have the same boundary conditions
as well as a regular horizon in the IR.

As the dilaton always has a nontrivial UV source, it will always have a nontrivial
profile in the black-hole solutions.
With the tachyon, things can be different. In the massless case, its source is zero.
Therefore there are two possible options
(as in the zero temperature configurations discussed above):
either it is identically zero
(if the vev $\langle\bar q q\rangle$ is also zero)
or it is non-zero (implying a non-zero vev).

Therefore we have two large classes of black holes in the massless case: those with
$\tau=0$ and those with $\tau\not=0$. We will first consider the tachyon-free class.

\subsection{Black holes without tachyon hair}

If $\tau=0$, we have
black holes in a single scalar theory, with potential $V(\l)=V_g(\l)-x_fV_{0f}(\l)$ from (\ref{i3}).
This is a potential with an extremum
for
$x_f\not=0$\footnote{The extremum may also exist only for $x_f$ above some fixed
$x_*$, see the discussion further below.}
and no extremum when $x_f=0$.

Black hole solutions for such potentials were discussed in generality in \cite{YM2}.
After fixing all invariances, they are characterised by a single IR constant, $\l_h$,
the value of the dilaton at the horizon.
The plot relating the temperature $T$ to $\l_h$  contains
important information about  the thermodynamics of such black holes.
Small values of $\l_h$ denote large black-holes whereas larger  values of $\l_h$
correspond to smaller black holes  (smaller horizon size and entropy).
In all plots of this paper, dilatonic black holes without tachyon hair are denoted by
red lines in the respective
$(\l_h,T)$-diagrams, and we shall call the corresponding function $T_u(\l_h)$.

When $x_f=0$, $\l$ can become arbitrarily large at zero temperature,
implying that $\l_h$ can also be arbitrarily large for the finite temperature configurations.
At finite temperature there are two branches:
 large black holes which are stable and small black holes which are unstable.
 If $T_u'(\l_h)<0$
 the black-hole thermodynamics is stable,
 otherwise it is unstable.
There is a minimum temperature above which such black holes exist as shown, for example,
by the black line in  Fig.~\ref{figPotxf0limit} (left or right).

When $x_f>0$, we have two possibilities. The first is that the potential
$V_\rmi{eff}(\l)
=V_g(\l)-x_fV_{0f}(\l)$ has an extremum at
$\l\to \l_*(x)$ for all $0<x_f<{11\over 2}$, with
$\l_*\left(x_f\to {11\over 2}\right)\to 0$ and $\l_*(x_f\to 0)\to \infty$.
The second is that such extremum
only exists for $x>x_*$, where $x_*<x_c$.
We shall
 denote these potentials with a star subscript.

At finite temperature, and when the potential $V_\rmi{eff}$ has no extremum, the black hole
without the tachyon hair exists for all positive $\l_h$. For the potentials studied here,
function $T_u(\l_h)$ is qualitatively similar to the YM case ($x_f \to 0$) \cite{YM2}.
As shown in Fig.~\ref{figPhasesModPotII} (top-left) and in Fig.~\ref{figPhasesModPotI} (left),
there are two black hole branches, which exist above some minimum temperature.
The branch at low $\l_h$ is thermodynamically stable, while the large-$\l_h$ branch is unstable.

When the extremum is present,
$0<\l_h<\l_*(x)$.   The temperature of the black-hole corresponding to
 $\l_h=\l_*(x)$ is $T=0$, while that of $\l_h\to 0$ has $T\to \infty$.
There is no minimum temperature here. For any temperature there is always at least one black-hole solution.
 There are several possibilities that are shown as red lines in Figs.~\ref{figTlah} (left),
 \ref{figTransitionsTend} (top), \ref{figTransitionsTs} (left) and
 \ref{figTransitionsUnstab} (left).

When $x_f$ is large, but still smaller than $x_c$, the $T=T_u(\l_h)$
relation is one-to-one but contains a bump (a change of concavity)
 like in Fig.~\ref{figTransitionsTend} (top). Then this is accompanied by a crossover behavior,
 signaled by a bump in the trace of the stress tensor $(\epsilon-3p)/T^4$, (aka interaction measure)
 as shown in Fig.~\ref{figTransitionsTend} (bottom-right).

At low enough $x_f$, the relation $T=T_u(\l_h)$ 
is not always one-to-one,
 as can be seen in Fig.~\ref{figTransitionsTs} (left) or in Fig.~\ref{figPotxf0limit}.
Then there are points where $T_u'(\l_h)=0$. 
 In such a case there can be a first order transition between
 the stable branches of the black hole solutions.
 This is a remnant of the deconfining transition at $x_f=0$ (pure YM).
 In Fig.~\ref{figPotxf0limit} both left and right several curves in the $(T,\l_h)$-plane
 for different $x_f$ indicate the successive structure of dilaton black holes (red lines).
 The black line corresponds to the pure YM ($x_f=0$) limit.

 When $x>x_c$ we are in the conformal window. The only black holes that exist here are those
 without tachyon hair.
  The relation $T=T_u(\l_h)$ 
is monotonic and there is a
 continuous transition to the black-hole phase at $T=0^+$, as in the AdS case in the Poincar\'e patch.
 The thermodynamic functions, especially the interaction measure,
   show a crossover maximum at a temperature that is moving towards the UV as $x_f\to \fra{11}2$.

\subsection{Black holes with tachyon hair and zero quark mass}

When $\tau\ne 0$ we have black holes in the two scalar theory.
The tachyon starts as $\sim r^3$
near the UV boundary as the source (quark mass) vanishes.
In all plots of this paper, such black holes (with both dilaton and tachyon hair)
are denoted by blue lines in the respective
$(\l_h,T)$-diagrams,
and we shall denote the corresponding functions by $T_b(\l_h)$.
They are still one parameter solutions and can be parametrized again by the value $\l_h$ of $\l$ at the horizon,
which translates into the temperature.
These black holes
usually
exist for all $x_f\in\, ]0,x_c[$
and our discussion below focuses in this region.

Because the presence of the nontrivial tachyon perturbs and annuls the
possible nontrivial IR fixed point,
for such black-holes,  $\l_h$ can take arbitrarily large values.
This can be seen from the blue lines in Figs.~\ref{figTlah} (left),
\ref{figTransitionsTend} (top), \ref{figTransitionsTs} (left) and
 \ref{figTransitionsUnstab} (left). For all such black holes, the chiral condensate is
 determined by the regularity of the black hole solution.
 It decreases as $\l_h$ decreases, and at some point it vanishes.
 At this point, the blue line in the
 ($\l_h,T$)-diagram merges
  with the red line corresponding to a $\l_h$ that we call $\l_\rmi{end}$ throughout the paper.
 This can be seen in all the figures mentioned above.

The shape of the blue line can vary as a function of $x_f$ and the type of potential.
There are three typical examples of shapes:
\begin{itemize}

\item Simple lines that are monotonic as the one depicted in Fig.~\ref{figTransitionsUnstab} (left).
This is an example of a monotonic blue branch where all such black-holes are thermodynamically unstable.
Moreover, they have a minimum temperature.
In such a case, they can never be thermodynamically dominant. At some temperature the vacuum thermal
solution is dominated by a dilaton black hole on the red line, and the
chiral restoration transition is 1st order.

\item Lines with two branches as the one depicted in Fig.~\ref{figTransitionsTs} (left). Here the blue
line has two parts one (to the left) that is thermodynamically stable and another
to the right that is thermodynamically unstable.  In such a case, the system is in the thermal vacuum
solution at low enough temperatures, then jumps with a 1st order transition to the tachyon-hairy
solution (the part of the blue line that is thick in Fig.~\ref{figTransitionsTs} (left)) which
still break chiral symmetry, and then eventually smoothly transits to the red line at the point
where the blue and red lines merge, via a chirally-restoring 2nd order
transition.\footnote{It may also happen that the thermodynamically stable branch is only metastable,
in which case the system jumps directly to the black hole branch without tachyon hair, and chiral
symmetry is thus restored at this 1st order transition.
The more complicated branch structure discussed in the next bullet may similarly contain metastable branches.}

\item Lines with more than two branches as the one depicted in Fig.~\ref{figTransitionsT12} (left).
In this example the blue line has four branches, two unstable and two stable.
There are in total three phase transitions here, first from the vacuum thermal solution to the
rightmost blue thick branch, then to the intermediate thick blue branch and finally a 2nd order
chirally restoring transition to the red branch at the point they touch. In this case there are two
1st order transitions between three
chirally breaking phases, and a 2nd order one to the chirally symmetric phase.

\end{itemize}

\begin{figure}[!tb]
\begin{center}

\includegraphics[width=0.6\textwidth]{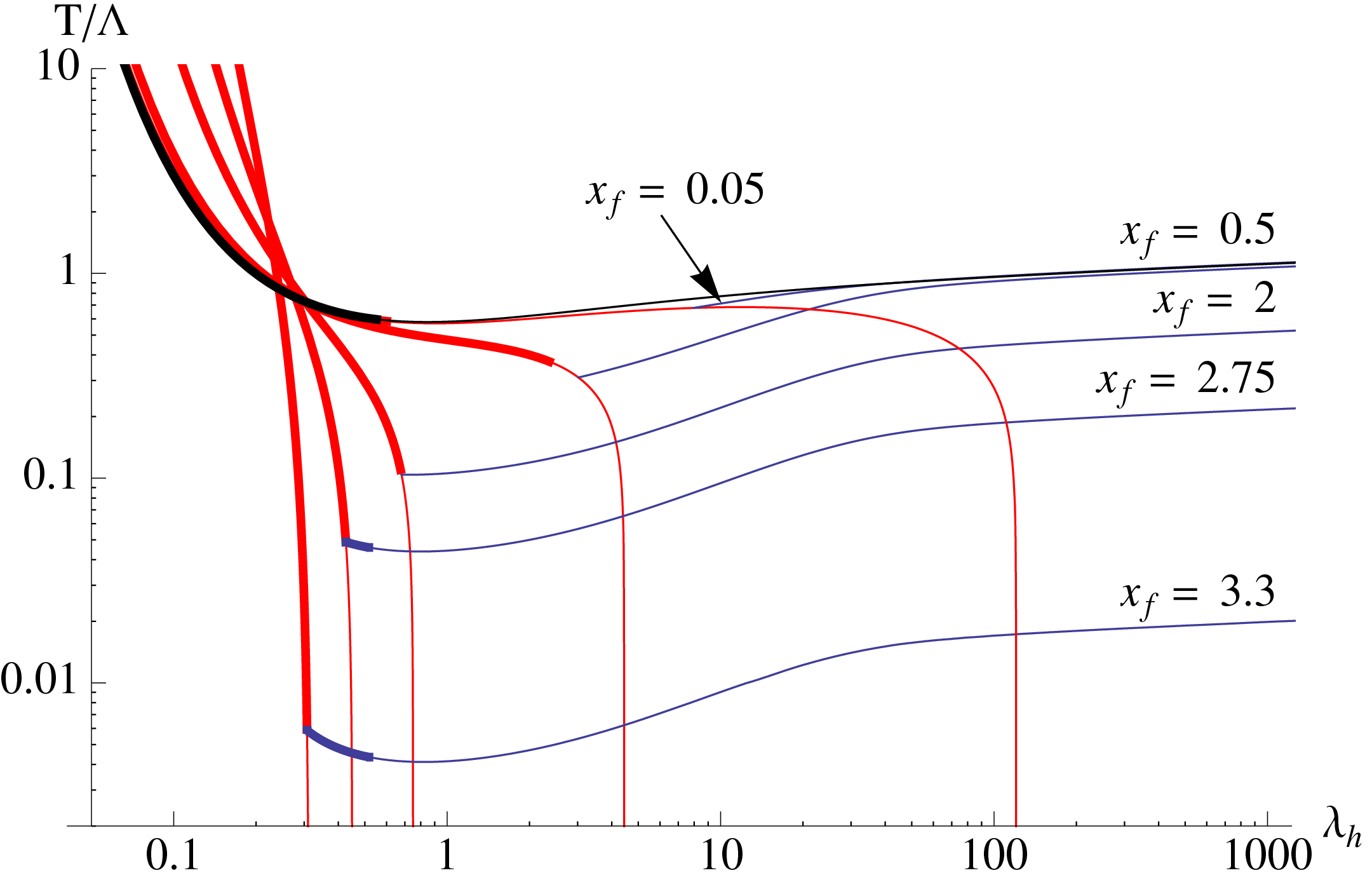}\hfill

\end{center}

\caption{\small $T(\l_h)/\Lambda$ plotted for potentials of type II with SB normalisation (definitions
specified in text) for various $x_f$ marked on figure. Thick curves represent stable phases. For
more details at small $x_f$, see Fig.~\protect\ref{figPotxf0limit}.
}
\label{figxfscan}
\end{figure}

A concrete overall view of the $x_f$ dependence is presented in Fig.~\ref{figxfscan}, in which
$T(\l_h)$ is plotted for potentials of type II with SB normalisation (definitions
specified later) for various $x_f$. One sees clearly how the pure (black) YM curve is
approached for $x_f\to0$. The thick curves represent stable phases; when a
thick curve ends, the system makes a 1st order transition to the low $T$ phase. When
thick curves change from red to blue curves, a 2nd order transition to a chirally
broken phase takes place. For a more accurate picture of small $x_f$, see Fig.~\ref{figPotxf0limit}.

\subsection{The phase structure of different V-QCD models}

There are three main ingredients that characterize a priori different QCD models which, however, have
the same phase structure and qualitative behavior at zero temperature:

\begin{itemize}

\item The asymptotics of the tachyon solution in the IR. This is controlled by the behavior of the
function $a(\l)$ in the tachyon potential in (\ref{i3}).
When $a(\l)$ is constant, the tachyon diverges exponentially in the IR, and we call such potentials
of type I. When $a(\l)$ diverges as $\l^{2\over 3}$ in the IR ($\l$ large)
then the tachyon diverges as a square root in the IR, and we call such potentials of type II.

\item For any potential,  the UV constant factor $W_0$ of $V_{0f}(\l)$ in (\ref{i3}), defined in
(\ref{Vcoeffeq}) can vary in finite range, which in appropriate units is
$]0,{24\over 11}]$,
as
in (\ref{LUVSB1}). We pick for each type of potential three
indicative values of ${\cal L}_0^2W_0$
that in general might give different physics, namely $0$, ${12\over 11}$, and
${24\over  11}$.\footnote{Notice that the exactly zero value of $W_0$ is actually excluded,
because it predicts wrong anomalous dimensions for quark mass or the chiral condensate \cite{jk}.
We anyhow consider it as the limiting case of the allowed solutions. Moreover, $W_0$ may exceed
the upper limit of $24/11$, if $x_f$ dependence is allowed.}
We also
consider $x_f$-dependent value, specified in (\ref{LUVSB}) that corresponds to the normalization
of the UV degrees of freedom of the free energy to the Stefan-Boltzmann limit of QCD.

\item A final variation can be obtained on all of the above by using a glue potential
$V_g(\l)-x_fV_{0f}(\l)$ in (\ref{i3}) that has

(a) an extremum
for all $x_f$ in the appropriate range,
$x_f\in ]0,{11\over2}[$.

(b) an extremum
only in part of this range,
$x_* < x_f < {11\over 2}$.
We will
denote the potentials in this case by a star subscript.

\end{itemize}

According to the above options PotI$_*(W_0=0)$ denotes a potential in the type I class, with $W_0=0$
and an IR critical point that exists only down to a finite $x_*$.

Let us then summarize the phase structure of the model as $x_f$ and the temperature are varied
(at zero quark mass). In general one expects the phase diagram of
Fig.~\ref{figxTphases},
so that for $0<x_f<x_c$ there is the 1st order
transition at finite
temperature,
which also separates the chirally symmetric and broken phases.
For $x_f>x_c$  the low temperature and high temperature
 configurations correspond to a tachyonless black holes, and,
one expects a
continuous crossover between these two.

For the various potentials presented above, this phase diagram is indeed obtained
in the zeroth approximation, but for $x_f<x_c$ there are additional details which
depend on the choice of potentials as follows.
\begin{itemize}
 \item For potentials I the phase structure depends strongly on the choice for $W_0$
 (see Fig.~\ref{figPhasesPotIW0}). For the lowest value $W_0=0$, there is only one 1st
 order transition at\footnote{$T_d=T_h$ in Fig.~\ref{figPhasesPotIW0}.}
$T=T_d$
for all $0<x_f<x_c$, except possibly for $x_f$ very close to
 $x_c$, where solving the phase diagram numerically becomes demanding.
As $W_0$ is increased, a complicated structure
 appears near $x_f=x_c$, where we have two 1st order transitions between chirally broken
 phases, and the restoration of chiral symmetry at a 2nd order transition at even higher
 temperature.
At even higher $W_0$ the 1st order transitions combine again into a single one, but the separate
2nd order transition continues to exist for $x_f$ close to $x_c$.
At low $x_f$, there is also a surprising change as $W_0$ increases. The chiral
 symmetry breaking phases disappear, but there is a 1st order transition between two chirally
 symmetric
black hole phases at a finite temperature instead.

 \item For potentials II the dependence on $W_0$ is milder
 (see Figs.~\ref{figTTransitionsSB} --~\ref{figPhasesPotIIW0_0}).
At high $W_0$,
for low $x_f$ up to some
 value $x_\chi$, there is only the 1st order transition at\footnote{$T_\chi=T_\rmi{end}$ of
 Figs.~\ref{figTTransitionsSB} --~\ref{figPhasesPotIIW0_0} when it is in the stable brach.}
$T=T_d$.
When $x_\chi<x_f<x_c$,
 the chiral symmetry restoration takes again place at a 2nd order transition at
$T_\chi$
 such that $T_\chi>T_d$.
For decreasing $W_0$, $x_\chi$ increases, and finally disappears by joining with $x_c$.

 \item For the potentials I$_*$, the phase structure is the standard one for high $x_f$,
 i.e., a 2nd order transition and a 1st order one with critical temperatures
 $T_\chi>T_d$
within a range $x_\chi<x_f<x_c$, with the former separating the
 chirally symmetric and broken phases (see Fig.~\ref{figPhasesModPotI}). For lower $x_f$
 there is only one 1st order transition. For $x_f \lesssim 2$, in the region where the
 effective potential does not admit an extremum,
 chiral symmetry is intact at all temperatures.
We find a single 1st order transition between chirally symmetric thermal gas and black hole phases.

 \item For potentials II$_*$,  the phase structure is
simple
 (see Fig.~\ref{figPhasesModPotII}):
there is a single 1st order transition for all $x_f \in ]0,x_c[$.
In particular, the system is in a
 chirally broken phase at low temperatures, even in the region of low $x_f$ where
the effective potential does not have an extremum.

\end{itemize}

\section{Defining V-QCD}

\subsection{Gravity action of the model}
The action of V-QCD is \cite{jk}
\be
S=M^3N_c^2\int d^5x\,\CL\equiv {1\over16\pi G_5}\int d^5x\,\CL,
\label{S}
\ee
where\footnote{Notice that for notational simplicity we have absorbed a factor of $x_f$,
which is visible in Eq.~\eqref{i2}, into $V_f(\l,\tau)$. See also Eq.~\eqref{Vf0def} below.}
\ba
\CL&=&\left[\sqrt{-g}\left(R-{4\over3}{
(\partial\lambda)^2\over\lambda^2}+V_g(\lambda)\right)
-
\,V_f(\lambda, \tau)\sqrt{\det\left(g_{ab}+\f(\lambda, \tau) \partial_a \tau\,\partial_b
\tau\right)}\right]\nn\\
&=&\sqrt{-g}\biggl[ R+\left[-\fra43g^{\mu\nu}\partial_\mu\phi\partial_\nu\phi+V_g(\lambda)\right]
-\, V_f(\lambda,\tau)\sqrt{1+g^{rr}\kappa(\lambda(r)) \tau'(r)^2}\biggr].
\label{lagrang}
\ea
The metric Ansatz is
\be
ds^2=b^2(r)\left[-f(r)dt^2+d{\bf x}^2+{dr^2\over f(r)}\right],
\quad b(r)=e^{A(r)} \underset{r\to0}{\sim}{\CL_\rmi{UV}\over r},
\label{bg}
\ee
and the two scalar functions, $1/\l$ sourcing $F^2$ and $\tau$ sourcing $\langle \bar q q\rangle$, are
\be
\lambda=\lambda(r)=e^{\phi(r)},\quad \tau=\tau(r).
\ee
In the second form $\sqrt{-g}$ has been factored out of the DBI action.
The Gibbons-Hawking counter term would be
\be
S_{GH}=-\int d^4x\,\sqrt{-\gamma}\,\biggl[2K+{6\over\CL_\rmi{UV}}+{\CL_\rmi{UV}\over2}R(\gamma)\biggr],
\ee
with, for a hypersurface $r=$const,
\be
K={\sqrt{f}\over 2b}\biggl(8{ b'(r)\over b}+{f'(r)\over f}\biggr).
\label{extK}
\ee
Notice also that we have set the gauge fields $A_{L,R}$, which are dual to the left and right handed
fermion currents, to zero, and neglected the Wess-Zumino terms.
These terms do not affect the thermodynamics of the models.

The background solution of the dilaton $\l(r)$ and the warp factor $A(r)$ are identified as the 't Hooft
coupling and the logarithm of the energy scale of the dual field theory, respectively \cite{YM1}.
As a matter of convention, we shall fix the normalisation of
$\lambda(r)$
so that its relation to the perturbative QCD coupling is
\be
\lambda(r)={g^2(b(r))\over 8\pi^2}.
\ee
The results of the model are independent of this normalisation, changing $\l\to\l/\l_0$
one simply has to change the potentials by $V(\l)\to V(\l_0\l)$.
The convention of \cite{jk}, for example, is obtained by shifting by $\l_0=1/(8\pi^2)$.

Important ingredients of the model are the relation of the bulk fields at $r$ to the
QCD beta and quark mass anomalous dimension functions evaluated for a coupling at scale
$b(r)$.
Motivated by the connection to field theory,
one defines
\be
\beta={d\l\over db/b}=\l'(A)=-b_0\l^2-b_1\l^3-b_2\l^4\dots,\quad \gamma=\tau'(A).
\label{defbeta}
\ee
Matching with the perturbative expansion of the QCD beta function gives
\be
b_0=\fr13(11-2x_f),\quad b_1=\fr16(34-13x_f).
\ee

The
action
contains the gluonic potential $V_g(\l)$ and the fermionic potential $V_f(\l,\tau)$,
which will be specified to the form
\be \label{Vf0def}
 V_f(\l,\tau) = x_fV_{f0}(\l)e^{-a(\l)\tau^2} \ .
\ee
The detailed form of these
and the functions $\kappa(\l),\,a(\l)$ will be discussed in the following subsections.

\subsection{Construction of the potentials}
The potentials can be constructed in stages.
First one fixes the potentials $V_g(\l)$ and $V_{f0}(\l)$
up to order $\lambda^2$ in the UV, using the two scheme independent coefficients of
the beta function.
This analysis is simplified by the fact that the tachyon decouples in the UV.
Next one fixes the UV behavior of the functions $a(\l)$ and $\f(\l)$, which parametrize the
tachyon dependence of the action
using the similarly scheme independent UV running
properties of the quark mass and the condensate. Finally, one fixes the
large $\lambda$ behavior of the potentials
by requiring that the model reproduces known features of QCD in the IR, such as confinement,
linear Regge trajectories, and reasonable zero-temperature phase structure.
We shall discuss the various steps in detail below (see also \cite{jk}).

\subsubsection{The potentials from the beta function in the UV}
In the UV, since the tachyon vanishes much faster than the dilaton, we can first set it to zero.
Then the dilaton profile can be linked to the effective potential
$V_\rmi{eff}(\l)=V_g(\l)- x_f V_{f0}(\l)$ \cite{jk} by using Einstein's equations \cite{YM1}.
Defining
$\beta=d\lambda/d\ln b=-b_0\lambda^2-b_1\lambda^3$, to order $\lambda^2$,
\ba
V_g-x_f V_{0f} &=&
{12\over \CL_\rmi{UV}^2}
\exp\biggl[-\fr89\int_0^\lambda d\lambda{\beta\over\lambda^2}\biggr]
\biggl(1-{\beta^2\over9\lambda^2}\biggr)\\
&=&{12\over \CL_\rmi{UV}^2}\biggl[1+\fr89b_0\lambda+(\fr{23}{81}b_0^2+\fr49b_1)\lambda^2
\biggr]\label{bcoeffeq} \\
&=&V_0-x_fW_0+(V_1-x_fW_1)\lambda+(V_2-x_fW_2)\lambda^2, \label{Vcoeffeq}
\ea
where we expanded
\be
V_g=V_0+V_1\lambda+V_2\lambda^2 +\mathcal{O}(\l^3),\quad V_{f0}=W_0+W_1\lambda+W_2\lambda^2 +\mathcal{O}(\l^3),
\ee
and where we have introduced an $x_f$ dependent AdS radius
\be
\CL_\rmi{UV}=\CL(x_f).
\ee

Applying equation~\eqref{bcoeffeq} for $x_f=0$ we have for the gluonic potential
\ba \label{Vgluonic}
V_g&=&{12\over \CL_0^2}\biggl(1+\fr89 b_0^{\rmi{YM}}\lambda+
{23(b_0^{\rmi{YM}})^2+36b_1^{\rmi{YM}}\over81}\lambda^2\biggr)\\
&=&{12\over \CL_0^2}\biggl(1+\fr{88}{27}\lambda+\fr{4619}{729}\lambda^2\biggr),
\ea
where $b_i^{\rmi{YM}}$ are the values of $b_i$ for $x_f=0$ and $\CL_0 = \CL(x_f=0)$.
In practice, one usually sets the (dimensionful) quantity $\CL_0=1$.

By using equations~\eqref{bcoeffeq} and~\eqref{Vcoeffeq}
one can now solve for the coefficients of the fermionic potential:
\be \label{WW0}
x_f\CL_0^2W_0=12\biggl(1-{\CL_0^2\over\CL_\rmi{UV}^2}\biggr),
\ee
\be
 x_f\CL_0^2W_1=\fr{32}3
\biggl(b_0^{\rmi{YM}}-b_0{\CL_0^2\over\CL_\rmi{UV}^2}\biggr)=\fr{12\cdot 8}{27}
\biggl[11-(11-2x_f){\CL_0^2\over\CL_\rmi{UV}^2}\biggr],
\ee
\ba \label{WW2}
x_f\CL_0^2W_2&=&\fr{12}{81}\biggl[23(b_0^{\rmi{YM}})^2+36b_1^{\rmi{YM}}-(23b_0^2+36b_1){\CL_0^2\over\CL_\rmi{UV}^2}
\biggr]\\\nn
&=&\fr{12}{729}\biggl[4619-(4619-1714x_f+92x_f^2){\CL_0^2\over\CL_\rmi{UV}^2}\biggr].
\ea
These equations still involve one free parameter, which can be taken to be either $W_0$ or $\CL_\rmi{UV}$.
We shall study two ways to fix this parameter. First, we can take $W_0$ to be constant. In this case \cite{jk}

\be \label{LUVSB1}
0\le \CL_0^2W_0
\le {24\over11},
\ee
and the $x_f$-dependent AdS radius is given by
\be
 \CL_\rmi{UV}={\CL_0\over\sqrt{1-\fra1{12}\CL_0^2W_0\cdot x_f}}.
\ee
Second, we can make a special $x_f$-dependent choice, which (as we shall show later) automatically
normalises the free energy at large $T$ to Stefan-Boltzmann:
\be
\CL_\rmi{UV}=\CL_0(1+\fra74 x_f)^{1/3}.
\label{LUVSB}
\ee

Further, we have to fix the $\lambda$ dependence of the functions $a(\l)$ and
$\f(\l)$ in the tachyon part
\be
x_f V_{f0}(\lambda)e^{-a(\l)\tau^2}\sqrt{1+g^{rr}\f(\lambda(r))\dot \tau^2},
\ee
of the action,
where $g^{rr}=f/b^2$.
The leading logarithmic term to the UV expansion of the tachyon should be
(remember that the energy dimension of $\tau$ is $-1$)
\be
\tau(r)/\CL_\rmi{UV}
=mr\,(-\ln\Lambda r)^{-\frac{\gamma_0}{b_0}}
=mr\,(-\ln\Lambda r)^{-\fra3{2b_0}}
\ee
to satisfy the scheme independent UV running of the quark mass.
Here $\gamma_0=3/2$ is the leading coefficient of the anomalous dimension of the
quark mass in QCD, $\gamma(\l)= \gamma_0 \l + \cdots$. By using the
tachyon equation of motion one sees that this requires that for small $\l$,
\be
{\f(\l)\over a(\l)}=\fr23\CL_\rmi{UV}^2\biggl[1-\biggl(\fr89 b_0+1\biggr)\l+\l^2+\cdots \biggr].
\label{massanomdim}
\ee

\subsubsection{Large $\lambda$ behavior of the potentials}\label{sec2pt2pt2}
To specify the full potential $V_g(\l)-x_f V_{f0}(\lambda)e^{-a(\l)\tau^2}$
we have to continue the small $\l$ expansions to large $\l$. The guideline is
quark confinement and chiral symmetry breaking at small $x_f$ and the appearance of
an infrared fixed point at some $x_f=x_c$
(see \cite{jk}). Since there is no unique path to the
result, we present the final forms of the potentials we use and motivate them.

We use the gluonic potential
\be
V_g(\lambda)={12\over\CL_0^2}\biggl[1+{88\lambda\over27}+{4619\lambda^2
\over 729}{\sqrt{1+\ln(1+\lambda)}\over(1+\lambda)^{2/3}}\biggr]
\label{VgSB}
\ee
which is constructed from the expansion
\nr{Vgluonic} by simply multiplying the $\l^2$ term by the confinement factor
\be
{\sqrt{1+\ln(1+\l)}\over(1+\l)^{2/3}} \ .
\label{conffactor}
\ee
Then $V_g$ has the proper large-$\l$ behavior \cite{YM1} but the small-$\l$ behavior is left intact.
One could add scale factors of type $\l/\l_0$ containing more parameters.

For the fermionic potential $V_{f0}$ in
\be
V_f (\lambda, \tau) = x_f V_{f0} (\lambda) e^{-a (\lambda) \tau^2}
\label{VfSB}
\ee
we consider two different choices. The first one is obtained directly using \nr{WW0}-\nr{WW2}
\bea\label{Vf0SB}
V_{f0}&=&{12\over\CL_\rmi{UV}^2x_f}\biggl[{\CL_\rmi{UV}^2\over\CL_0^2}
-1+{8\over27}\biggl(11{\CL_\rmi{UV}^2\over\CL_0^2}-11+2x_f\biggr)\lambda\\\nn
&&+{1\over729}\biggl(4619{\CL_\rmi{UV}^2\over\CL_0^2}-4619+1714x_f-92x_f^2\biggr)\lambda^2\biggr].
\eea
Here one could as well use the parameter $W_0$ which is related to $\CL_\rmi{UV}$ by
\be
{\CL_0^2\over\CL_\rmi{UV}^2}=1-{x_f\CL_0^2W_0\over12}.
\ee
For this choice the effective potential
\be
 V_\rmi{eff}(\l) = V_g(\l)-x_f V_{f0}(\l)
\ee
has a single maximum at finite positive $\l=\l_*$ for all $0<x_f<11/2$, indicating a (possible)
infra-red fixed point.

The second choice is obtained  introducing the confinement
factor \nr{conffactor} also for the fermionic potential, i.e.,
\bea\label{Vf0mod}
V_{f0}&=&{12\over\CL_\rmi{UV}^2x_f}\biggl[{\CL_\rmi{UV}^2\over\CL_0^2}
-1+{8\over27}\biggl(11{\CL_\rmi{UV}^2\over\CL_0^2}-11+2x_f\biggr)\lambda\\\nn
&&+{1\over729}\biggl(4619{\CL_\rmi{UV}^2\over\CL_0^2}-4619+1714x_f-92x_f^2\biggr)\lambda^2
{\sqrt{1+\ln(1+\lambda)}\over(1+\lambda)^{2/3}}\biggr].
\eea
Now the effective potential has a maximum only at large $x_f$. To see this concretely,
consider again the case \nr{LUVSB}.
The asymptotic large-$\lambda$ behavior of $V_g-x_fV_{f0}$ now is $\lambda^{4/3}\sqrt{\ln\lambda}$
times the function
\be
{18476\over 243}-4{4619(1+\fra74 x_f)^{2/3}-4619+1714x_f-92x_f^2\over 243(1+\fra74 x_f)^{2/3}}.
\ee
This function is positive for small $x_f$, negative at large $x_f$ ($<11/2$) and has a zero at
$x_f=3.26817$. Thus there is a (possible) fixed point $\l_*$ only for $3.26817<x_f<11/2$.

Let us then discuss the IR behavior of the potentials $a$ and $\f$ which appear in the tachyon DBI action.
For the function $\f$ we will consider the large-$\l$ asymptotics
\be
\f(\l)\underset{\l \to \infty}{\sim} \l^{-4/3} \ .
\ee
This is motivated by the fact
that
in the action the combination $\f(\l)/b^2$
has the same asymptotics as
$1/b_s^2 $ at large
$\l$, where $b_s=b\l^{2/3}$ is the metric factor $b$ in the string
frame.
To ensure that the fractional
exponent limit at large $\l$ does not spoil analyticity at small $\l$, we replace
$\l^{4/3}$ by $(1+\#\l)^{4/3}$ in the expression for $\kappa(\l)$.

More precisely, two qualitatively different, acceptable choices for the IR asymptotics
of $a$ (and $\f$) were identified in \cite{jk}. These are produced by the following two choices.
The first choice has
\be
a (\lambda) = \fr32\,{1\over\CL_{\rm UV}^2},\quad
 \kappa(\lambda) = {1\over [1+\fra34(\fra89 b_0+1)\lambda]^{4/3}}=
  \frac {1} {\left (1 + \frac {115 -
          16 x_f} {36}\lambda \right)^{4/3}},
          \label{akappa}
\ee
and leads to tachyon growing exponentially at large $r$,
\be
\tau(r)\underset{r \to \infty}{\sim}  \tau_0e^{Cr}
\label{exptachyon}
\ee
where $C$ is a known constant (see Appendix~\ref{AppIR}) and $\tau_0$ parametrises the solutions.
The second choice is given by
\be
\kappa(\lambda)={1\over(1+\lambda)^{4/3}},\quad
a(\lambda)=\kappa(\lambda){3\over 2\CL_\rmi{UV}^2}\biggl[1+\biggl(\fr89 b_0+1\biggr)
\lambda+\lambda^2\biggr]
\label{akappa2}
\ee
and for them the leading divergence is
\be
\tau(r)\underset{r \to \infty}{\sim} C\sqrt{r-r_1},
\ee
where the constant $C$ is again known and now $r_1$ parametrises the solutions. To select this solution,
it is required that the last term in the square brackets  in \nr{akappa2} grows faster than $\l^{4/3}$.

Finally, let us summarize our choices for acceptable potentials. We always keep $V_g$ fixed to the
expression~\eqref{VgSB} and choose $V_{f0}$, $a$, and $\f$ as follows:
\begin{itemize}

 \item {\bf Potentials I}: We take $V_{f0}$ as in equation~\eqref{Vf0SB}, so that the fixed point
 $\l_*$ exists for all $0<x_f<11/2$. For $a$ and $\f$ we use the choice of equations~\eqref{akappa},
 which lead to exponentially diverging tachyon in the IR.

 \item {\bf Potentials II}: We take again $V_{f0}$ from equation~\eqref{Vf0SB}, but use the other
 choice \eqref{akappa2} for $a$ and $\f$. Then the tachyon diverges as $\tau \sim \sqrt{r}$ in the IR.

 \item {\bf Potentials I$_*$}: We use now the fermionic potential $V_{f0}$ of equation~\eqref{Vf0mod},
 which contains the confinement factor. Thus the extremum exists only within the
 interval
 $3.26817<x_f<11/2$. For $a$ and $\f$ we use the choice of equations~\eqref{akappa}, which lead to
 exponentially diverging tachyon in the IR.

 \item {\bf Potentials II$_*$}: We use $V_{f0}$ with the confinement factor,
 but use the other choice \eqref{akappa2} for $a$ and $\f$. Then the fixed point exist only
 for large $x_f$, and the tachyon diverges as $\tau \sim \sqrt{r}$ in the IR.
\end{itemize}

To fully pin down the potentials, we also need to specify the value of $W_0$ (or $\CL_\rmi{UV}$)
which is used. We choose four reference values:
\begin{itemize}
 \item $W_0=0$ (and constant). This is the lower bound of $W_0$. Actually, exactly zero $W_0$
 is not acceptable because the anomalous dimensions of the quark mass and the chiral condensate
 do not sum up to zero.
This case is nevertheless interesting as it is the limit of acceptable solutions.
 \item $W_0=12/11$. This is the standard choice studied in \cite{jk}.
 \item $W_0=24/11$. For constant $W_0$, this is the largest possible value, for which
 $\CL_\rmi{UV} \to \infty$ as $x_f \to 11/2$.
 \item $W_0$ (and $\CL_\rmi{UV}$) fixed such that the free energy
automatically
matches with the standard Stefan-Boltzmann
 result at high temperature with the correct number of degrees of freedom
(see Eq.~\eqref{LUVSB} and the discussion in Sec.~\ref{secthermo} below).
\end{itemize}

An ongoing work \cite{mesons} studies the meson spectra in this model. As it turns out,
the potentials I  and I$_*$ admit linear ``Regge'' trajectories, so that the quadratic
masses are asymptotically linear in the excitation number, $m_n^2 \sim n$,
independently of the other quantum numbers.
Potentials II and II$_*$, however, have linear trajectories only in the glueball sector,
while the other trajectories are quadratic,  $m_n^2 \sim n^2$. As linear trajectories are
expected in QCD, this observation favors potentials I and I$_*$.

\subsubsection{IR fixed point and the BF bound for the tachyon}

Now that the potentials are defined, one can check that they satisfy an important requirement:
they permit the determination of the bulk dilaton mass and, equating this with the
Breitenlohner-Freedman (BF)
instability
bound, the determination of the start of the conformal window.
Take $\tau(r)=0$ (there is no chiral
 symmetry breaking in the conformal window) and note that at small $\l$,
 $V_g(\l)-x_f V_{f0}(\lambda)>0$. However, $V_{f0}(\l)$ grows faster and the conformal
 window starts at the value $\l_*$ defined by the vanishing derivative
 \be
 V_g'(\l_*)-x_fV_{f0}'(\lambda_*)=0.
 \label{defoflastar}
 \ee

 Given $\l_*$ one defines an IR AdS radius
 \be
 {12\over \CL_\rmi{IR}^2}= V_g(\l_*)-x_f V_{f0}(\l_*), \quad \CL_\rmi{UV}>\CL_\rmi{IR}.
 \ee
 The tachyon mass at $\l_*$ in units of $\CL_\rmi{IR}$ becomes
 \be
 -m_\rmi{IR}^2\CL_\rmi{IR}^2={24a(\l_*)\over
 \kappa(\l_*)[V_g(\l_*)-x_fV_{f0}(\lambda_*)]}.
 \label{emel}
 \ee
 Gravity solutions
with $\tau=0$
are stable when $m_\rmi{IR}^2\CL_\rmi{IR}^2>-4$;
 the conformal window thus starts when \nr{emel}, as a function of $x_f$, has the value 4.

\begin{table}[ht!]
\begin{center}
\begin{tabular}{ c || c | c | c | c}
 & PotI & PotI$_*$ & PotII & PotII$_*$ \\
\hline
$W_0=0$ & 4.10209  & 4.33334 & 4.17825 & 4.38493 \\
$W_0=12/11$ & 3.99591  & 4.33334 & 4.07968 & 4.38493 \\
$W_0=24/11$ & 3.71607 & 4.33334 & 3.80086& 4.38493 \\
$W_0$ SB & 3.59172  & 4.33334 & 3.70008 & 4.38493 \\
\end{tabular}
\caption{\label{tab:xcvalues} The critical values $x_c$ for the various potentials. Notice that for the types
I$_*$ and II$_*$, $x_c$ is independent of $W_0$. }
\end{center}
\end{table}

 Eq.~\nr{emel} can be evaluated for the two choices of $a,\kappa$ above.
For the choice \nr{akappa}
 (types I and I$_*$)
the equation becomes
 \be
 {36[1+\fra1{36}(115-16x_f)\l_*]^{4/3}\over
 \CL_\rmi{UV}^2[V_g(\l_*)-x_fV_{f0}(\lambda_*)]}=4 \ .
 \label{emelI}
 \ee
 For the choice \nr{akappa2}
(types II and II$_*$),
the $x_c$-equation \nr{emel} has the form
 \be
 {36[1+\fra1{27}(115-16x_f)\l_*+\lambda^{*2}]\over
 \CL_\rmi{UV}^2[V_g(\l_*)-x_fV_{f0}(\lambda_*)]}=4 \ .
 \label{emelII}
 \ee
The values of $x_c$ can then be calculated by inserting the potential $V_g-x_f V_{f0}$ and the
chosen value for $W_0$ in these equations. The critical values for the potentials listed above
are given in Table~\ref{tab:xcvalues}.

\begin{figure}[!tb]
\begin{center}

\includegraphics[width=0.49\textwidth]{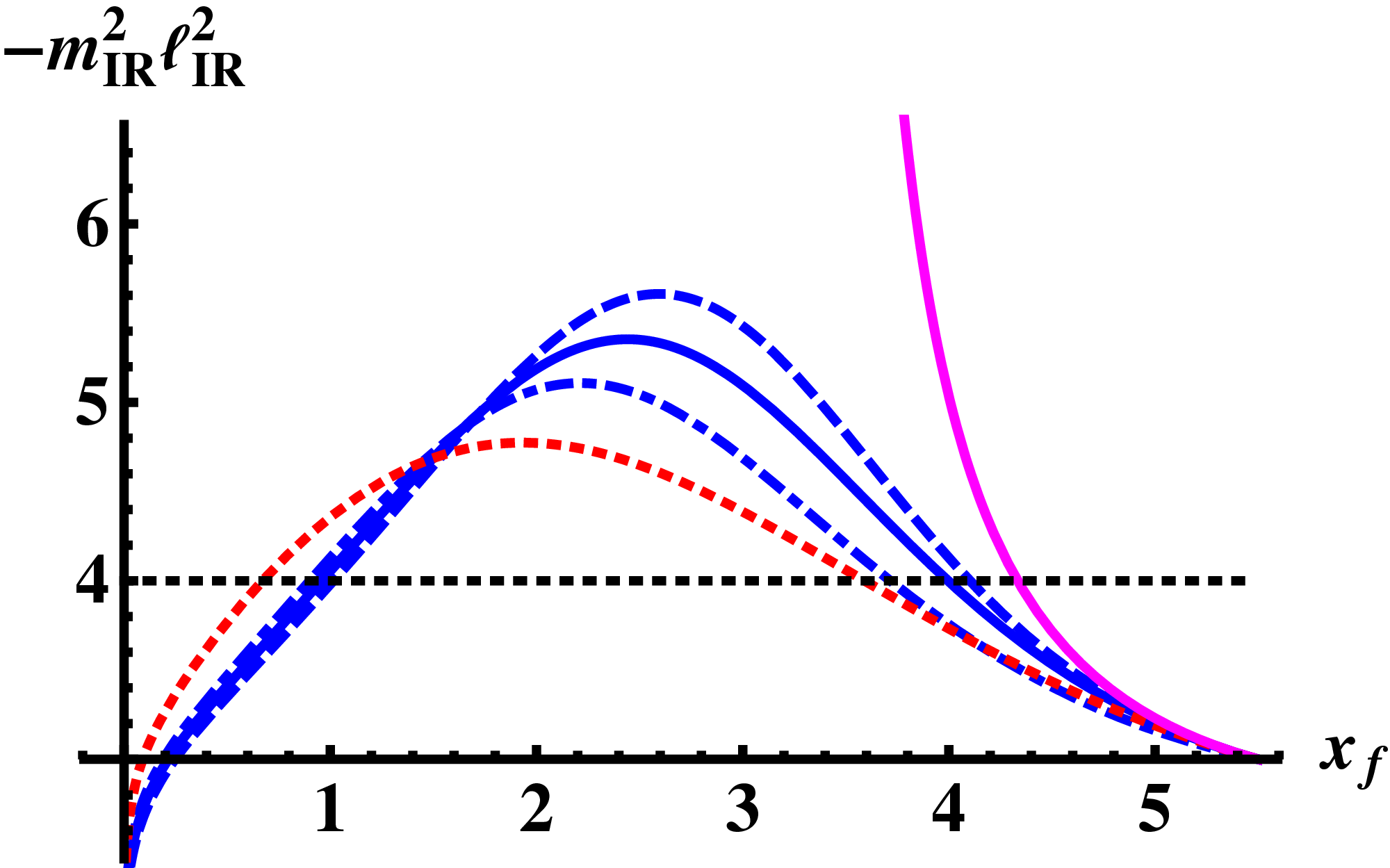}\hfill
\includegraphics[width=0.49\textwidth]{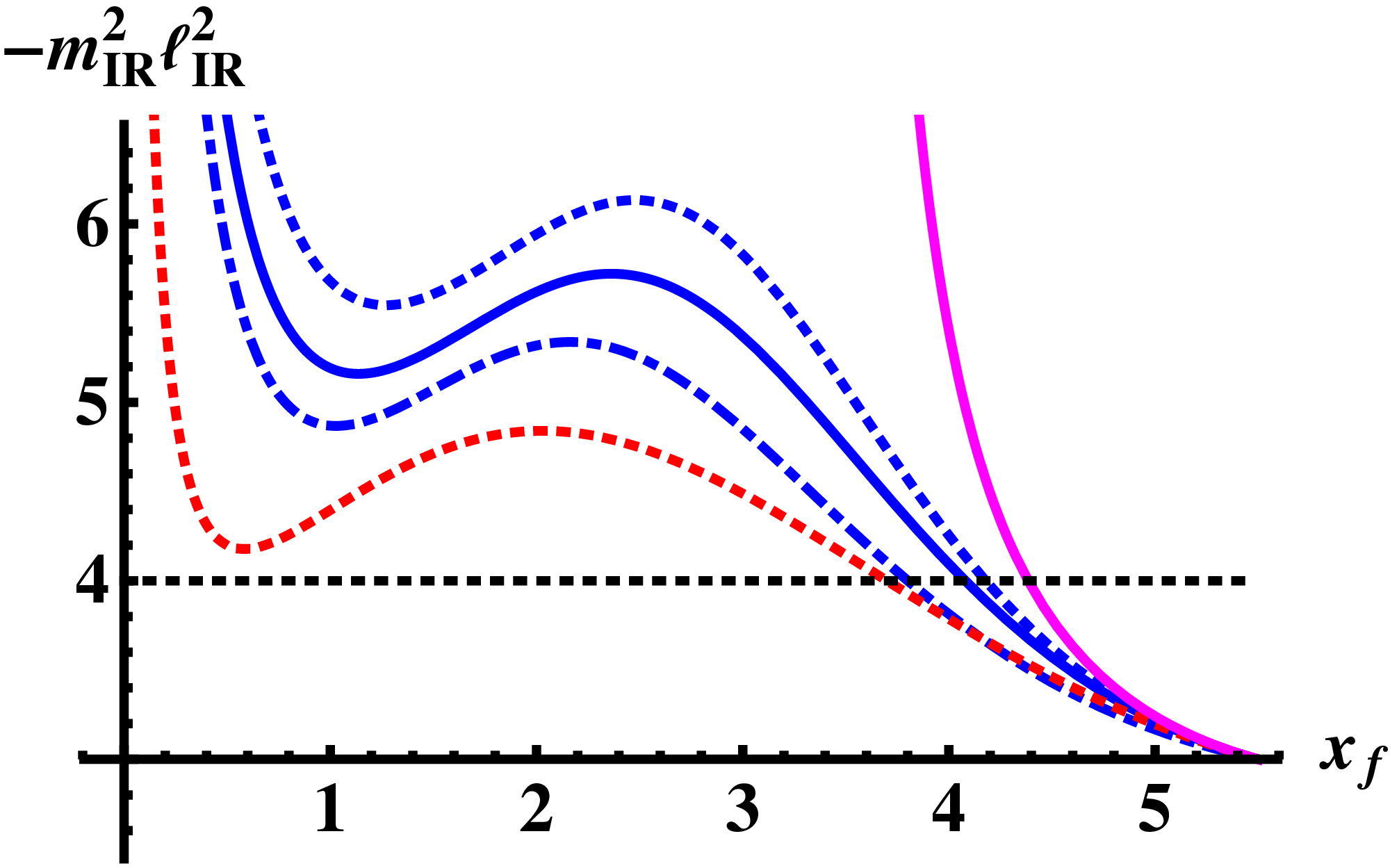}
\end{center}

\caption{\small The squared tachyon mass at the IR fixed point, see Eqs.\protect\nr{emelI} and
\protect\nr{emelII}.
Left: potentials I, right: potentials II.
The blue curves give the masses for constant $W_0$. The dashed, solid and dotdashed curves have
$W_0 = 0$, $12/11$, and $24/11$, respectively. The dotted red curves have $W_0$ fixed according
to the Stefan-Boltzmann normalization of the free energy in the UV. The solid magenta (uppermost)
curves are for potentials I$_\ast$ and II$_\ast$, for which the tachyon mass is independent of $W_0$.
The black dotted horizontal line marks the BF bound. }
\label{tachyonmassfig}
\end{figure}

The $x_f$-dependence of the tachyon mass for all the potential choices suggested above
is shown in Fig.~\ref{tachyonmassfig}. The critical value $x_c$ is the rightmost point
where the curve intersects the horizontal dashed line where the BF bound is saturated.
For potentials I$_\ast$ and II$_\ast$ (solid magenta curves) the fixed point only exists for
$x_*<x_f<11/2$ with $x_* \simeq 3.27$. In this case the tachyon mass diverges as
$x$ approaches $x_*$ from above.

{}From \nr{emelI} and \nr{emelII} one sees, using
the asymptotics of the potentials (see Eq.~\nr{limitlastar} below),
that
$-m_\rmi{IR}^2\CL_\rmi{IR}^2\sim 1/\sqrt{\ln(1/x_f)}$
for type I and $-m_\rmi{IR}^2\CL_\rmi{IR}^2\sim 1/x_f$
for type II
as $x_f \to 0$.
They thus behave completely differently in
this
limit, for type I the mass
vanishes, for type II it grows without bounds.
In particular, for potentials I and for low $x_f$ the (absolute value of the squared)
tachyon mass dives below the BF bound. This means that the existence of a solution with a
nontrivial tachyon profile and zero quark mass is not guaranteed \cite{jk},
which means that chiral symmetry could remain intact even at low temperatures.
However, in
most of the cases, such a solution anyhow exists all the way down to $x_f=0$,
and the expected picture with chiral symmetry breaking is obtained.
 We shall
discuss this issue in more detail below.

\section{V-QCD at finite temperature: equations and their solution}

The V-QCD action has two kinds of vacua at finite temperature, either with identically vanishing
tachyon or with nontrivial tachyon profile. The tachyonless black hole solutions can be constructed
in the same way as in the Yang-Mills case \cite{YM2}.
Below most of the discussion will in principle assume the presence of the tachyon, but the
construction for the solutions without the tachyon can be obtained simply by setting $\tau=0$ everywhere.

\subsection{Equations and numerical solution}
The goal now is to find numerical solutions of the Einstein's equations for
the metric functions $b(r)=e^{A(r)}, \,f(r)$ and the scalars $\lambda(r),\,\tau(r)$,
satisfying
\be
f(r_h)=0,\quad f(0)=1,\quad b(r) \underset{r\to0}{\sim} {\CL_\rmi{UV}\over r},
\ee
where $r_h$ marks the location of the horizon.

Due to the
singular behavior of the solutions near the UV boundary ($r\to0$),
it proves to
be convenient to use $A=\ln b$ as a coordinate instead of $r$ in the numerical solution.
Carrying out this transformation, one finds that the combination
\be
q(A)=e^A\frac{dr}{dA}=-\frac{1}{W}
\label{defofq}
\ee
appears naturally. This is just a rewriting of
the superpotential
\be
W=-{\dot b\over b^2}=-e^{-A}{dA\over dr}.
\ee

The equations of motion then become
\ba
12-6\frac{q^\prime}{q}+\frac{4}{3}\frac{\lambda^{\prime 2}}{\lambda^2}+3\frac{f^\prime}{f}
&=& \frac{q^2}{f}\left(V_g-V_f\sqrt{1+f\f\tau^{\prime 2}/q^2}\right),\label{A1}\\
12-\frac{4}{3}\frac{\lambda^{\prime 2}}{\lambda^2}+3\frac{f^\prime}{f} &=&\frac{q^2}{f}
\left(V_g-\frac{V_f}{\sqrt{1+f\f\tau^{\prime 2}/q^2}}\right),\\
4-\frac{q^\prime}{q}+\frac{f^{\prime\prime}}{f^\prime} &=& 0,\label{A3}\\
\tau^{\prime\prime}+\biggl(4-\frac{q^\prime}{q}+\frac{f^\prime}{f}+
\l^\prime \frac{\partial\ln \f }{\partial \l}
+\lambda'\frac{\partial\ln V_f}{\partial\lambda}\bigr)\tau^\prime &=&
-\frac{f\f}{q^2}\biggl(4+\frac{f^\prime}{2f}+ \frac{\l^\prime}{2} \frac{\partial\ln \f}{\partial \l}
+\lambda^\prime
\frac{\partial\ln V_f}{\partial\lambda}\biggr)\tau^{\prime 3}
\nn\\ &&+\frac{\partial\ln V_f}{\partial\tau}
\tau^{\prime 2}+\frac{q^2}{f\f}\frac{\partial \ln V_f}{\partial\tau},\\
\frac{\lambda^{\prime\prime}}{\lambda}+{f'\over f}{\lambda'\over\lambda}+
4\frac{\lambda^\prime}{\lambda}-\frac{\lambda^{\prime 2}}{\lambda^2}
-\frac{q^\prime}{q}\frac{\lambda^\prime}{\lambda} =
-\frac{3}{8}\frac{q^2\lambda}{f}&&\hspace{-0.6cm}
\biggl(\frac{\partial V_g}{\partial\lambda}-\frac{\partial V_f}{\partial\lambda}
\sqrt{1\!+\!\frac{f}{q^2}\f\tau^{\prime 2}}
-\frac{f}{2q^2}\frac{V_f\frac{d\f}{d\lambda} \tau^{\prime 2}}{\sqrt{1\!+\!\frac{f}{q^2}
\f\tau^{\prime 2}}}\biggr).\nonumber
\\
\ea
Here the prime denotes differentiation with respect to $A$. Near the UV boundary $r=0$,
\be
A=\ln b=\ln{\CL_\rmi{UV}\over r}\to+\infty.
\label{bAdS}
\ee
The range of $A$ thus is $A_h<A<+\infty$, where $A_h$ is the horizon,
\be
f(A_h)=0.
\label{fathor}
\ee

Numerical integration starts by solving $q',\,\lambda',\,f'',\,\tau''$
from the four first ones
in terms of lower derivatives; the fifth equation, the equation
for $\lambda$, will be used as a check and constraint.
For brevity we introduce two square root factors:
\be
R_1=\sqrt{1+{f\f\over q^2}\tau'^2},
\ee
and
\be
R_2=\sqrt{12+{3f'\over f}-{q^2\over f}\biggl(V_g-{V_f\over R_1}\biggr)}.
\ee
The equations to be solved numerically then are
\be
q'=q\biggl[4+{f'\over f}-{q^2\over 6f}\biggl(2V_g-V_fR_1-{V_f\over R_1}\biggr)\biggr],
\ee
\be
\l'=-{\sqrt3\over2}\l R_2,\label{A3num}
\ee
\be
f''=f'\biggl[{f'\over f}-{q^2\over 6f}\biggl(2V_g-V_fR_1-{V_f\over R_1}\biggr)\biggr]
=f'\biggl({q'\over q}-4\biggr),\label{A1num}
\ee
\ba
\tau''&=&-{q^2\over 6f}\biggl(2V_g-V_fR_1-{V_f\over R_1}\biggr)\tau'
-{f\f\over q^2}\biggl(4+{f'\over 2f}\biggr)\tau'^3
\\\nn&&
+{\sqrt3\over2}\biggl(\tau'+{f\f\over2q^2}\tau'^3\biggr){\l
\partial_\l\f
\over \f} R_2+
{\sqrt3\over2}\biggl(\tau'+{f\f\over q^2}\tau'^3\biggr){\l\partial_\l V_f\over V_f} R_2
+\biggl({q^2\over f\f}+\tau'^2\biggr){\partial_\tau V_f\over V_f}.
\ea
In the $\lambda$ equation the minus branch has to be chosen as
$\lambda(A)$ is a monotonically
decreasing function of $A$. The derivatives are with respect to $A$.
The equations are autonomous
in the sense that there is no explicit $A$ dependence.
Numerical integration then proceeds as follows:

1. Let us fix the horizon at $A=A_h=-\e$,
where $\e$ is a sufficiently small number, e.g., $\e=10^{-6}$.
the values of the functions at $A=0$, which is taken as the initial value of numerical
integration, are computed
by using the expansions \nr{horexpf}-\nr{horexptau} in Appendix~\ref{AppIR}.
These numbers can now be obtained by inserting the values of $\lambda_h,\,\tau_h,f'_h$.
Among these the horizon values of the scalars, $\lambda_h,\,\tau_h$, remain as parameters,
$f'_h$ can be given an arbitrary positive value, +1, say. One then finds a solution $q_1(A),\,f_1(A),\,
\lambda_1(A),\,\tau_1(A)$ valid from $A=0$ to some large upper limit $A_+$ by using NDSolve of
Mathematica. The spatial coordinate $r(A)$ can then, if needed, be computed by similarly integrating
the differential equation
\be
r'(A)= e^{-A}q(A)
\ee
with the initial condition $r(A=\infty)=0$.

2. The so obtained first-level solution $f_1(A)$ is scaled to one 
in the UV ($A\to\infty$) by writing $f_2(A)=f_1(A)/f_1(A_+)$.
Simultaneously $q_2(A)=q_1(A)/\sqrt{f_1(A_+)}$, which is
needed since Eq.~\nr{A1} demands that $q^2/f$ be
invariant. Finally, $\lambda_2=\lambda_1,\,\tau_2=\tau_1$.

3. The final scaling is performed to guarantee that all solutions use the same unit of energy
or, equivalently, have the same integration constant in
the integral of the definition \nr{defbeta}
of the beta function.
This implies
\be
A-\hat A_0=\ln(b)-\hat A_0= {1\over b_0\lambda(A)}+{b_1\over b_2^2}\ln(b_0\lambda(A))+
\biggl({b_2\over b_0^2}-{b_1^2\over b_0^3}\biggr)\lambda(A)+\CO(\lambda^2),
\label{rlambda}
\ee
where $\hat A_0$ is the integration constant. By inserting the UV expansions of $A$
and $\l$ from Appendix~\ref{AppUV}, we identify
$ \hat A_0 = \ln (\CL_\rmi{UV} \Lambda)$. We wish to scale $\Lambda$ to
one\footnote{After this, all quantities are expressed in units of $\Lambda$; omitting
the factor $\CL_\rmi{UV}$ would give a unit of energy depending on $x_f$}, and therefore define
\be
 A_0 = \hat A_0 -\ln  \CL_\rmi{UV} = \ln \Lambda \ ,
\ee
and shift solutions by $A_0$.
In practice, one implements this by
determining, for a given numerical solution (the $\CO(\lambda_2)$ term is optional),
\be
A_0=\lim_{A\to\infty} \biggl[A-\ln\CL_\rmi{UV}-{1\over b_0\lambda_2(A)}
-{b_1\over b_0^2}\ln(b_0\lambda_2(A))
-\biggl({b_2\over b_0^2}-{b_1^2\over b_0^3}\biggr)\lambda_2(A)
\biggr]
\label{A0det}
\ee
and then performing the scaling
\be
\lambda_3(A)=\lambda_2(A+A_0)
\ee
etc.
for all the functions at level 2. The set $q_3(A),\,f_3(A),\,
\lambda_3(A),\,\tau_3(A)$, parametrised by the values of $\lambda_h,\,\tau_h$, is the
final numerical solution. Note that the horizon has now been shifted to
\be
A_h=-A_0-\e\approx -A_0;
\ee
at level 2 it was defined by $f_2(-\e)=0$.

\subsection{Physical quantities}
The set of functions $q(A),\,f(A),\,\lambda(A),\,\tau(A)$ (leaving out the index 3) can now
be converted to various physical quantities:

The temperature is
\be
T=-{1\over4\pi}f'(r_h)=-{e^A\over4\pi q(A)}f'(A_h)\vert_{A_0=A_h}
={e^{-A_0}\over-4\pi q(-A_0)}f'(-A_0),
\ee
and the value of $b$ at the horizon is
\be
b_h=e^{-A_0}.
\ee
The quark mass $m_q$ is defined by the UV expansion of the tachyon:
\be
\tau(r)=\CL_\rmi{UV} m_q(-\ln\Lambda r)^{-\fra9{22-4x_f}}r
\ee
so that, using the relation \nr{rlambda},
\be \label{mqextra}
m_q=\lim_{A\to\infty}\CL_\rmi{UV}^{-1}\,\tau(A)\,\exp\biggl[{1\over b_0\lambda(A)}+
\biggl({b_1\over b_0^2}-{9\over 22-4x_f}\biggr)\ln(b_0\lambda(A))\biggr].
\ee
In practice, the extrapolation to $A=\infty$ can be carried out by measuring
$\tilde m_q(A)$, as defined by the right hand side of Eq.~\eqref{mqextra},
at two large values of $A$ and then linearly extrapolating to $\lambda=0$:
\be
m_q={\tilde m_q(A_1)\lambda(A_2)-\tilde m_q(A_2)\lambda(A_1)\over \lambda(A_2)-\lambda(A_1)}.
\label{mqextra2}
\ee
Linear extrapolation is chosen, because the leading neglected terms in the expansion of
Eq.~\eqref{mqextra} are (up to logarithmic corrections) linear in $\l$.

\begin{figure}[!tb]
\begin{center}

\includegraphics[width=0.49\textwidth]{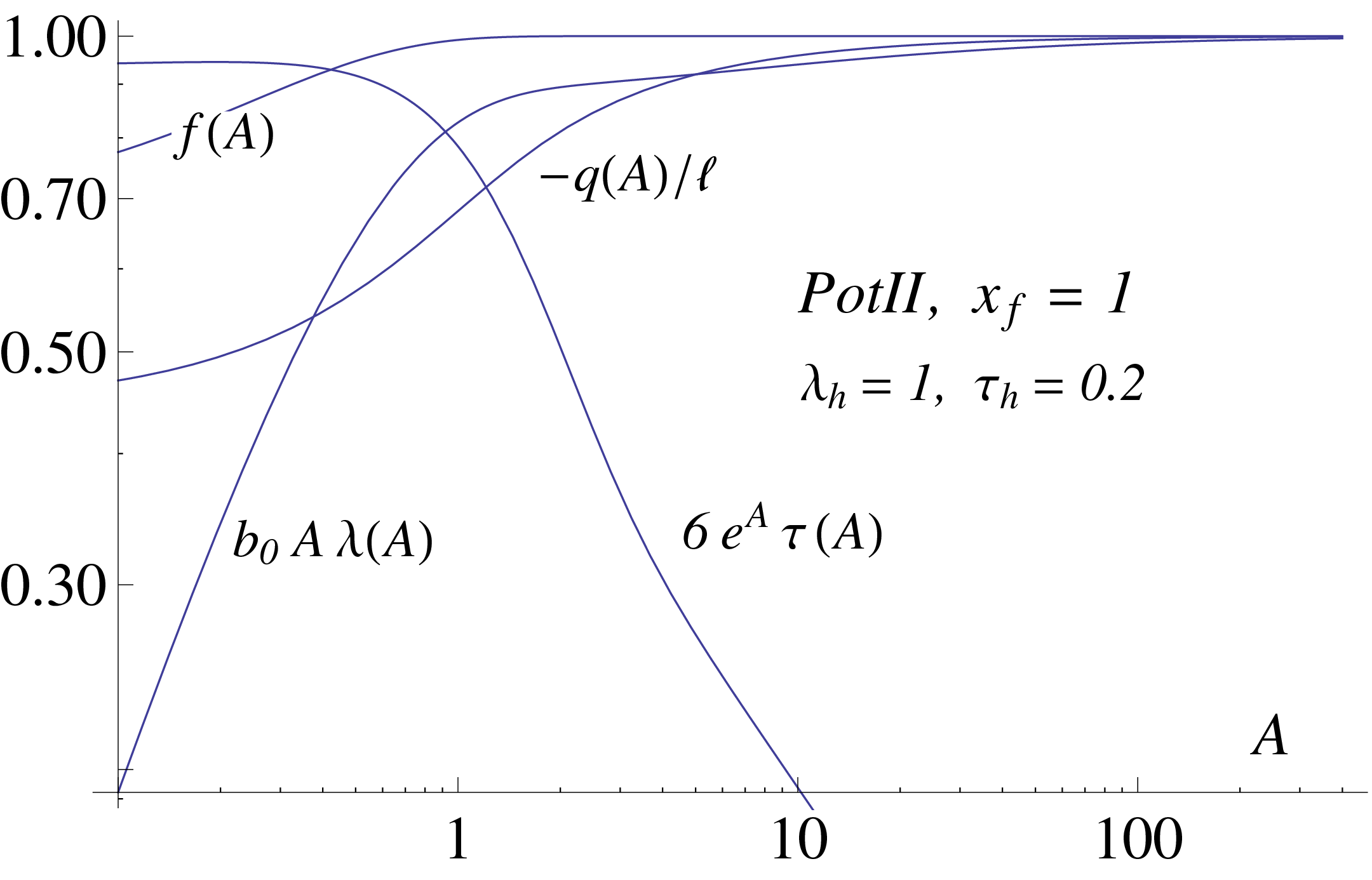}\hfill
\includegraphics[width=0.49\textwidth]{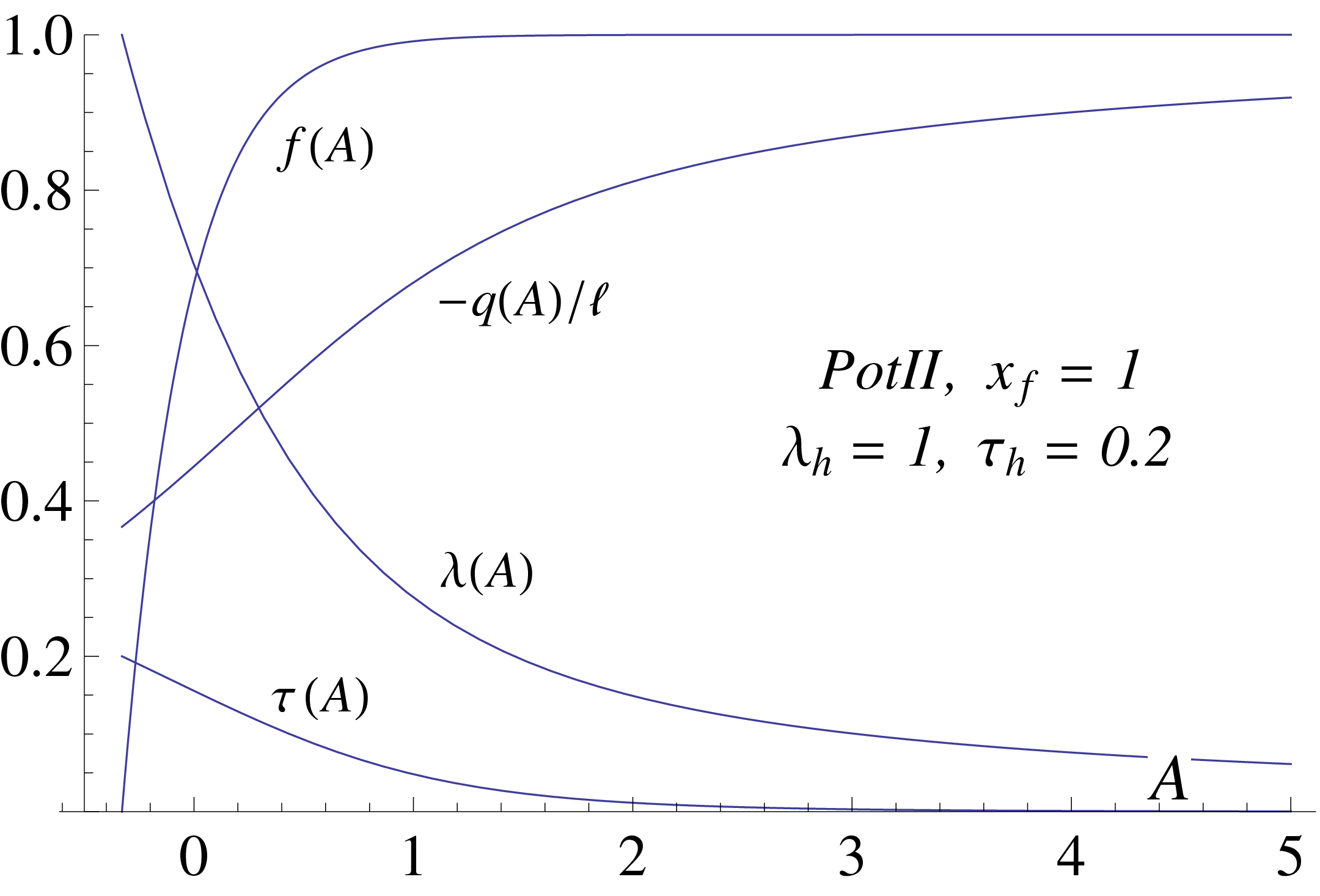}
\end{center}

\caption{\small Explicit bulk configurations. Left: UV large-$A$ region. Right: Near horizon region.
For this configuration $T=0.3839,\,b_h=0.7200,\,m_q=0.05422$.}
\label{configs}
\end{figure}

\subsection{Fixing quark mass}
The above is for fixed $\lambda_h,\,\tau_h$. The really demanding task is to find the
field configurations at fixed $m_q$. For this one needs the curves $\tau_h(\lambda_h,m_q)$.
The quark mass is determined by the UV behavior of the tachyon:
$\tau(r)/\CL_\rmi{UV} \simeq m_q (-\ln r)^{-\gamma_0/b_0}r$.
To fix $m_q$ at fixed $\lambda_h$ we have to solve the equations of motion at various $\tau_h$ and
find that value of $\tau_h$ which leads to the desired UV behavior of $\tau(r)$.

\subsubsection{Zero quark mass}
In particular, we are interested in $m_q=0$. This case splits in two parts: either $\tau(r)=0$ identically
(chiral symmetry holds) or $\tau(r)$ nonzero (chiral symmetry broken).

If $\tau=0$, solutions with $m_q=0$ are obtained simply by setting $\tau_h=0$ above. The solution is
then controlled by the effective potential $V_g(\lambda)-x_fV_{f0}(\lambda)$.
For classes I and II, this increases monotonically
from $\lambda=0$, but since $V_{f0}$ grows faster, the derivative
decreases and becomes finally zero
at some $\lambda=\lambda_*(x_f)$ (see Eq.~\nr{defoflastar}).
The extremum of the potential marks the location of the IR fixed point,
which is screened by the horizon at finite temperature.
Indeed, the tachyonless black holes have $0<\l_h<\l_*$, and for $\l_h$ very close to $\l_*$
we obtain configurations where the dilaton is approximately constant, $\l \simeq \l_h \simeq \l_*$
for a long range of the coordinate before the horizon is reached in the deep IR.

For classes I$_*$ and II$_*$, the effective potential $V_g(\lambda)-x_fV_{f0}(\lambda)$ does not have
an extremum for $x_f$ below $x_* \simeq 3.27$.
In this case the fixed point is absent, and the tachyonless black hole solutions are qualitatively
similar to Yang-Mills ($x_f \to 0$) \cite{YM1}. In particular $\l_h$ can take any value.

\begin{figure}[!tb]
\begin{center}

\includegraphics[width=0.49\textwidth]{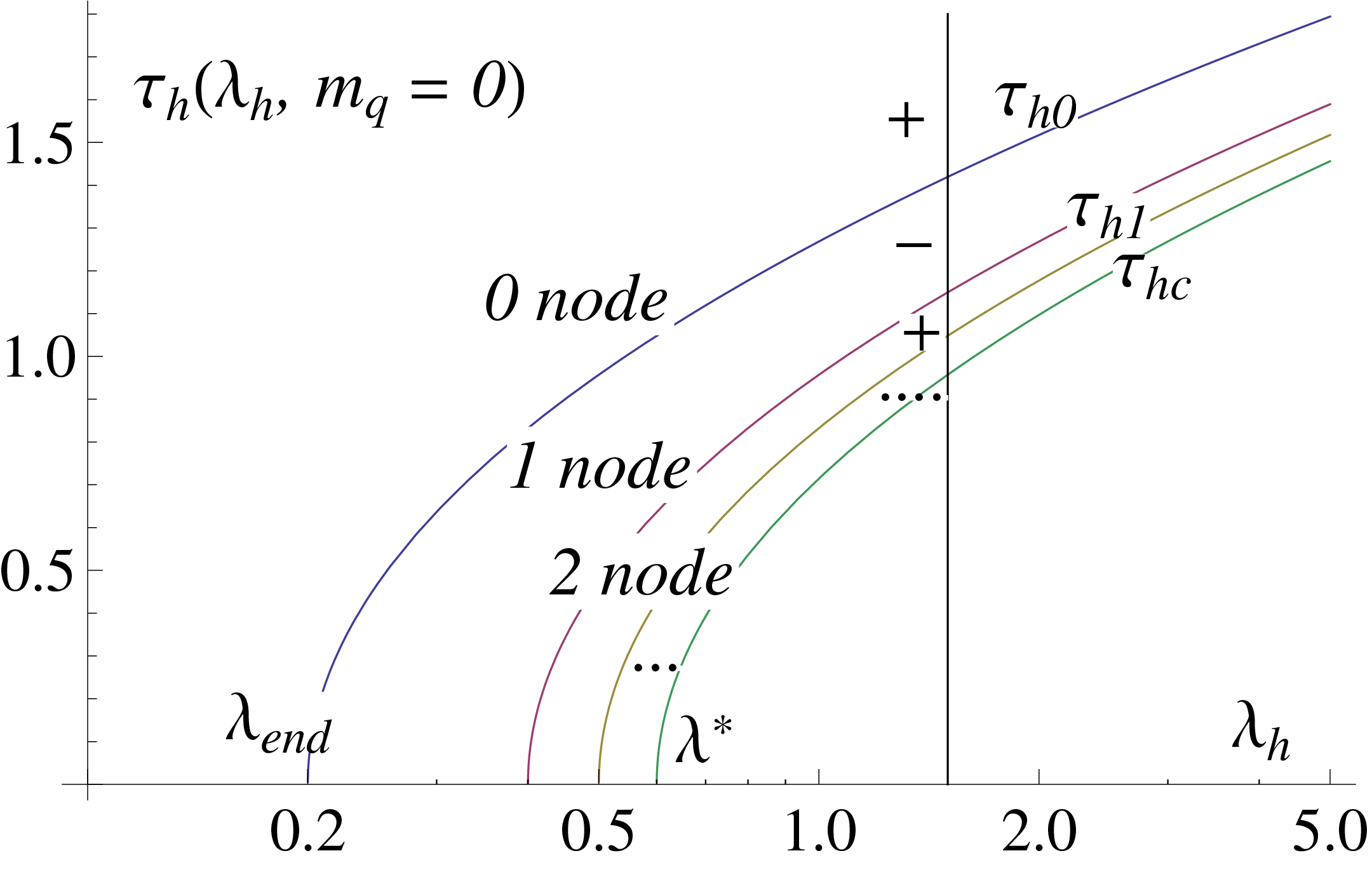}\hfill
\includegraphics[width=0.49\textwidth]{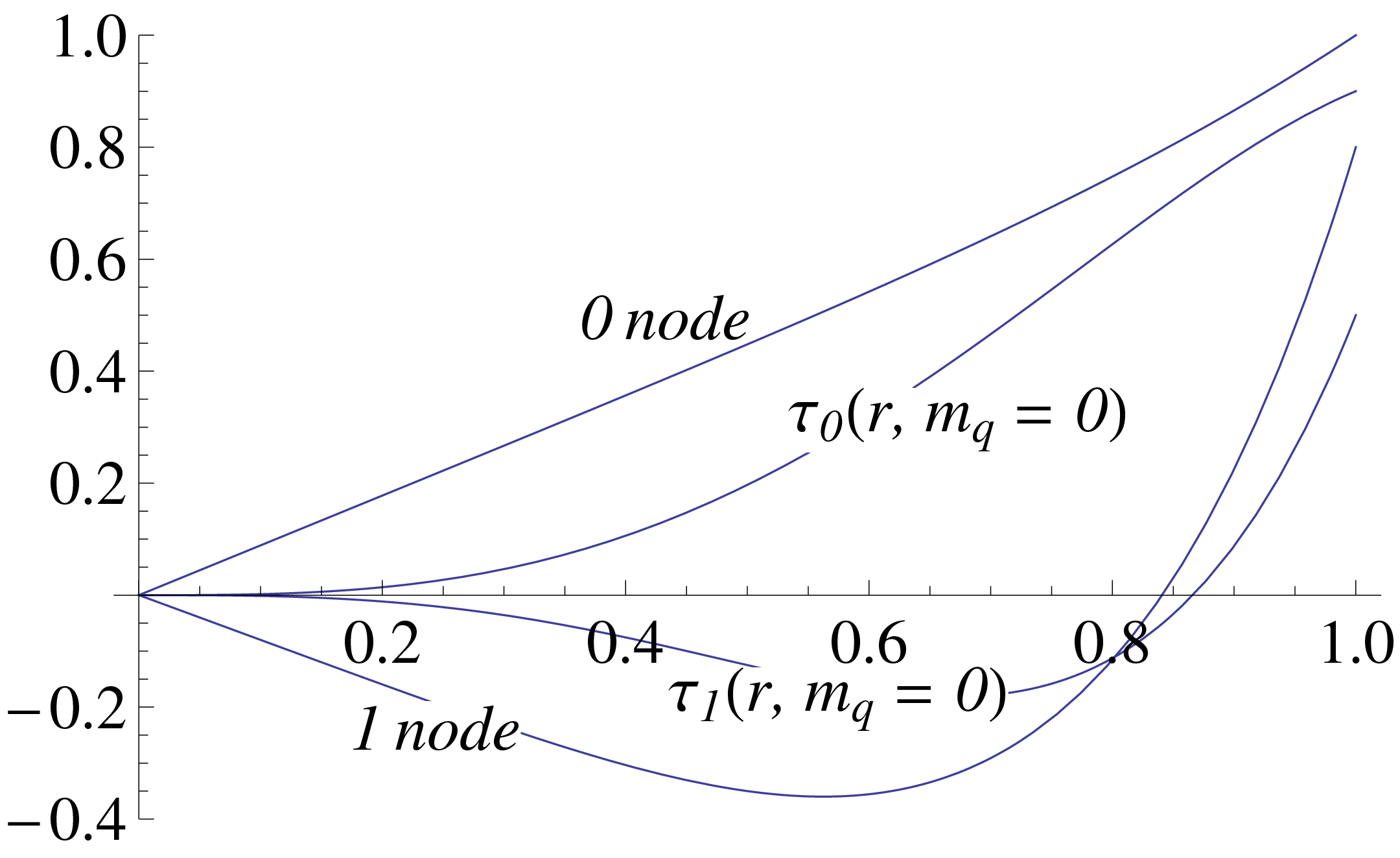}
\end{center}

\caption{\small Left: Schematic presentation of the curves $\tau_h(\lambda_h,m_q=0)$, i.e., those values
of the horizon value $\tau_h$ of the tachyon which lead to configurations with $m_q=0$, at that
particular $\lambda_h$. Choosing
$\tau_h$ on the curve $\tau_{hi}$ leads to a $\tau(r)$ without a linear term and with $i$  zeros at some
$r$. The zero mass solutions with vanishing tachyon live on the line $0<\lambda_h<\lambda_*$.
The plus and minus signs indicate the sign of the quark mass in each region limited by the curves $\tau_{h0}$,
$\tau_{h1}$,\ldots
Right: Schematic presentation of the $r$ dependence of the bulk tachyon for low node numbers.
Tachyon solutions for top to bottom are: a generic solution with $m_q>0$ and no nodes (``0 node''),
the standard solution with zero quark mass ($\tau_0$), the solution with zero quark mass and one node
($\tau_1$), and a generic solution with $m_q<0$ and one node (``1 node'').
See the text for a more detailed explanation.
 }
\label{mass0pic}
\end{figure}

For non-zero $\tau(r)$, the discussion of $m_q=0$ configurations has to take into account
the existence of Efimov zeroes, oscillatory behavior when approaching $r=0$,
which was discussed above in the introduction.
We discuss here the standard picture which is seen in most cases for $x_f<x_c$.
A rough description of more complicated cases is given in Appendix~\ref{AppEfimov}.
The situation is
summarised in Fig.~\ref{mass0pic}. For large $\tau_h>0$, $\tau(r)$ decreases monotonically
from $\tau_h$ towards $r=0$ and ends with positive $m_q$.
We evaluate $m_q $ using (\ref{mqextra2}) with two large values of $A$
(corresponding to a small UV cutoff $\e$ in the $r$-coordinate).
When $\tau_h$ is decreased, ultimately 
an (approximate) $m_q=0$ configuration ($\tau_{0}(r,m_q=0)$ in Fig.~\ref{mass0pic})
with monotonically decreasing $\tau(r)$ is obtained.

This defines the curve $\tau_{h0}(\lambda_h)$ in Fig.~\ref{masspic}. One finds that these
solutions are possible only if $\lambda_h$ is larger than a fixed positive value, which we
call $\lambda_\rmi{end}$. Decreasing $\tau_h$ further,
$\tau(r)$ first develops a zero so that $m_q<0$. 
Continuing even further we find a second location where
 $m_q=0$ vanishes.
This is a configuration with one tachyon node ($\tau_1(r,m_q=0)$ in Fig.~\ref{mass0pic}).
The pattern
continues with an ever increasing number of nodes, until one ends up with the curve
$\tau_{hc}(\lambda_h)$,
below which a solution with the standard UV boundary does not exist.
Numerically, the curves $\tau_{h0}(\lambda_h)$ and $\tau_{h1}(\lambda_h)$
can be separated, but already $\tau_{h2}(\lambda_h)$ would require so much effort that
we have not embarked on computing it.
As we approach the conformal window, the curves $\tau_{h0},\, \tau_{h1},\ldots$ get closer and
closer to $\tau_{hc}$ and finally vanish for $x_f \ge x_c$.

We expect that increasing the number of nodes increases the free energy so that to
study equilibrium states it is enough to compute $\tau_{h0}(\lambda_h,m_q=0)$.
This was checked at zero temperature in \cite{jk} numerically for potentials I, and
analytically in the limit $x_f \to x_c$ as well as in the limit of large number of tachyon nodes.

\subsubsection{Nonzero quark mass}
\label{subsubsect_nonzeromq}
\begin{figure}[!tb]
\centering

\includegraphics[width=0.49\textwidth]{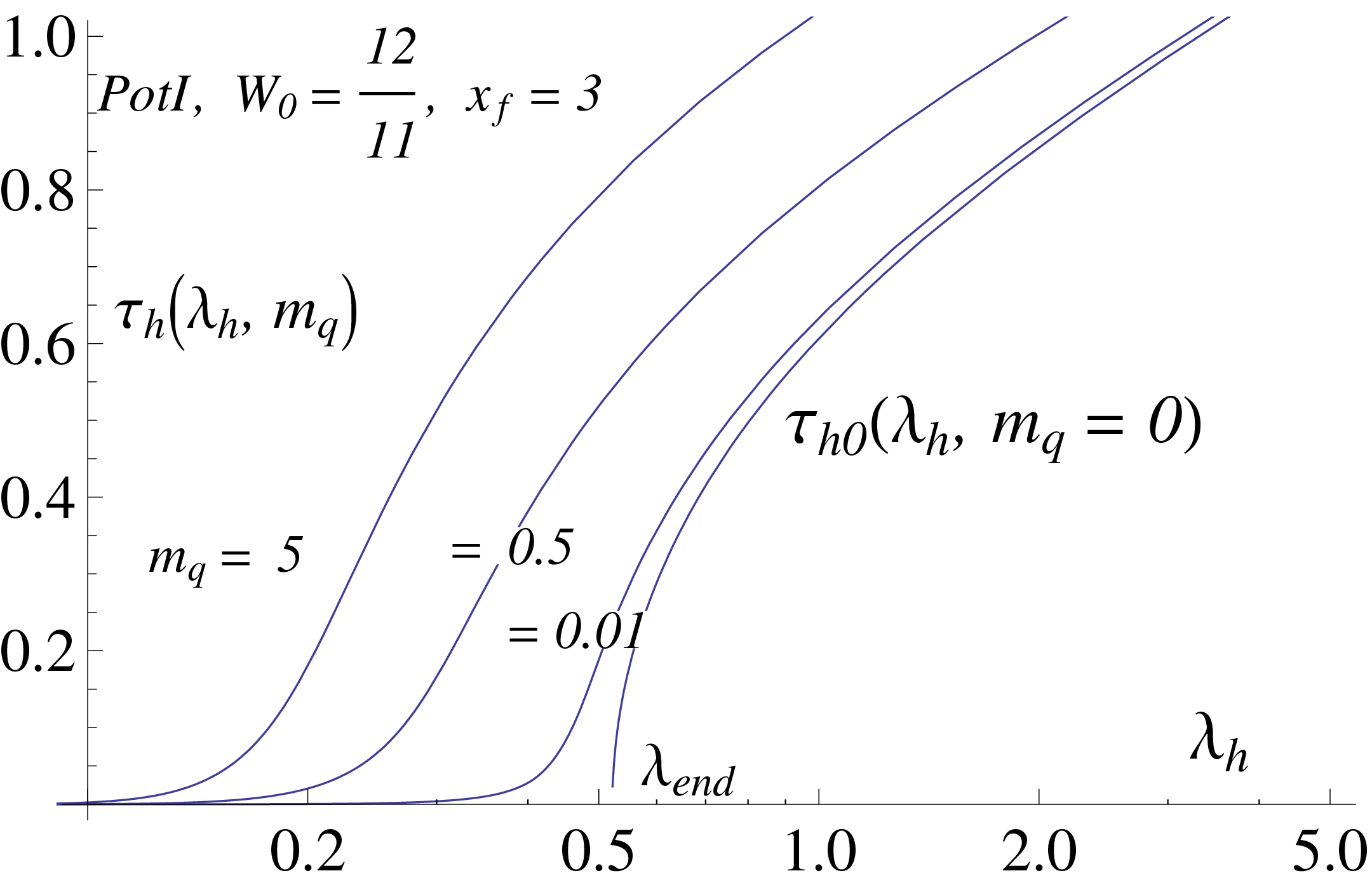}

\caption{\small
The
curves $\tau_h(\lambda_h,m_q)$ for values of $m_q$ marked in the figure, computed for Potential I with
$W_0=12/11$. Here $\lambda_\rmi{end}=0.5221$ and $\lambda_*=0.6467$.
}
\label{masspic}
\end{figure}

For nonzero quark mass
the special solution with identically vanishing tachyon profile is missing.
However, there are solutions of various types for $\tau_h>0$, as suggested by Fig.~\ref{mass0pic}.
We shall here restrict to the ``standard'' solutions which have monotonic tachyon, i.e.,
the region above $\tau_{h0}$ in Fig.~\ref{mass0pic} (left).
Below this curve there can be Efimov type solutions where the tachyon has nodes.
As for $m_q=0$, we expect that these solutions have higher free energies than the standard one.
In the region of standard solutions, the dependence of quark mass is smooth
(see Fig.~\ref{masspic}). We have found numerically that for fixed $\l_h$ the
correspondence between $m_q$ and $\tau_h$ is one-to-one.
Therefore $m_q$ can be kept fixed by following a set of well-defined curves on the
($\l_h,\tau_h$)-plane, some of which are sketched in Fig.~\ref{masspic}.

It is also interesting to notice how the $m_q=0$ solution is obtained from the ones
having finite quark masses as $m_q \to 0$.
What happens for nonzero quark mass is shown in Fig.~\ref{masspic} for a concrete computation.
If $0<\lambda_h<\lambda_\rmi{end}$ (and $\l_h$ fixed), the curve $\tau_h(\l_h,m_q)$ approaches
zero as $m_q\to0$, indicating that $\tau(r)$ approaches the chiral symmetry conserving solution
($\tau(r)\equiv 0$) uniformly.
If $\lambda_h>\lambda_\rmi{end}$, $\tau_h(\l_h,m_q)$ approaches $\tau_{h0}(\l_h)$ instead,
which implies that $\tau(r)$ converges to the standard chiral symmetry breaking solution $\tau_0(r)$.

\subsection{Thermodynamics} \label{secthermo}
We now want to compute minus free energy density or pressure $p(T,m_q;x_f)$ of the gravity dual,
assuming that all the quarks have the same mass $m_q$. In particular, we are interested in $m_q=0$.
The chemical potential is zero, there is an equal number of quarks and antiquarks. The equilibrium
phase has the largest pressure.

The basic strategy is to compute the temperature and entropy density from the
formulas
\be
T=-{1\over4\pi}f'(r_h),\quad s={1\over 4G_5}b^3(r_h),
\label{Ts}
\ee
where $f$ and $b$ are obtained by solving Einstein's equations. The pressure is then
obtained by integrating $s(T)=p'(T)$. The key technical issues are keeping track of
the quark mass and specifying the integration constant in the pressure integral.

\begin{figure}[!tb]
\centering

\includegraphics[width=0.49\textwidth]{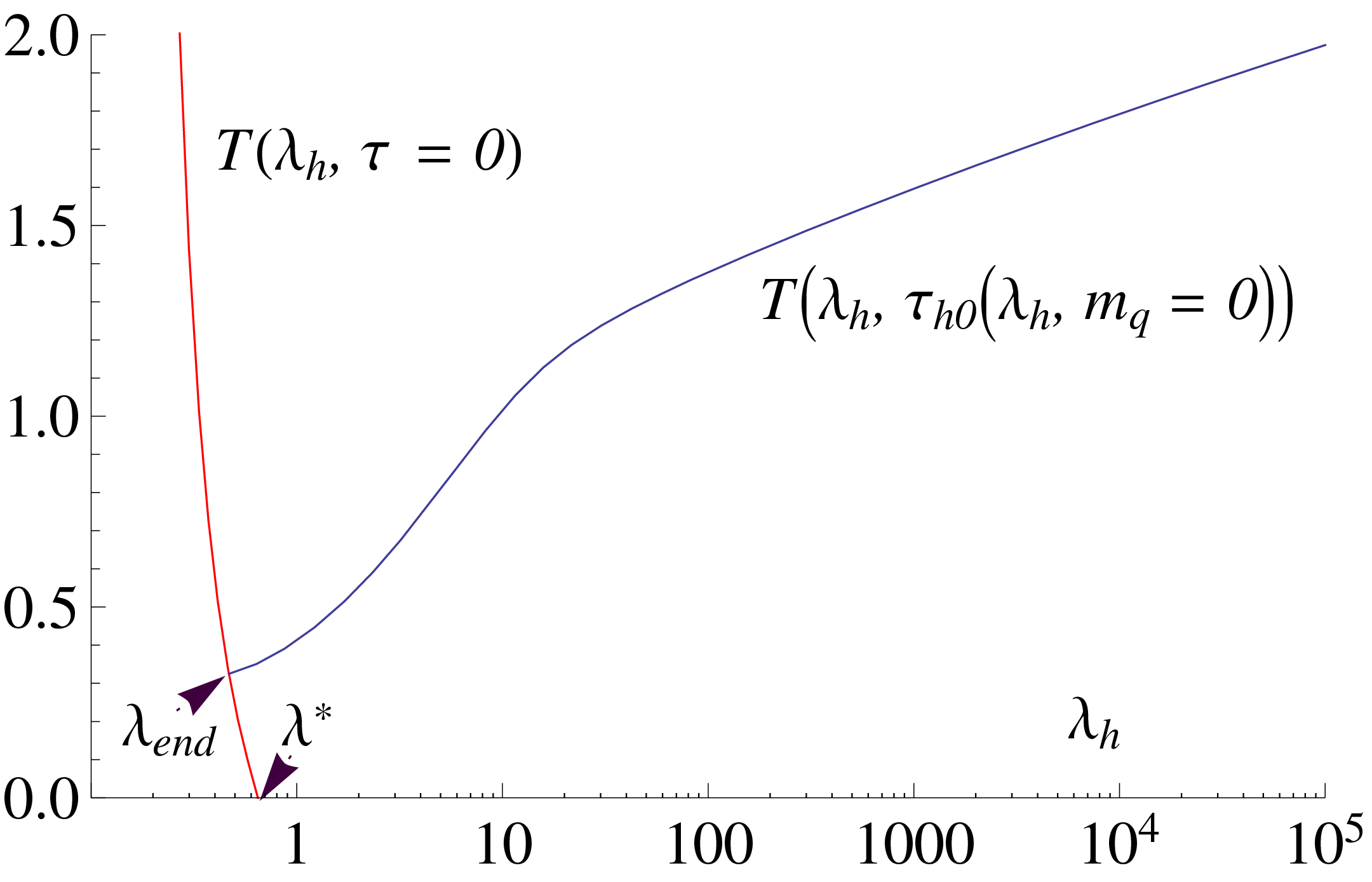}
\includegraphics[width=0.49\textwidth]{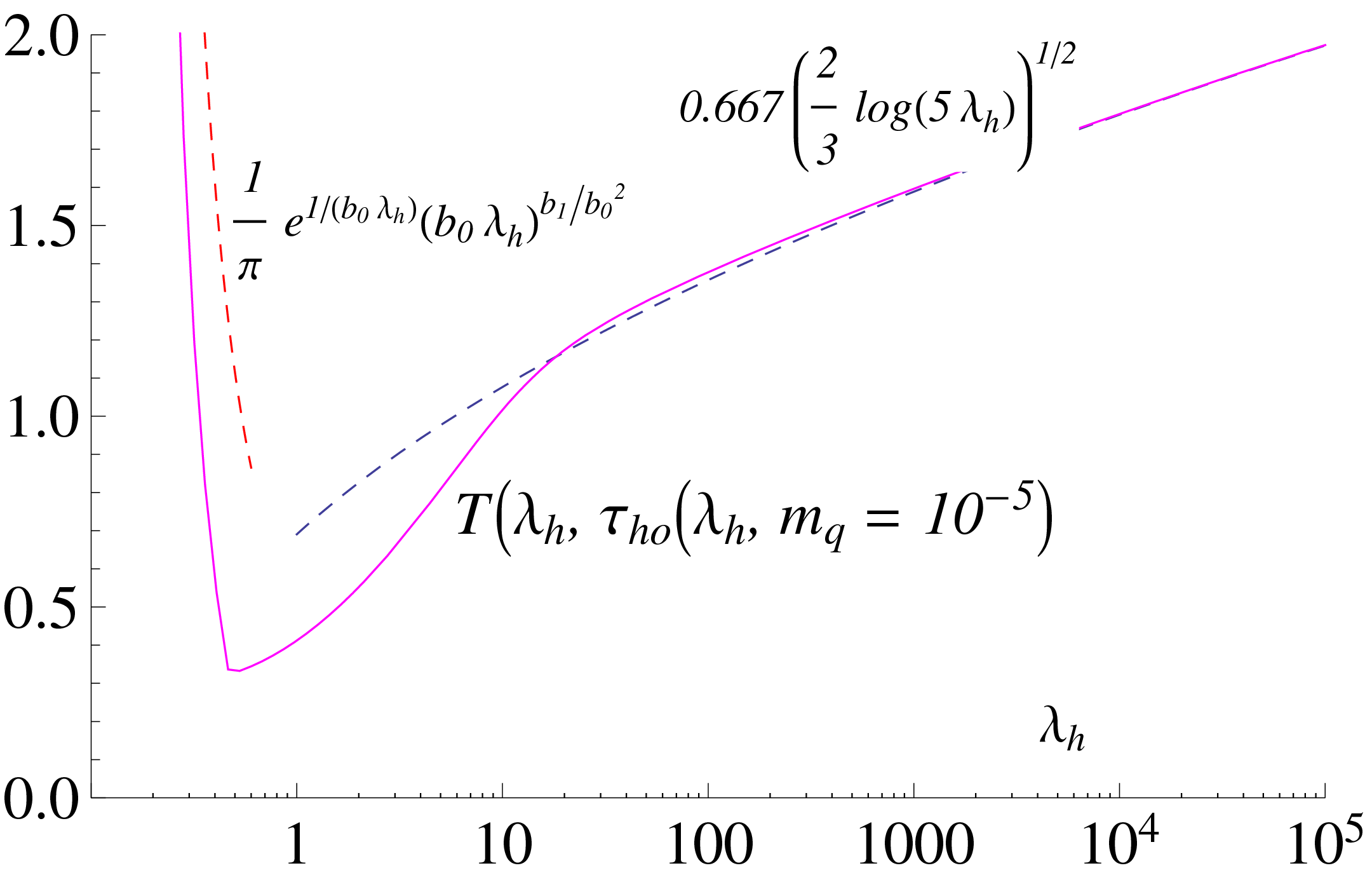}

\caption{\small The temperature as a function of $\lambda_h$ for solutions for Pot II at
$x_f=3$ and $W_0=12/11$,
both for zero (left) and very small mass (right). The asymptotic limits
(\protect\ref{TlahUV}) and (\protect\ref{TlahIR})
are also shown for $m_q=10^{-5}$, in the range of the figure the UV limit is not yet accurate. The
maximum value $\l^*$ of $\l$ for the $\tau=0$ curve is defined in (\protect\ref{defoflastar}).
See also Fig.~\protect\ref{figxfscan}.
}
\label{figTlah}
\end{figure}

The general structure of temperature
(for a case containing a fixed point)
is shown in Fig.~\ref{figTlah}, to be consulted in
association with Figs.~\ref{mass0pic} and \ref{masspic}. For $m_q=0$ two branches
separate. Firstly, for $0<\lambda_h<\lambda_*$ there is the temperature computed for
chirally symmetric vanishing tachyon solutions.
We shall use the notation $T_u(\l_h) \equiv T(\l_h,\tau_h=0)$ for this temperature below.

The chiral symmetry breaking solution
exists for $\lambda_\rmi{end}<\lambda_h<\infty$ and
as $\l_h \to \l_\rmi{end}$, the corresponding temperature curve ends precisely on the curve
which has identically vanishing tachyon. The temperature curve is computed by using the zero node zero mass
curve $\tau_{h0}(\lambda_h)$ in Fig.~\ref{mass0pic}.
We shall use the notation $T_b(\l_h) \equiv T(\l_h,\tau_{h0}(\l_h,m_q=0))$ for this temperature.
If we computed the temperature for
the one node solution $\tau_{h1}(\lambda_h)$, we would get a curve which lies
significantly
below the zero node curve in Fig.~\ref{figTlah} and again ends on the zero tachyon curve.
These solutions will have a higher free energy and we can thus neglect them.

Whenever the quark mass is nonzero, the tachyon cannot be vanishing and that branch
disappears. However,
as seen from Figs.~\ref{masspic} and~\ref{figTlah},
the small-mass curve very closely approximates
the zero tachyon curve, also at small $\lambda_h$.

Analytic approximations are often useful. In the UV $f(r)\to 1-r^4/r_h^4$ so that
\be
\pi T(\lambda_h)={1\over r_h}=e^{1/(b_0\lambda_h)}(b_0\lambda_h)^{b_1/b_0^2}
={b(\lambda_h)\over\CL_\rmi{UV}}.
\label{TlahUV}
\ee
Similarly, in the IR (see \nr{blaIR} in Appendix B),
\be
T(\lambda_h)\sim(\fra23\ln\lambda_h)^{1/2},\quad
b(\lambda_h)\sim{1\over\lambda_h^{2/3}}(\fra23\ln\lambda_h)^{1/4}.
\label{TlahIR}
\ee
For a numerical check, see Fig.~\ref{figTlah}.
The interesting physics takes place in the region connecting these two limits.

The function $b(\lambda_h)$ decreases monotonically while the function $T(\lambda_h)$ decreases
in the UV but starts increasing in the IR. The physics of the UV increase is obvious, this is
the weak coupling limit which naturally corresponds to large $T$ of a thermal fluid.
The (extremely slow) increase in the IR is a quantitative fact but does not correspond to a
stable phase. This is simplest seen by computing the sound velocity
\be
c_s^2={dp\over d\epsilon}={s\over Ts'(T)}={b(\lambda_h)\over 3T(-b'(\lambda_h))}
\biggl(-{dT\over d\lambda_h}\biggr).
\ee
A stable phase has $c_s^2>0$ (equivalently, has a positive specific heat) and this requires
$T'(\lambda_h)<0$. Thus only the UV decreasing part can correspond to a stable phase, the
IR part is the unstable small black hole region, small since $s\sim b^3(\lambda_h)\to0$ there.
It is, nevertheless, crucially important for the phase structure.

To compute the pressure, we have to integrate the entropy density \nr{Ts} over $T$. Taking
$\lambda_h$ as a variable, we have integrals over the two branches in Fig.~\ref{figTlah}:
\ba
p_b(T)&=&{1\over 4G_5}\int^\infty_{\lambda_h(T)}d\lambda_h(-T_b'(\lambda_h))\,b_b^3(\lambda_h)
+p_b(\infty),\label{peeb}\\
p_u(T)&=&{1\over 4G_5}\int^{\lambda_*}_{\lambda_h(T)}d\lambda_h(-T_u'(\lambda_h))\,b_u^3(\lambda_h)
+p_u(\lambda_*),\label{pees}
\ea
where $b,\,u$ refer to the chiral symmetry broken ($\tau_h=\tau_{h0}(\lambda_h,m_q=0))$ and chirally
symmetric (or unbroken, $\tau=0$) phases. The continuity of pressure at $T_\rmi{end}=T(\lambda_\rmi{end})$ leads
to a rather remarkable consistency check of the entire scheme: it demands
\be
{1\over 4G_5}\int^{\lambda_*}_{\lambda_\rmi{end}}d\lambda_h(-T_u'(\lambda_h))\,b_u^3(\lambda_h)-
{1\over 4G_5}\int^\infty_{\lambda_\rmi{end}}d\lambda_h(-T_b'(\lambda_h))\,b_b^3(\lambda_h)
=p_b(\infty)-p_u(\lambda_*).
\label{consist}
\ee
However, the difference on the RHS is nothing but the difference between the free energies
of the broken and symmetric phases at $T=0$:
\be
p_b(\infty)-p_u(\lambda_*)=-F_b(T=0)+F_u(T=0).
\ee
This difference was computed in \cite{jk} from the $T=0$ solutions, with no black hole. Here they
are computed in \nr{consist} from the black hole solutions and
we have checked numerically that
the results agree within the numerical precision.

The computation of the free energy now proceeds as follows, first for the simple structure of
$T(\lambda_h)$ in Fig.~\ref{figTlah}:
\bi
\item
Start by integrating \nr{peeb} from some large value of $\lambda_h$
down to $\lambda_\rmi{end}$, choosing $p_b(\infty)=0$.
Since $T'(\lambda_h)>0$ in Fig.~\ref{figTlah}, this leads to a negative pressure.
This is not the stable phase, the physical stable phase is not described by this metric.
The stable phase with the largest pressure is the thermal gas phase with $p=0$.
\item
At $\lambda_\rmi{end}$ move to the chirally symmetric $\tau=0$ branch and fix the constant
$p_u(\lambda_*)$ by demanding continuity of pressure. Since now $T'(\lambda_h)<0$, $p$ starts increasing.
At first $p$ is still negative and the stable phase is the thermal gas phase with $p=0$.
\item
At some $\lambda_h\equiv \lambda_c$ pressure passes through 0. This defines a transition
temperature $T_h$ since from now on the black hole metric has the largest pressure.
Since $\tau=0$ this black hole phase is chirally symmetric.
\item
The latent heat of the transition is
\be
{L\over T_h^4}={s(T_h)\over T_h^3}={1\over 4G_5}\biggl({b(\lambda_h)\over T_h}\biggr)^3<
 N_c^2{4\pi^2\over45}(1+\fra74 x_f),
\ee
where the maximum value is obtained taking normalisation
from \nr{pnormal} and using the UV approximation \nr{bhoverTApp}.
Counting degrees of freedom one has $N_f^2$ Goldstone bosons in the low
$T$ phase (for which we do not have a $T$ dependent gravity dual) and $2N_c^2+\fra72N_cN_f$ degrees
of freedom in the high $T$ phase. These are equal at $x_f=4$ and if latent heat is naively
assumed to be proportional to the jump in the number of degrees of freedom, one might rather
expect $L$ to decrease when $x_f$ increases.
\item
Asymptotically, for large $T$, $\lambda_h\to0$ we have $\pi T=1/r_h=b(\lambda_h)/\CL_\rmi{UV}$
so that
\be
4G_5p=(\pi\CL_\rmi{UV})^3\int^\infty_{\lambda_h} dx (-T'(x))\,T^3(x)=\fra14 (\pi\CL_\rmi{UV})^3T^4.
\label{plimit}
\ee
If one for large $T$ assumes that the system becomes
a gas of non-interacting bosons and fermions one should have
\be
{p\over T^4}=(1+\fra74 x_f){\pi^2\over45}N_c^2.
\label{SBpressure}
\ee
This is obtained from \nr{pees} if
\be
{1\over 4G_5}={4\over45\pi}{1+\fra74 x_f\over \CL_\rmi{UV}^3}N_c^2,
\label{pnormal}
\ee
which can be used to normalise the pressure.
\item
The above was for the simple $T(\lambda_h)$ in Fig.~\ref{figTlah}. Depending on the potentials,
more complex structures can appear, as analysed in the following section.

\item
To present results for $p/T^4$ we choose to normalise it so that it approaches
at large $T$ the ideal gas Stefan-Boltzmann pressure according to \nr{SBpressure}.
However, we have no dynamical argument for fixing the $x_f$ dependence of $\CL_\rmi{UV}$
in \nr{pnormal}. We shall present the
phase diagrams for two choices, for the automatically SB-normalised case (see Eq.~\nr{pnormal})
\be
\CL_\rmi{UV}=(1+\fra74 x_f)^{1/3},\quad W_0={12\over x_f}
\biggl[1-{1\over(1+\fra74 x_f)^{2/3}}\biggr],
\label{SBnormalisedL}
\ee
and for the $W_0$ fixed case
\be
\CL_\rmi{UV}={1\over\sqrt{1-\fra1{12}W_0x_f}},\quad W_0=0,\,\fr{12}{11},\,
\fr{24}{11}
\label{W0fixed}
\ee
In the former case one simply has
\be
{1\over 4G_5}={4\over45\pi}N_c^2
\ee
and in the latter case\footnote{Notice that in this case the glue part of the V-QCD action will
also depend on $x_f$ through the normalization factor $1/4G_5$.}
\be
{1\over 4G_5}={4\over45\pi}{1+\fra74 x_f\over (1-\fra1{12}x_fW_0)^{2/3}}N_c^2;
\ee
the factor $N_c^2$ is furthermore often implied, i.e., results for $p/(N_c^2T^4)$
are given.
\ei

\begin{figure}[!tb]
\centering

\includegraphics[width=0.49\textwidth]{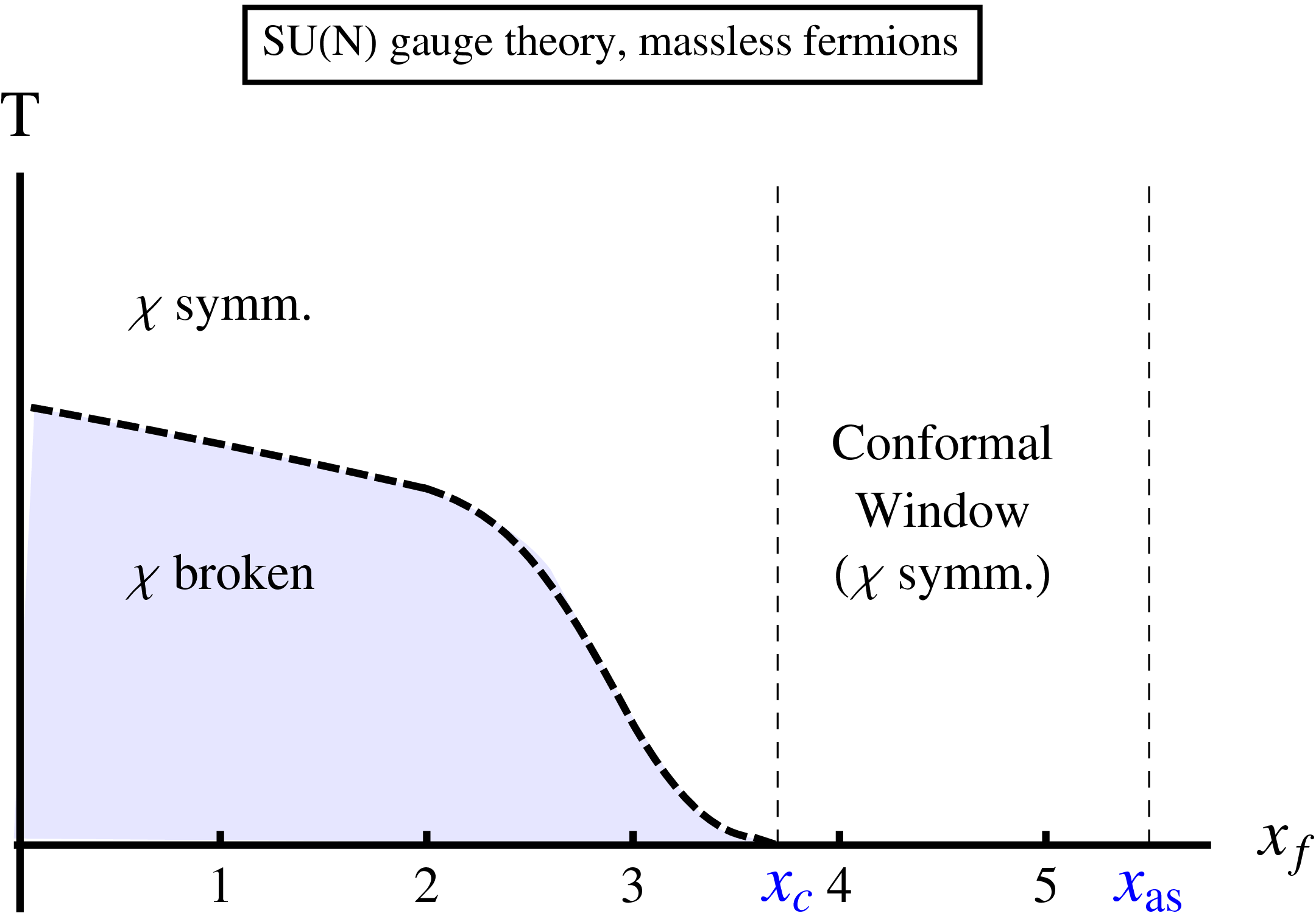}

\caption{\small Qualitative behavior of the transition temperature between the low and high
$T$ phases of V-QCD matter.
 }
\label{figxTphases}
\end{figure}

\begin{figure}[!tb]
\centering

\includegraphics[width=0.49\textwidth]{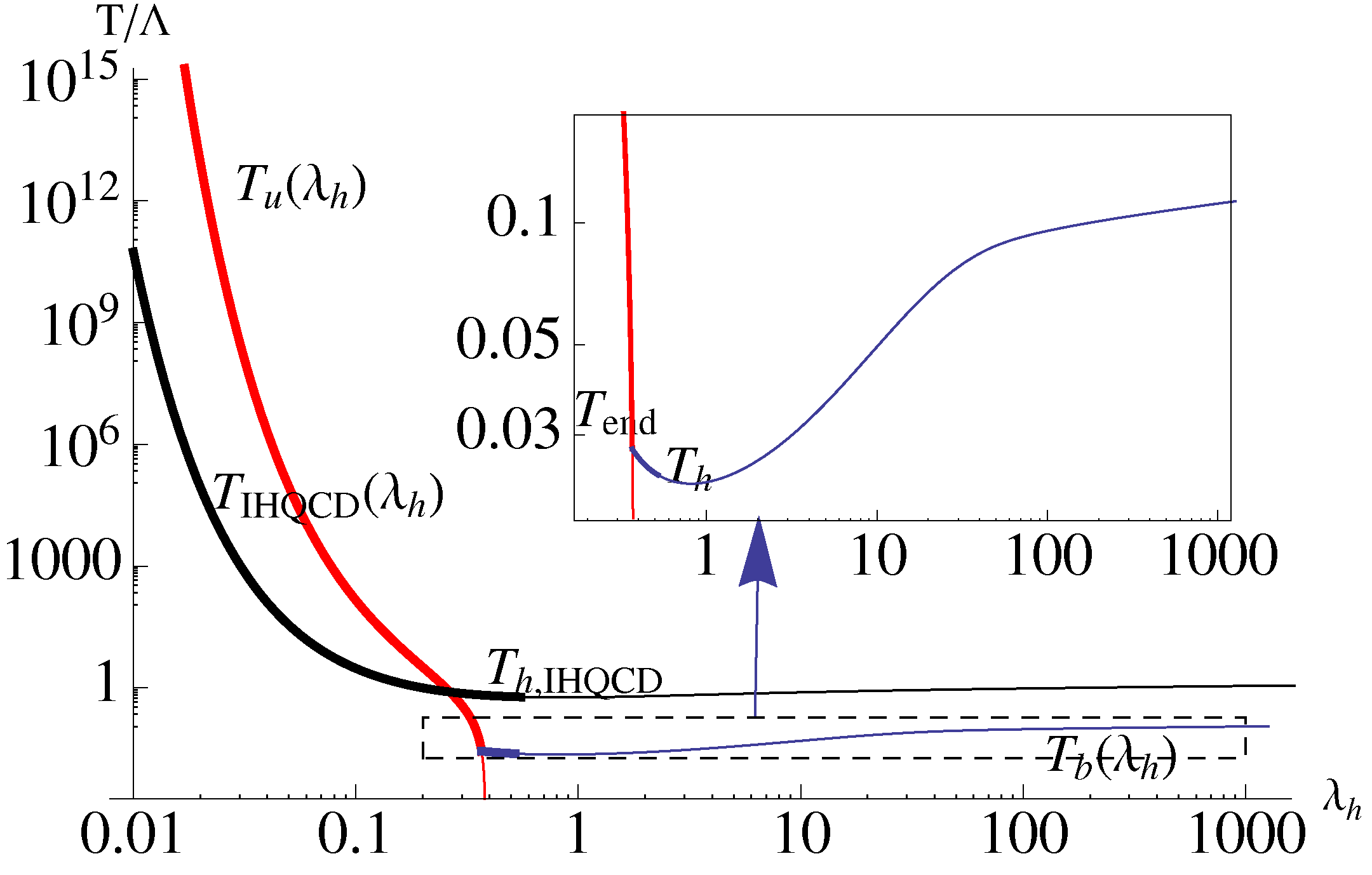}

\vspace{0.1mm}
\includegraphics[width=0.49\textwidth]{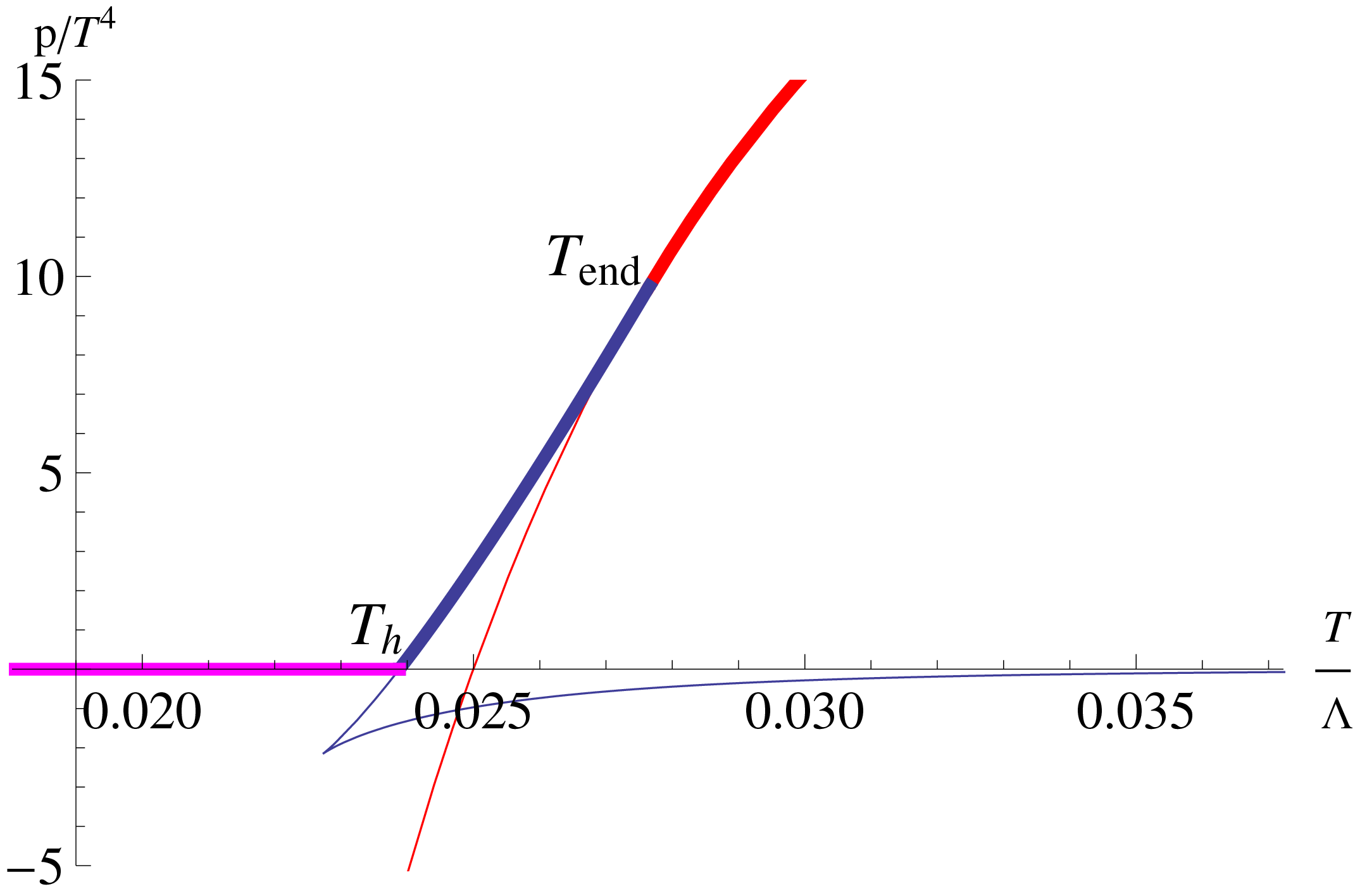}%
\includegraphics[width=0.49\textwidth]{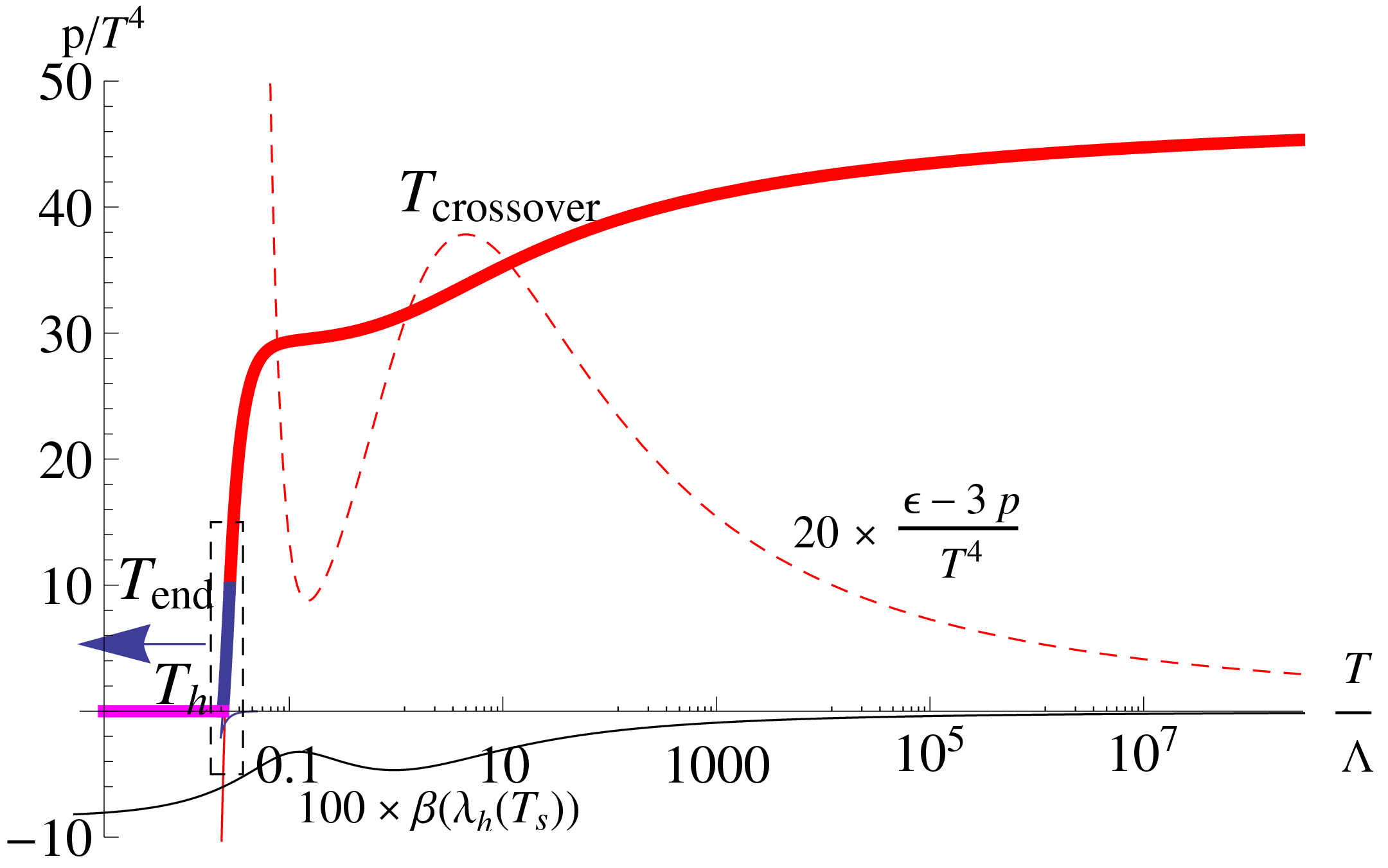}

\caption{\small
Examples of the $T_\rmi{end}$, $T_{h}$ and $T_\rmi{crossover}$ transitions in potential II
with Stefan-Boltzmann -normalization of $\CL_{\rmi{UV}}$ and with $x_f = 3$
(see also Fig.~\protect\ref{figTTransitionsSB}).
\emph{Upper}: The temperature $T(\lambda_h)$ . The curving of
$T_u(\lambda_h)$ at $\lambda_h \sim 0.2$, $T \sim 2$ is related to the crossover.
The inset shows the minimum of $T_b(\lambda_h)$, which causes $p_b$ to be positive between
$T_h$ and $T_{\rmi{end}}$. For comparison, we also plot $T(\l_h)$ for IHQCD with $x_f=0$.
\emph{Lower left}: $p/T^4$ in a close-up around the region of the $T_h$
and $T_{\rmi{end}}$ -transitions. \emph{Lower right}: an overview of the pressure in the same case, also
showing the interaction measure, the peak of which determines the position of $T_{\rmi{crossover}}$. The black
curve shows the vacuum beta function, scaled to fit, as a function of temperature in the symmetric phase,
so that $\beta(T) = \beta(\lambda_u(T))$, where $\lambda_u(T)$ is the inverse function of $T_u(\lambda_h)$.
The walking maximum of the beta function clearly coincides with the plateau related to $T_\rmi{crossover}$,
confirming that the $p/T^4 \sim \mathrm{constant}$ phase below $T_\rmi{crossover}$ is indeed the
quasi-conformal phase related to walking dynamics.
}
\label{figTransitionsTend}
\end{figure}

\begin{figure}[!tb]
\centering

\includegraphics[width=0.49\textwidth]{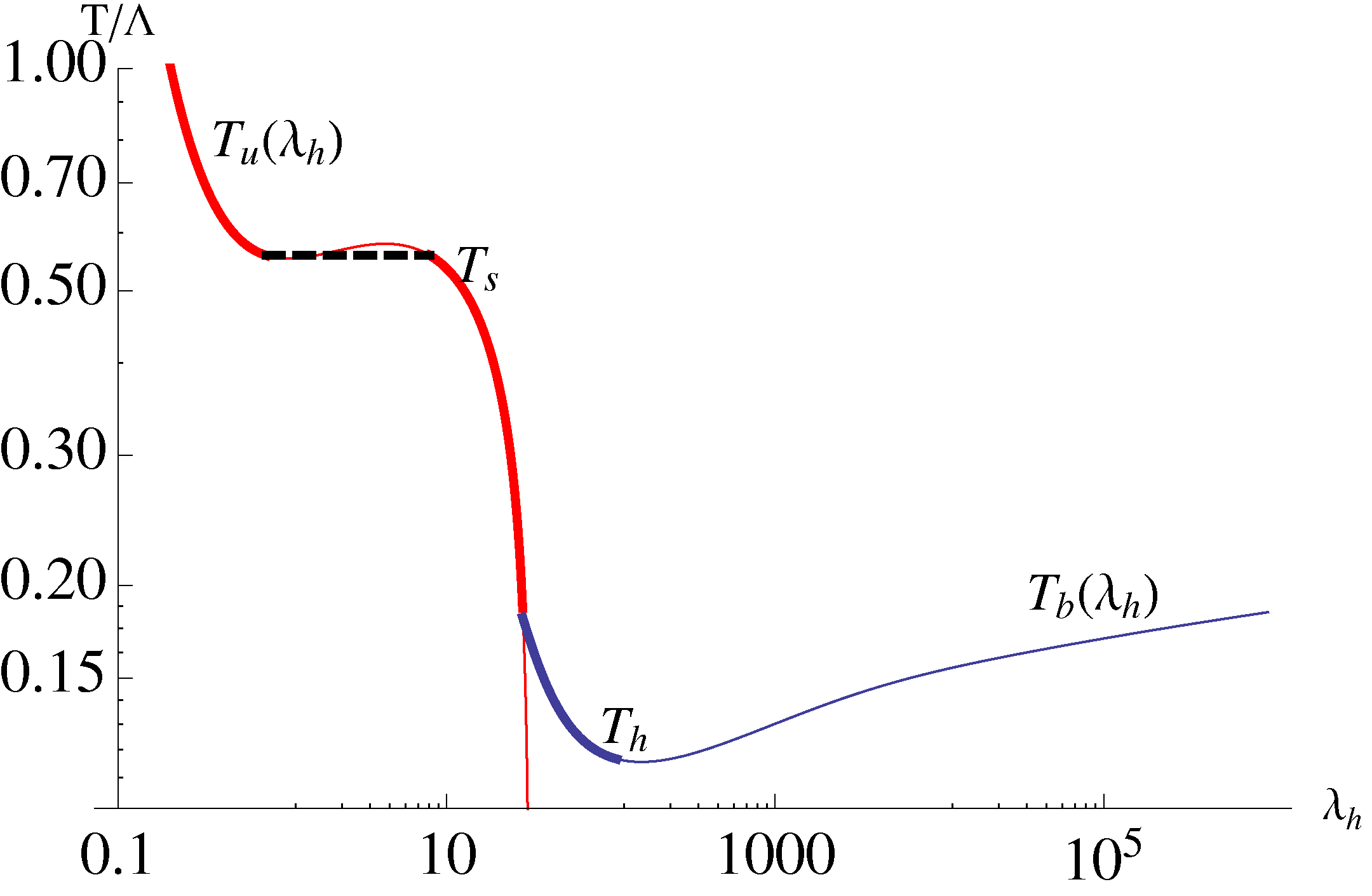}
\includegraphics[width=0.49\textwidth]{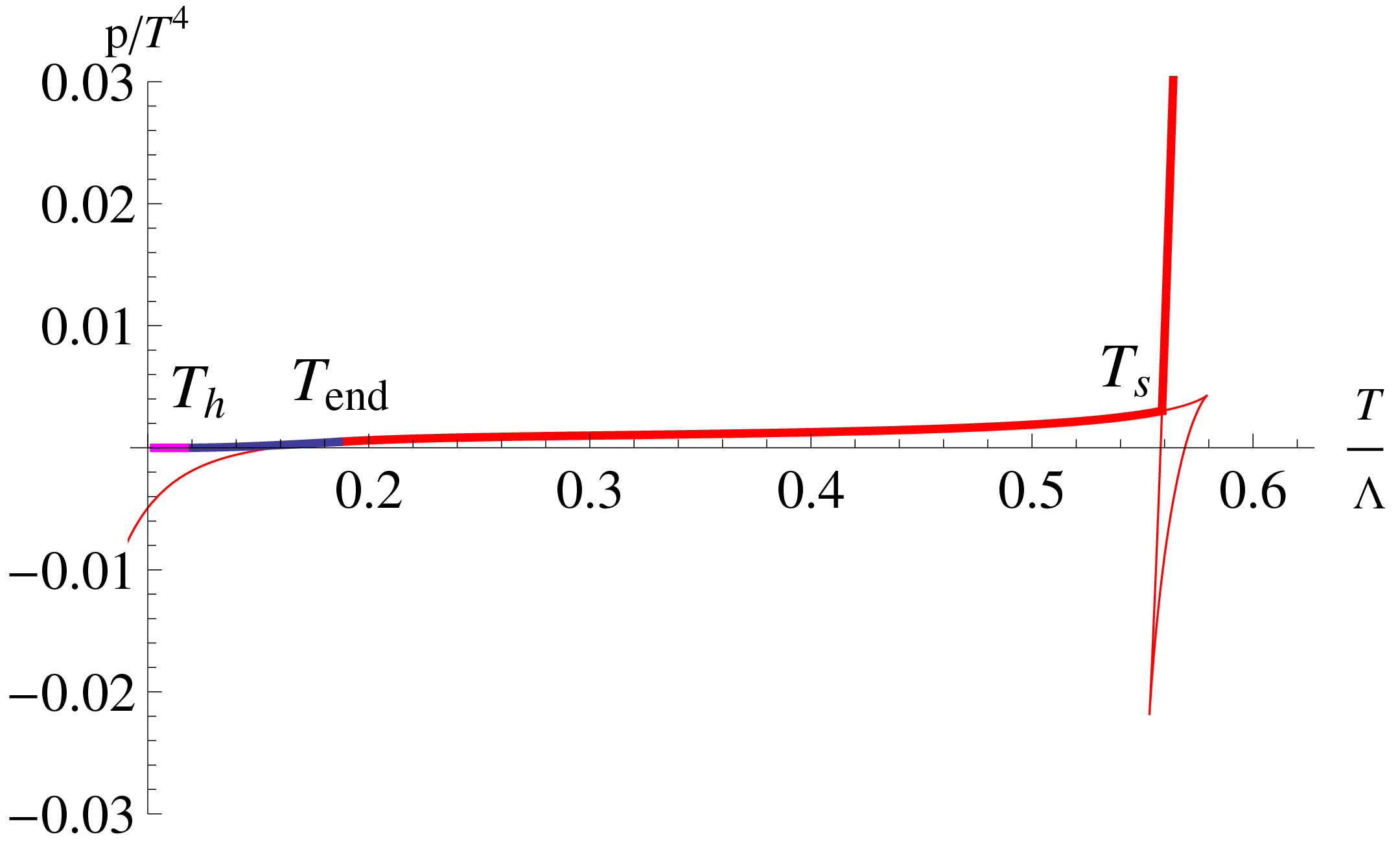}

\caption{\small An example of the $T_\rmi{s}$ transition in potential I
with $W_0 = 24/11$ and with $x_f = 0.3$ \emph{Left}: The local maximum and minimum which generate the
1st order $T_{\rmi{s}}$ -transition.  \emph{Right}: $p(T)/T^4$ in the region around which the
1st order $T_\rmi{s}$ transition takes place, extending to smaller $T$ in order to show the
relation to the $T_h$ and $T_\rmi{end}$ transitions.}
\label{figTransitionsTs}
\end{figure}

\section{Results for the phase structure} \label{secphases}
\subsection{Phase transitions}
Let us first review what one qualitatively expects for the phase structure of V-QCD when
the number of (massless) fermions is changed \cite{tuominen}. This is shown in Fig.~\ref{figxTphases}, where the
transition temperature between a low $T$ and a high $T$ phase is plotted as a function
of $x_f$.

A few reminders are in order. In the absence of quarks, YM has a $Z_{N_c}$ center symmetry that
is central in the definition of the confined and deconfined phases.
The relevant order parameter is the Polyakov loop that transforms nontrivially under $Z_{N_c}$.
If its expectation value is zero, we are in the confined phase, while the expectation value
becomes non-zero in the deconfined phase.

This expectation value is simple to calculate holographically, \cite{noro}. It corresponds to
a string world-sheet along the time circle, and hanging down straight in the
holographic (radial) direction. The important difference is where it ends.
At zero temperature, this worldsheet extends to $r\to \infty$ and is the world-sheet of a
free quark. Standard renormalization
subtracts its contribution completely and therefore
the Polyakov loop vev is zero (to leading order in $1/N_c$) in the zero temperature phase.

In a regular black-hole phase, the worldsheet terminates at the horizon and after subtraction
the Polyakov loop expectation value is non-zero.
This is in agreement with the identification of black-hole phases generically as deconfined phases.

In the presence of massless quarks, the center symmetry is not a symmetry any more, and the
Polyakov loop is not an order parameter.
However at large $N_c$, there is alternative order parameter for a deconfined phase, namely
 the $N_c$ dependence of the free energy, $F$. In the confined phases $F\sim {\cal O}(1)$ while
 in deconfined phases, $F\sim {\cal O}(N_c^2)$. Again, with this criterion, the vacuum solutions
(without horizons) are ``confining"   ($F\sim {\cal O}(1)$) while any black hole solution
with regular horizon is ``deconfined" ($F\sim {\cal O}(N_c^2)$).
It is therefore natural to use this criterion in our analysis in order to define deconfined phases.

The true symmetry in the case of massless quarks is chiral symmetry. This
always has an order parameter, the chiral condensate, that
distinguishes  chirally symmetric from chirally broken phases.

Given the remarks above, we summarize what we would expect.

\bi
\item
For $x_f=0$ one has the Yang-Mills 1st order phase transition between a confined and
deconfined phase. In the high $T$ deconfined phase, the $Z_{N_c}$ symmetry is broken.
\item
For a somewhat higher  $x_f$ one expects that there still is a 1st order transition. However,
now this transition will involve chiral symmetry
breaking/restoration.

\item
For $x_f$ approaching $x_c$ one expects the transition temperature to decrease rapidly
as follows from Miransky scaling.

\item
For $x_f$ in the conformal window, $x_c<x_f<11/2$, both the low and high $T$ phases are
conformal ones, which can be separated by a crossover. The only transition happens at $T=0^+$
like in the AdS black hole in Poincar\'e coordinates.
\ei

The models we consider contain the full fermion backreaction and therefore predict a somewhat more detailed
phase structure. New phase transitions of different orders can take place, lines can
split in two, etc.
The behavior in the conformal window ($x_c<x_f<11/2$) is nonetheless always simple: there are
no transitions, but a crossover between the low and high temperature conformal phases.
Therefore we concentrate first on the phase structure in the region below the conformal transition ($x_f<x_c$).

While the details of the phase structure depend on the choice of potential, the various phase transitions
encountered appear in certain systematic ways. We will define a consistent
notation, and describe the classes
of transitions, assuming the system is heated up and we go from low temperatures to high temperatures.

To motivate the notation, we
first list the various transitions and the corresponding temperatures.
\bi
\item $T_h$ is the analogue of the QCD hadronisation transition
if it is the chiral restoration transition (chirally symmetric $\to$  chirally broken).
\item $T_{\rmi{end}}$ is the end point of the curve
$T_b(\l_h)=T(\l_h,\tau_{h0}(\l_h,m_q=0))$, 
which contains
the black holes with tachyon hair. For values of $\lambda_h$ smaller than at this endpoint,
the black-holes have no tachyon hair.
\item $T_{\rmi{crossover}}$ marks the position of  a crossover.
This crossover is defined by the position of the peak in the equation-of-state ($(\epsilon-3p)/T^4$)
as a function of temperature.
\item $T_\rmi{s}$ takes place at small $x_f$ within the chirally symmetric phase when one can jump
from one decreasing branch of $T_u(\l_h)$ (no tachyon hair) to another.
\item Finally $T_{12}$ involves the splitting of one 1st order line to two.
\ei

With this notation we may now describe
in detail
the various types of transitions and crossovers we have found,
and show examples of each case. In the figures we denote the stable phases with thick lines and meta- and unstable
phases with thin lines.

\begin{figure}[!tb]
\centering

\includegraphics[width=0.49\textwidth]{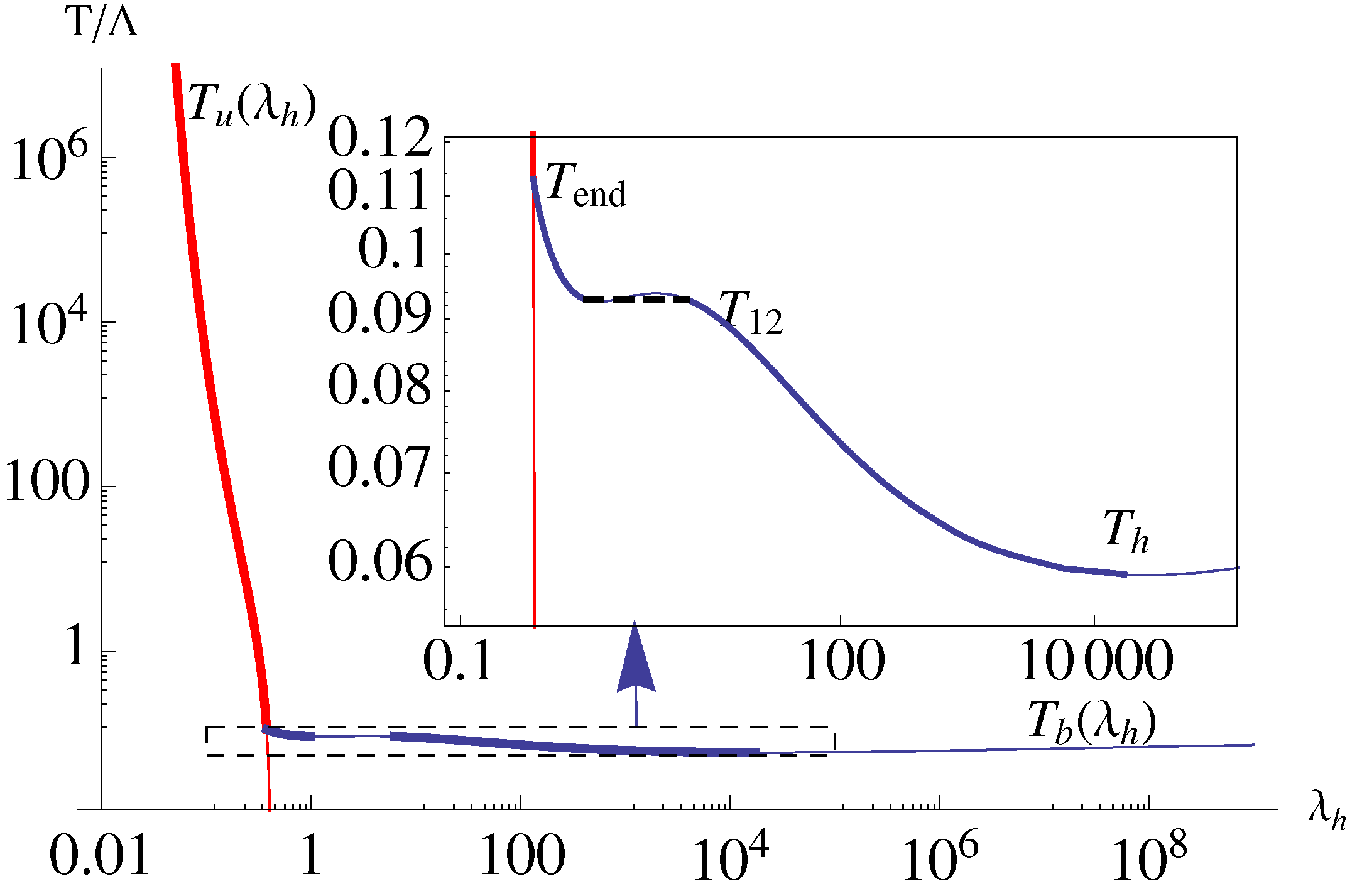}
\includegraphics[width=0.49\textwidth]{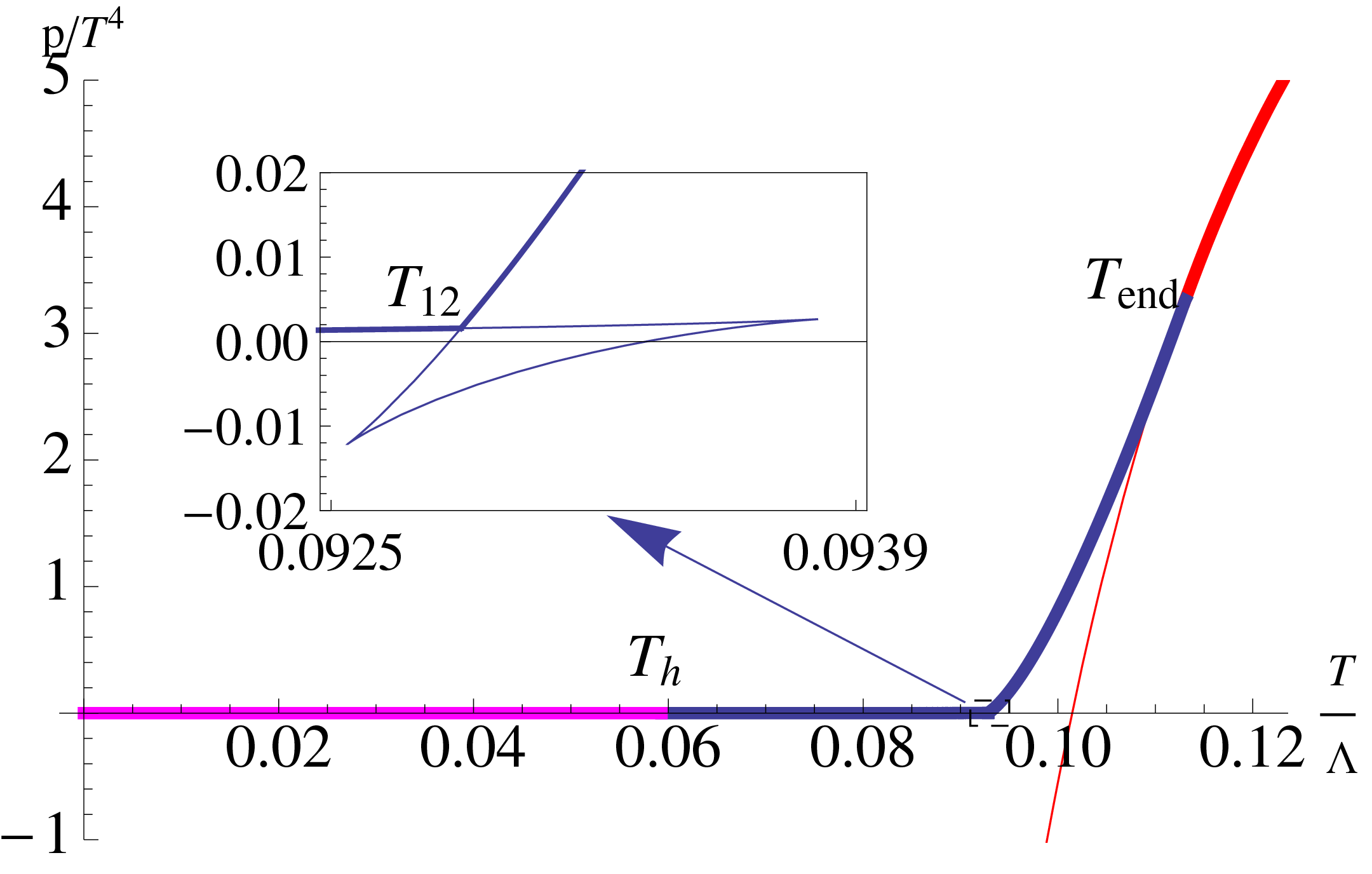}

\caption{\small An example of the $T_\rmi{12}$ transition in potential I
with $W_0 = 12/11$ and with $x_f = 3.5$ \emph{Left}: The overall structure of $T(\lambda_h)$,
with an inset showing the maximum and minimum in more detail.
\emph{Right}:  a close-up of $p(T)/T^4$ in the region where
the $T_{12}$ -transition happens, with an inset showing further detail.}
\label{figTransitionsT12}
\end{figure}

\begin{itemize}
\item The 1st order hadronisation transition at $T_h$, happens either between the chirally broken  $\to$
a  chirally symmetric phase (see Fig.~\ref{figTlah}) or from a chirally broken $\to$
a chirally broken phases
(see Fig.~\ref{figTransitionsTend}).\footnote{There is also the special case of potentials
I$_*$ at low $x_f$ where the transition analogous to $T_h$ takes place from a chirally
{\em symmetric} thermal gas to chirally symmetric black hole phase (see Fig.~\ref{figPhasesModPotI}).}
As described above, our normalization for pressure is
such that the pressure of the ($T=0$) hadron gas phase is zero.
In the holographic setup, this transition is between that of the black hole phases, whose pressure remains
positive down to the lowest temperature, and the hadron gas phase. The transition takes place at the
temperature $T_h$ where the pressure of the BH phase reaches zero. Whether this phase is chirally symmetric or
non-symmetric depends on the potential choices and $x_f$.
For an example, see Fig.~\ref{figTTransitionsSB}.

\item The 2nd order chirally broken  $\to$
chirally symmetric transition at $T_{\rmi{end}}=T(\l_\rmi{end},0)=T(\l_\rmi{end},\tau_{h0}(\l_\rmi{end},m_q=0))$,
see Figs.~\ref{figTlah} or \ref{figTransitionsTend}.
Since the chiral symmetry breaking
solution starts to exist only above some $\lambda_{\rmi{end}}$, the system makes at that point a transition
to the chirally symmetric phase.
However,  this transition may be absent in the thermodynamic limit:
if $p_b(\lambda_h)$ is everywhere negative, the
transition is between two thermodynamically metastable phases, and the relevant saddle point is never dominant.
We denote the temperature of the transition by $T_{\rmi{end}}$.
Since this transition takes place at one single value $\lambda_h=\lambda_\rmi{end}$, both pressure
and entropy density  are
continuous ($b(\lambda_h)$ does not jump).
Therefore, only $p''(T)$ or $c_s^2$ are
discontinuous, and the transition is of second order.

\item The high-$T$ chirally symmetric $\to$ chirally symmetric crossover at $T_{\rmi{crossover}}$,
see Fig.~\ref{figTransitionsTend}.
This is a crossover which is expected on general grounds when $x_f$ is near but below $x_c$.
It reflects the change of the dynamics from the walking region, where the QCD coupling constant evolves slowly,
to the region in the deep UV where it runs.
In this sense, above the crossover it is the nontrivial fixed point theory that controls the thermodynamics, while
below the crossover it is the YM-like theory that controls the dynamics.

The thermodynamics behaves as follows:
At first $p/T^4$ stabilizes to some intermediate value, before eventually increasing very slowly
toward the Stefan-Boltzmann limit. For the potentials studied here,
this creates a clear, although very broad, peak in the interaction measure,
and the position of that peak can be used to define the temperature $T_{\rmi{crossover}}$ at
which there is a crossover.
The peak of the interaction measure is also observed at low values of $x_f$.
In this region, however, $T_{\rmi{crossover}}$ is typically relatively close to $T_h$. Note also that
for SU($N_c$) YM theory, $N_f=0$ the interaction measure starts decreasing immediately at $T_h$
\cite{panero},
$T_{\rmi{crossover}} \simeq T_h$.

\item The 1st order high-T chirally symmetric $\to$ chirally symmetric transition at $T_{\rmi{s}}$,
see Fig.~\ref{figTransitionsTs}:
With some choices of potential, at low
$x_f$, $T_u(\lambda_h)$ in the chirally symmetric (unbroken) part of the solution develops a local maximum
and minimum. There are then two values of $\lambda_h$ between which both the pressure and the
temperature of the solution match, and there is a 1st order transition between these two branches
of the chirally symmetric solution.
Interestingly, $T_s$ approaches the temperature of the YM transition in IHQCD as $x_f \to 0$
(see the discussion in Section \ref{xfto0}).

\item The 1st order chirally broken $\to$ chirally broken transition at $T_{12}$,
see Fig.~\ref{figTransitionsT12}.
This happens in the chirally
non-symmetric phase, with potential I and $W_0 = 12/11$,
$T(\lambda_h)$ which
develops a local minimum and maximum at
large $x_f$. This again induces a 1st order transition, which we denote by $T_{12}$.
In this case the single 1st order transition at $T_h$ splits into two 1st order transitions as
$x_f$ increases above some critical value.
Above this value, the transition with higher (lower) temperature is identified as $T_{12}$ ($T_h$).
\end{itemize}

\begin{figure}[!tb]

\centering

\includegraphics[width=0.49\textwidth]{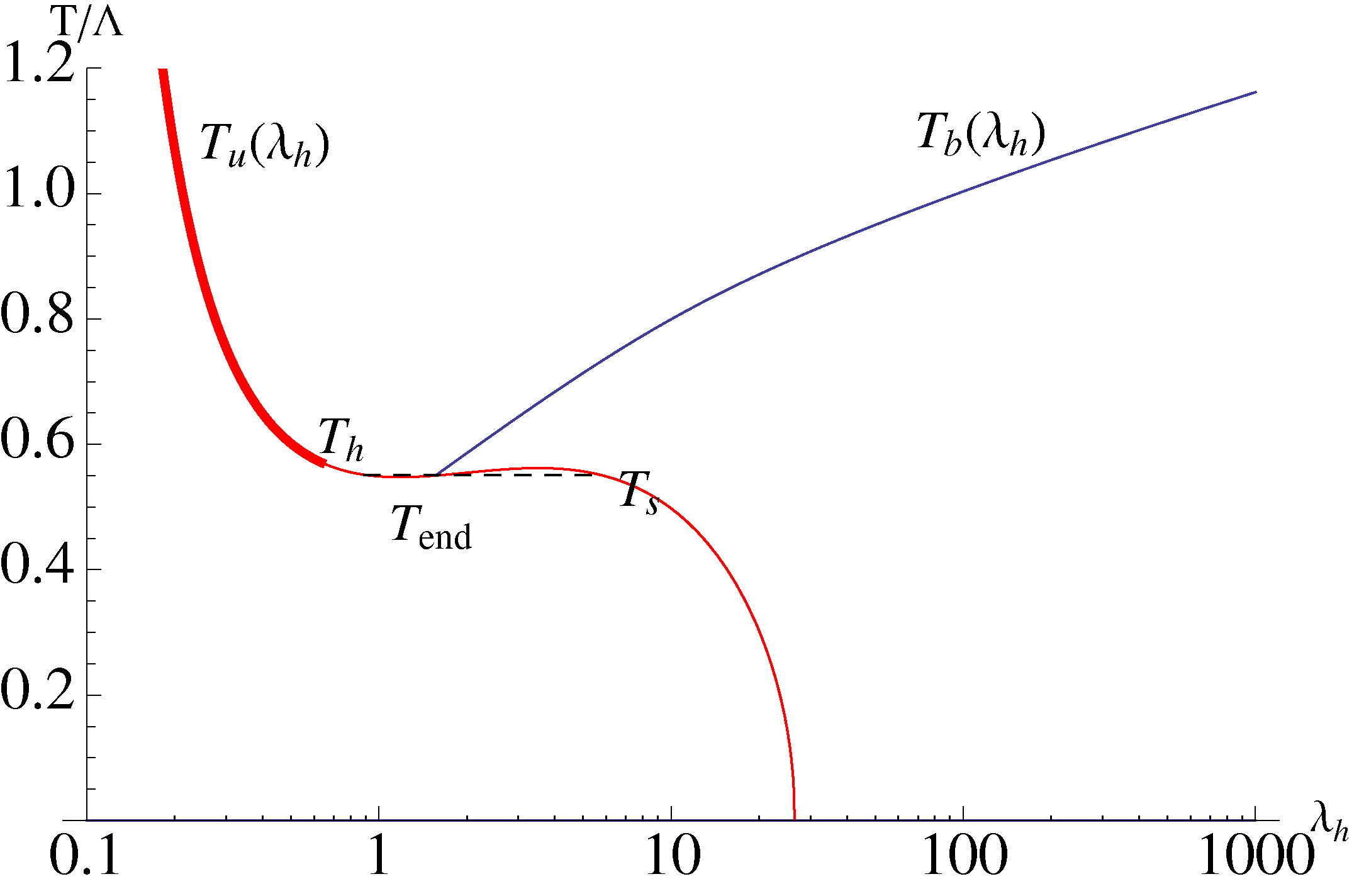}
\includegraphics[width=0.49\textwidth]{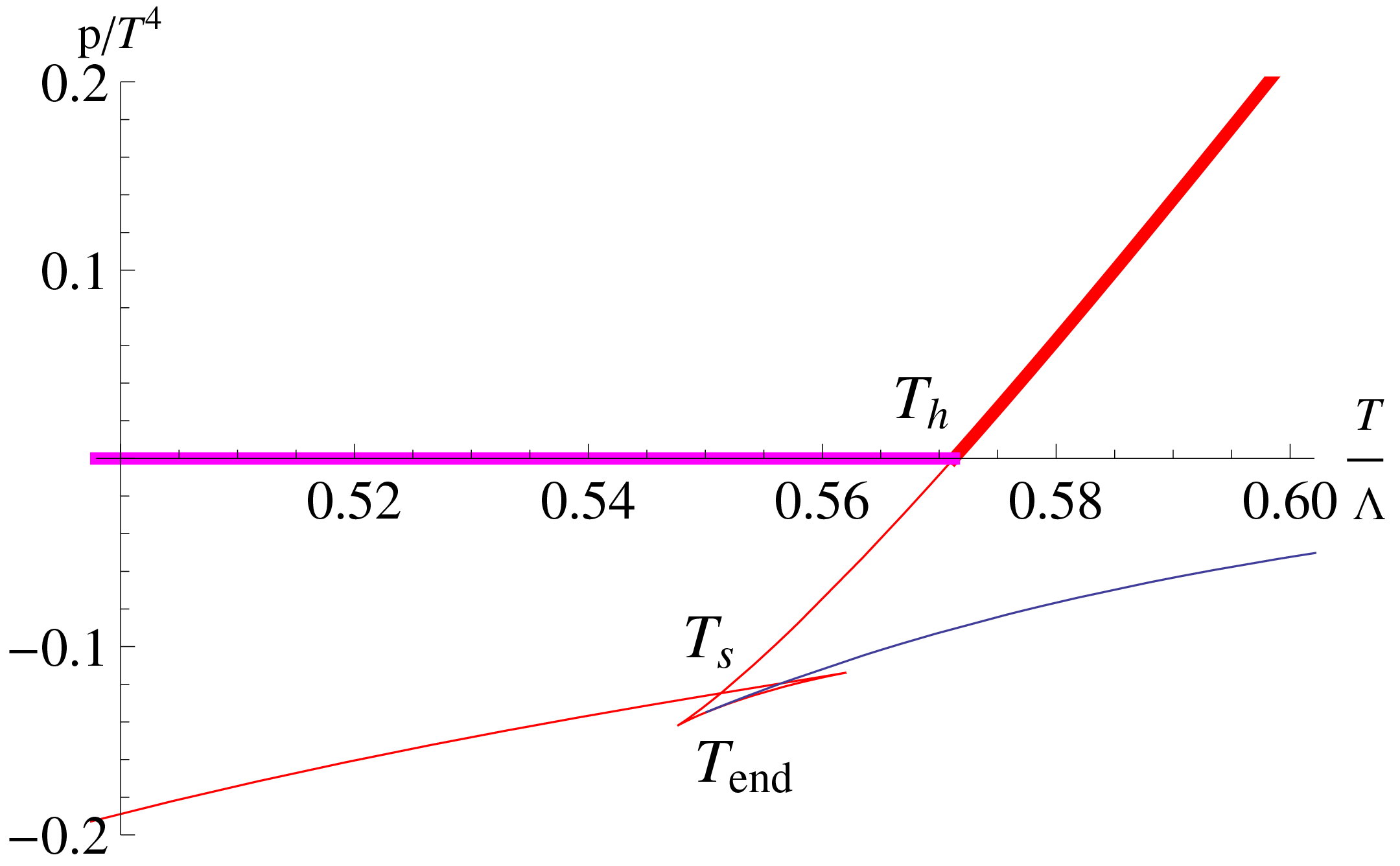}

\caption{\small An example of a configuration where all but the crossover and hadronisation
transitions $T_\rmi{crossover}$, $T_h$, are in the thermodynamically unstable region, in
the initial stages of the approach to the IHQCD limit. The potential is II
with $W_0 = 12/11$ and with $x_f = 0.4$ \emph{Left}:The temperature $T(\lambda_h)$.
Note that everything to the right of the $T_h$ transition is in the unstable phase.
\emph{Right}:  $p(T)/T^4$ in the region where
the $T_h$ transition and the unstable $T_\rmi{end}$ and $T_\rmi{s}$ -transitions happen.}
\label{figTransitionsUnstab}
\end{figure}

\subsection{Class-II Potentials}
Let us then discuss the details of the phase structure for the various potentials and
choices of $W_0$ defined in Sec.~\ref{sec2pt2pt2}.

We take Class-II first since it leads systematically to a simple phase structure.
We observe two possibilities: First,
for $x_f$ up to some value $x_\chi<x_c$ the 1st order deconfinement and chiral transition
temperatures coincide, $T_d=T_\chi$, from this value up to $x_c$ one has $T_\chi>T_d$
and the higher chiral transition is of 2nd order.
Second,
$T_d=T_\chi$ all the way up to $x_c$ and $x_\chi$ is absent.

For this choice of potentials
the tachyon diverges $\sim\sqrt{r-r_1}$ at large $r$.
The part $V_{f0}(\l)$ of the fermionic potential is given by
 Eq.~\eqref{Vf0SB} and $a(\l)$ and $\f(\l)$ are given in \eqref{akappa2}.
Notice that the deconfinement temperature $T_d$ always equals the temperature of the
``standard'' 1st order transition $T_h$ in the holographic framework.
The temperature of the chiral symmetry restoration $T_\chi$ can be either $T_\rmi{end}$ or
$T_h$ depending on the order of the transitions, see examples below.

\begin{figure}[!tb]
\centering

\includegraphics[width=0.5\textwidth]{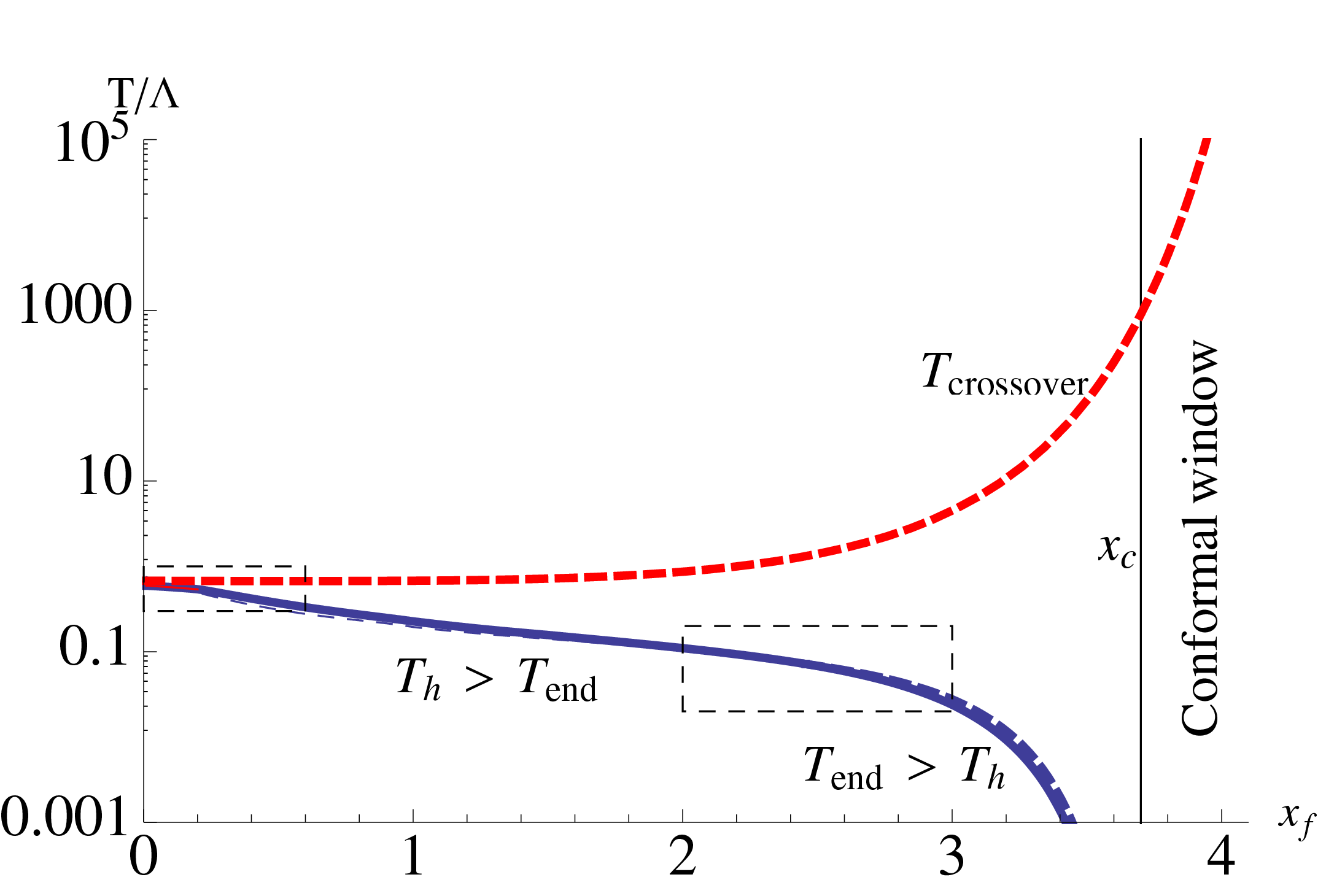}

\vspace{0.01mm}
\includegraphics[width=0.4\textwidth]{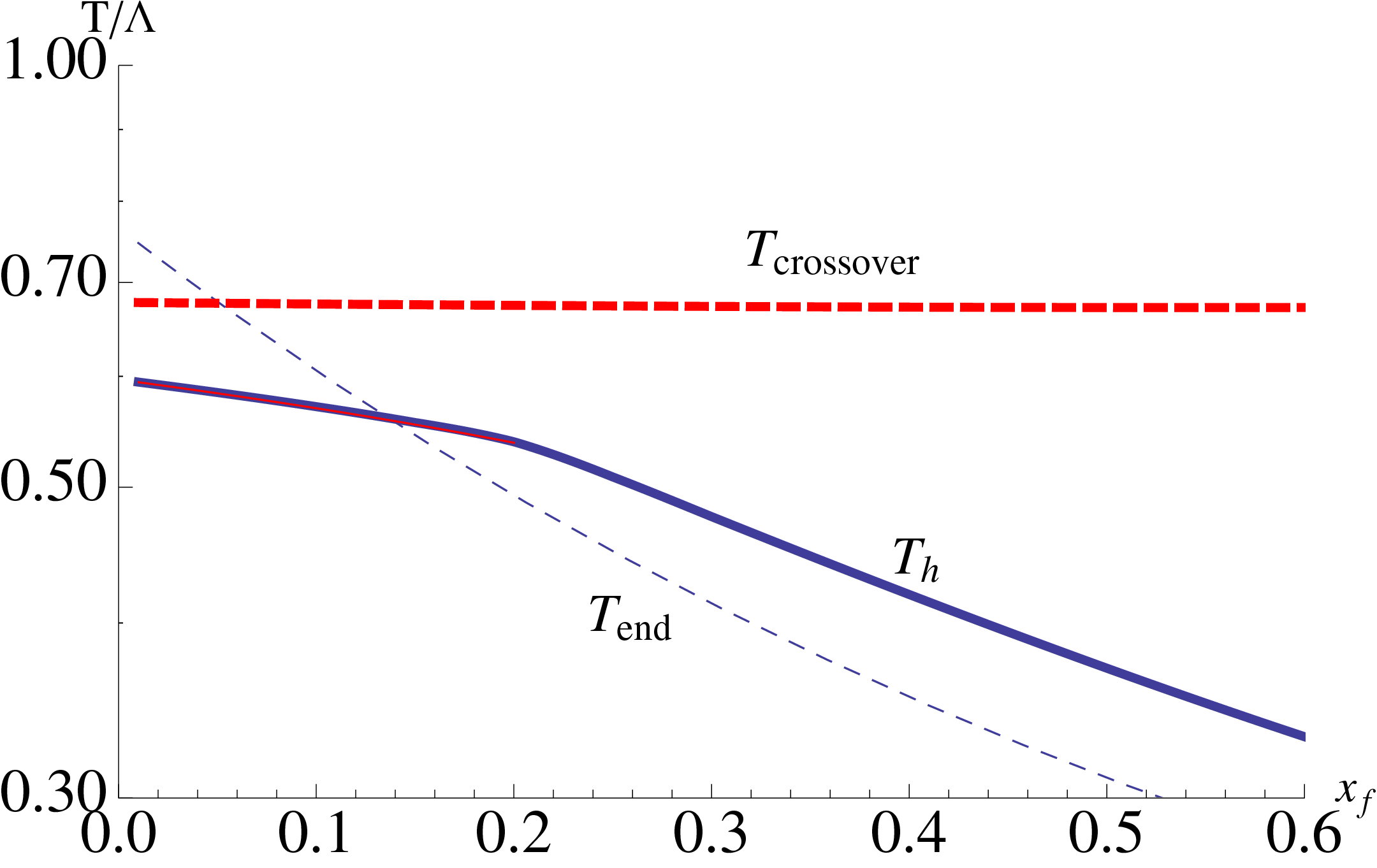}
\includegraphics[width=0.4\textwidth]{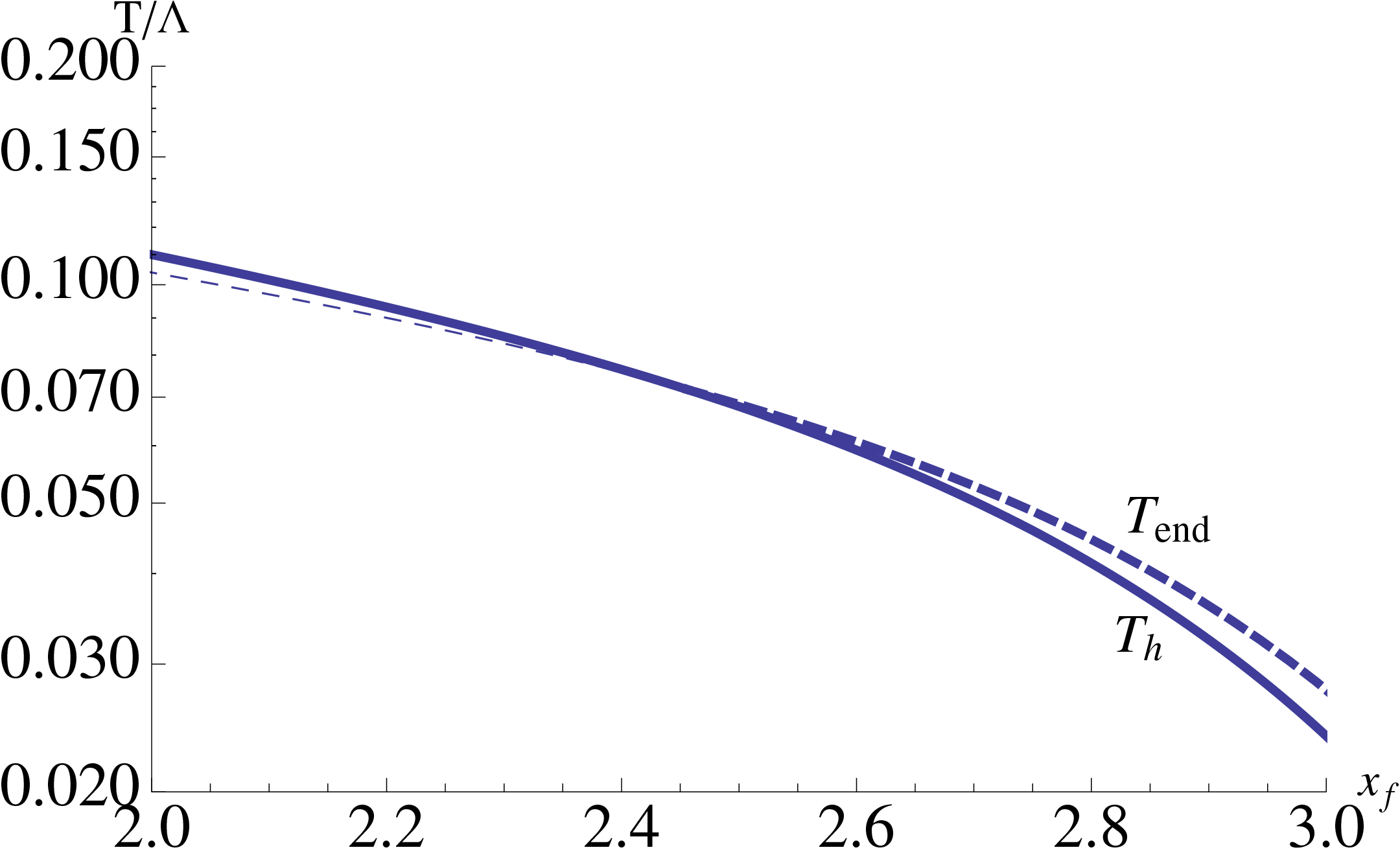}

\caption{\small \emph{Upper:} The phase diagram for potential $\mathrm{II}$,
$W_0$ Stefan-Boltzmann normalized, $x_c=3.70$. The dashed boxes show the
regions detailed in the bottom two plots. In the bottom left plot
$T_s\lta T_h$ at $x_f\lta0.2$ is not visibly separated. For discussion, see text.
}
\label{figTTransitionsSB}
\end{figure}

The result for the SB-normalised case is shown in Fig.~\ref{figTTransitionsSB}.
For $0<x_f< x_\chi \simeq 2.46$
we find that $T_\rmi{end}<T_h$,
but $T_\rmi{end}$ is
in the metastable branch of the solution.
Thus the deconfinement and chirality transitions coincide here, $T_d=T_\chi =T_h$.
In other words, if one could sufficiently supercool the
system below $T_h$ in the high-$T$ chirally symmetric phase, the symmetry
breaking transition could take place at $T_\rmi{end}<T_h$. In the thermodynamic limit
there is no supercooling and only $T_h$ is seen.

Above $x_f \simeq 2.46$, 
the second order $T_\rmi{end}$ moves above $T_h$ and becomes stable,
as seen in the bottom right plot of Fig.~\ref{figTTransitionsSB}.
Therefore, we first have  a 1st order  $T_h$ transition from the thermal gas solution
to a chirally breaking black-hole phase, and then
 a 2nd order transition
from the chirally broken low-$T$ phase to the chirally symmetric high-$T$ phase.
In other words, $T_\chi>T_d$ with a 2nd order chiral and 1st order deconfinement transition.
For a more detailed view of the thermodynamics in this region at $x_f=3$,
the reader is guided to  the left panel of Fig.~\ref{thermosigma} where the chiral condensate
as well as the energy and the pressure are plotted as functions of $T$.
The chirally symmetric crossover transition
$T_\rmi{crossover}$ is for all $x_f$,
the highest temperature transition.

For $x_f\to x_c$ both $T_\rmi{end}$ and $T_h$ are expected to approach zero as specified
by Miransky scaling. Numerical results are compatible with this.

When $x_f\to0$ one would expect that the $T_h$
transition smoothly approaches the transition temperature of large $N_c$ hot Yang-Mills
theory. Note, however, that strictly speaking the limit of YM theory demands $N_f=0$ and falls outside
the Veneziano limit $N_f\to\infty$ of QCD. Thus it is not surprising that nontrivial
metastable structures appear at $x_f\to0$.
What happens is that the curve $T=T_u(\l_h)$
of the chirally symmetric phase suddenly at $x_f \sim 0.2$
develops a local minimum similar to the one shown in red in Fig.~\ref{figTransitionsTs}. Further evolution
of this minimum is shown in Fig.~\ref{figPotxf0limit}. Associated with this there
is a first order $T_\rmi{s}$ transition in the metastable branch. It is so slightly below
$T_h$ that it is not visibly separated in the bottom left plot of Fig.~\ref{figTTransitionsSB}.
As discussed in section \ref{xfto0},
both $T_h$ and $T_\rmi{s}$ approach the transition temperature of YM as $x_f \to 0$.
$T_\rmi{end}$ crosses above all of the other transitions for low $x_f$,
but it is also in the metastable branch, see Fig.~\protect\ref{figTransitionsUnstab}
for details.

\begin{figure}[!tb]
\centering

\includegraphics[width=0.5\textwidth]{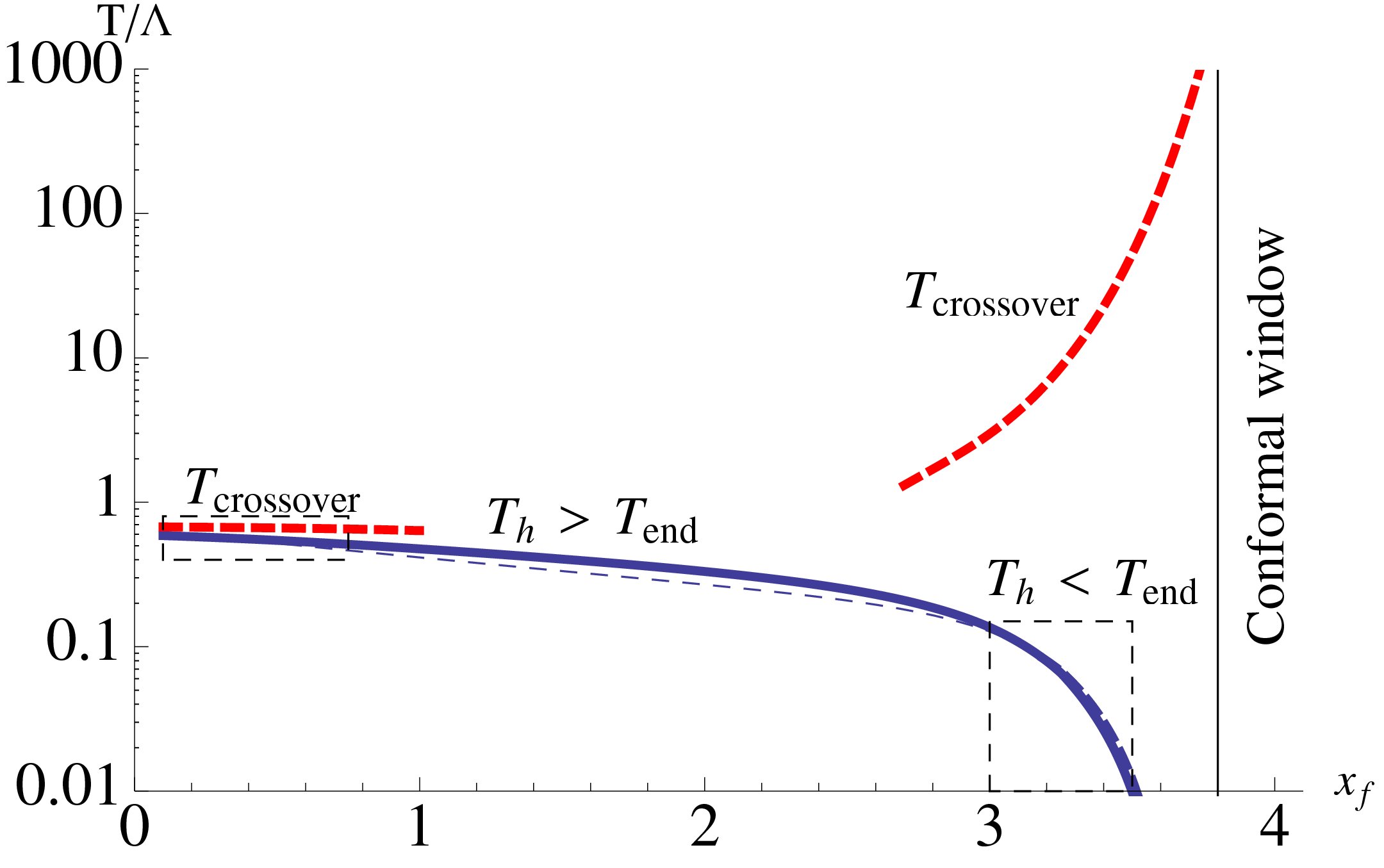}

\vspace{0.01mm}
\includegraphics[width=0.4\textwidth]{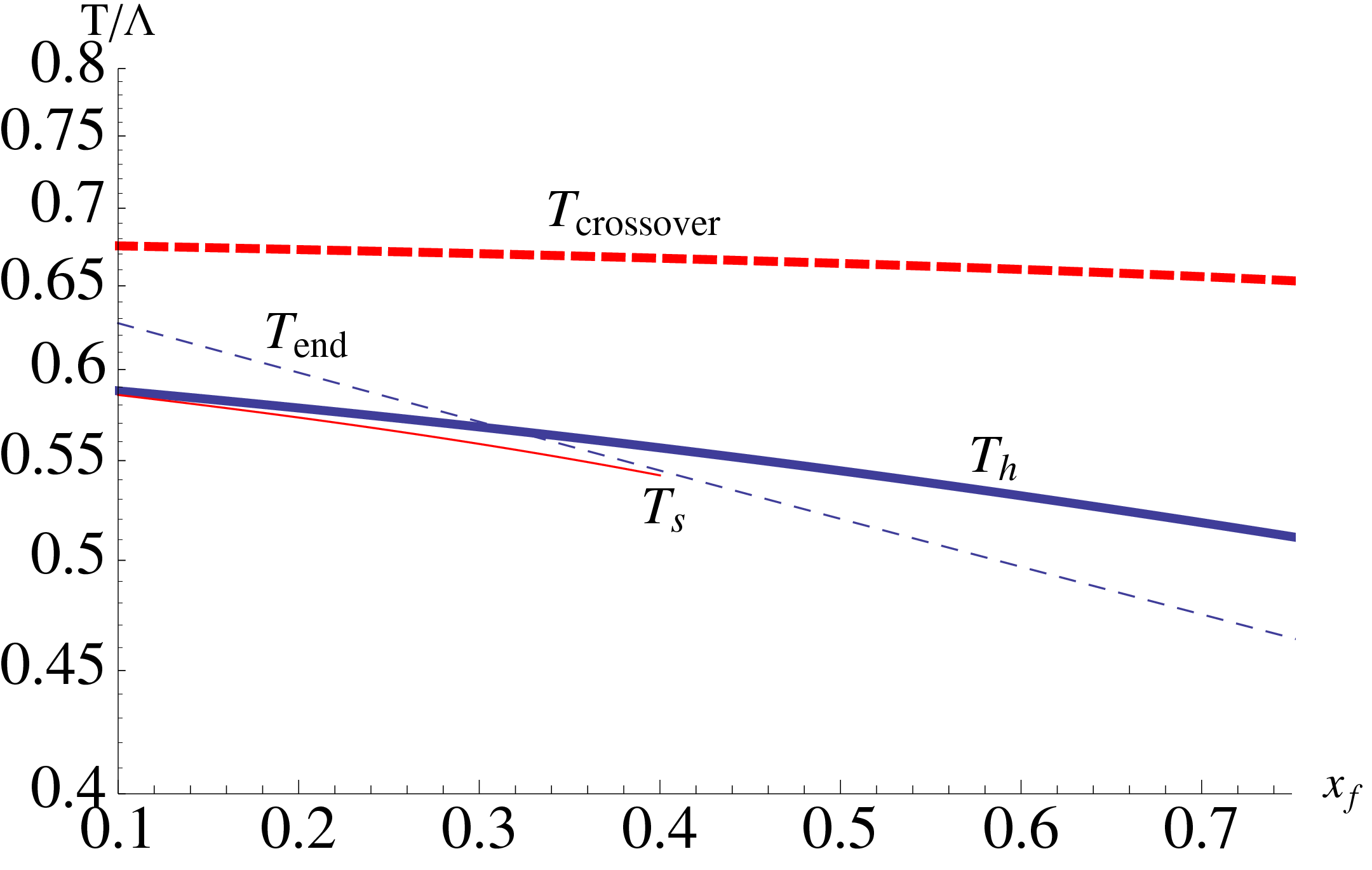}
\includegraphics[width=0.4\textwidth]{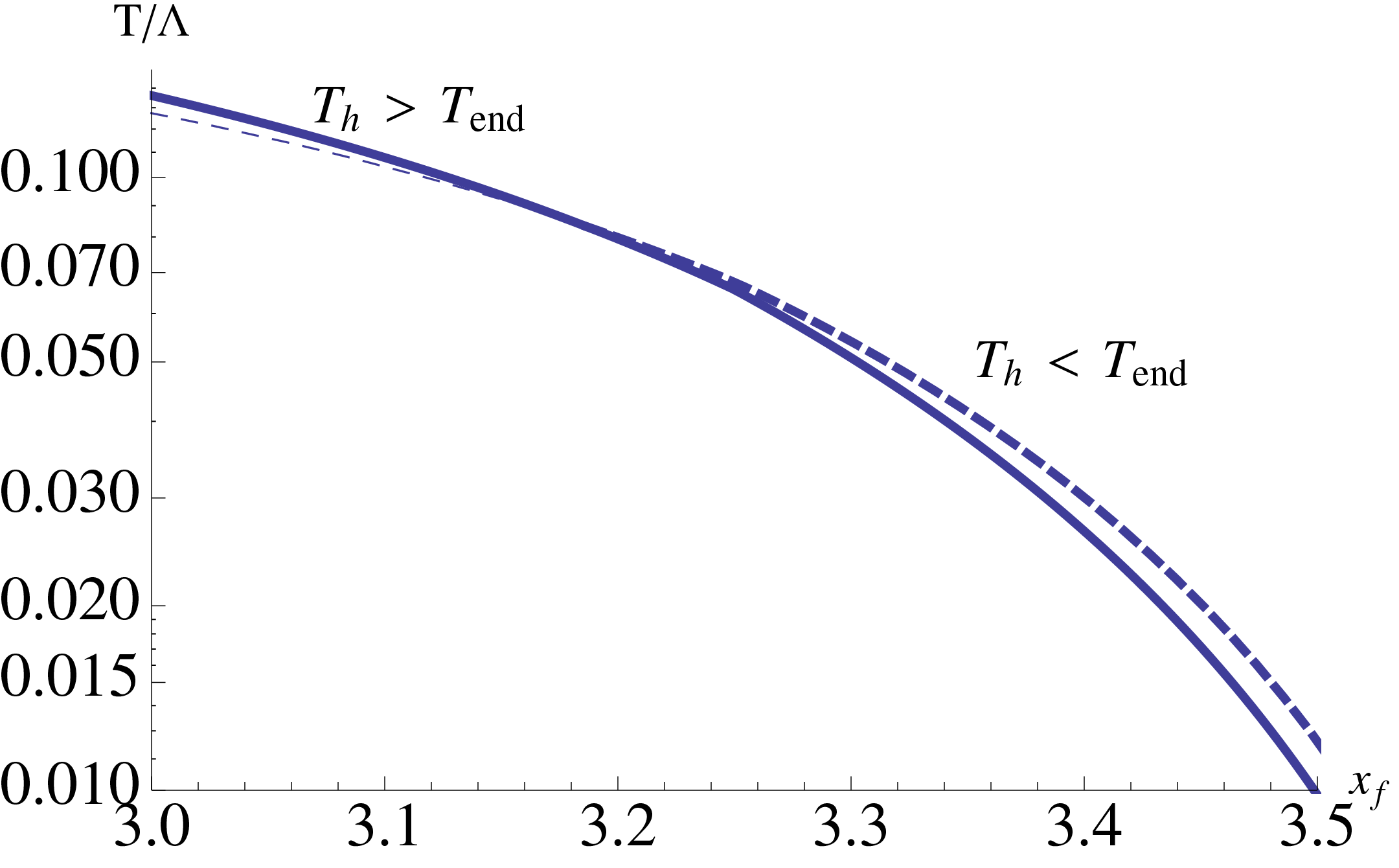}

\caption{\small \emph{Upper:} The phase diagram for potential
$\mathrm{II}$, $W_0 = 24/11$, $x_c=3.80$.
The dashed boxes show the
regions detailed in the bottom two plots. For discussion, see text.
}
\label{figTTransitions}
\end{figure}

The phase diagram for potential $\mathrm{II}$ at $W_0 = 24/11$ is shown
in Fig.~\ref{figTTransitions}.
The phase structure is qualitatively similar to the SB-normalized case.
For $x_f<x_\chi \simeq 3.19$
the stable $T_h$ transition is the only one in the thermodynamic limit,
with
$T_\rmi{end}<T_h$ in the metastable branch of the solution. Thus again $T_d=T_\chi$.
Above $x_f \simeq 3.19$,
the second order $T_\rmi{end}$ moves above $T_h$ and becomes
stable, see bottom right plot of Fig.~\ref{figTTransitions}.
Thus we again have $T_\chi>T_d$ with a 2nd order chiral and 1st order deconfinement transition.
The chirally symmetric crossover transition
$T_\rmi{crossover}$ is for all $x_f$
the highest temperature stable transition,
except between $x_f \sim 1$ to $x_f \sim 2.7$, where the interaction measure does not have a
maximum and the crossover therefore does not exist.

Now $T_s$
which appears in the metastable branch slightly below $T_h$ in Fig.~\ref{figTransitionsTs} (bottom-left)
visibly separates from $T_h$. Again $T_s$ and $T_h$ approach the temperature of the
YM-transition in the $x_f \rightarrow 0$ -limit, as discussed in section \ref{xfto0}.
$T_\rmi{end}$ crosses above the $T_h$ transition for $x_f<0.34$,
but it is also in the metastable branch, see
Fig.~\ref{figTransitionsUnstab} for details.

\begin{figure}[!tb]
\centering

\includegraphics[width=0.49\textwidth]{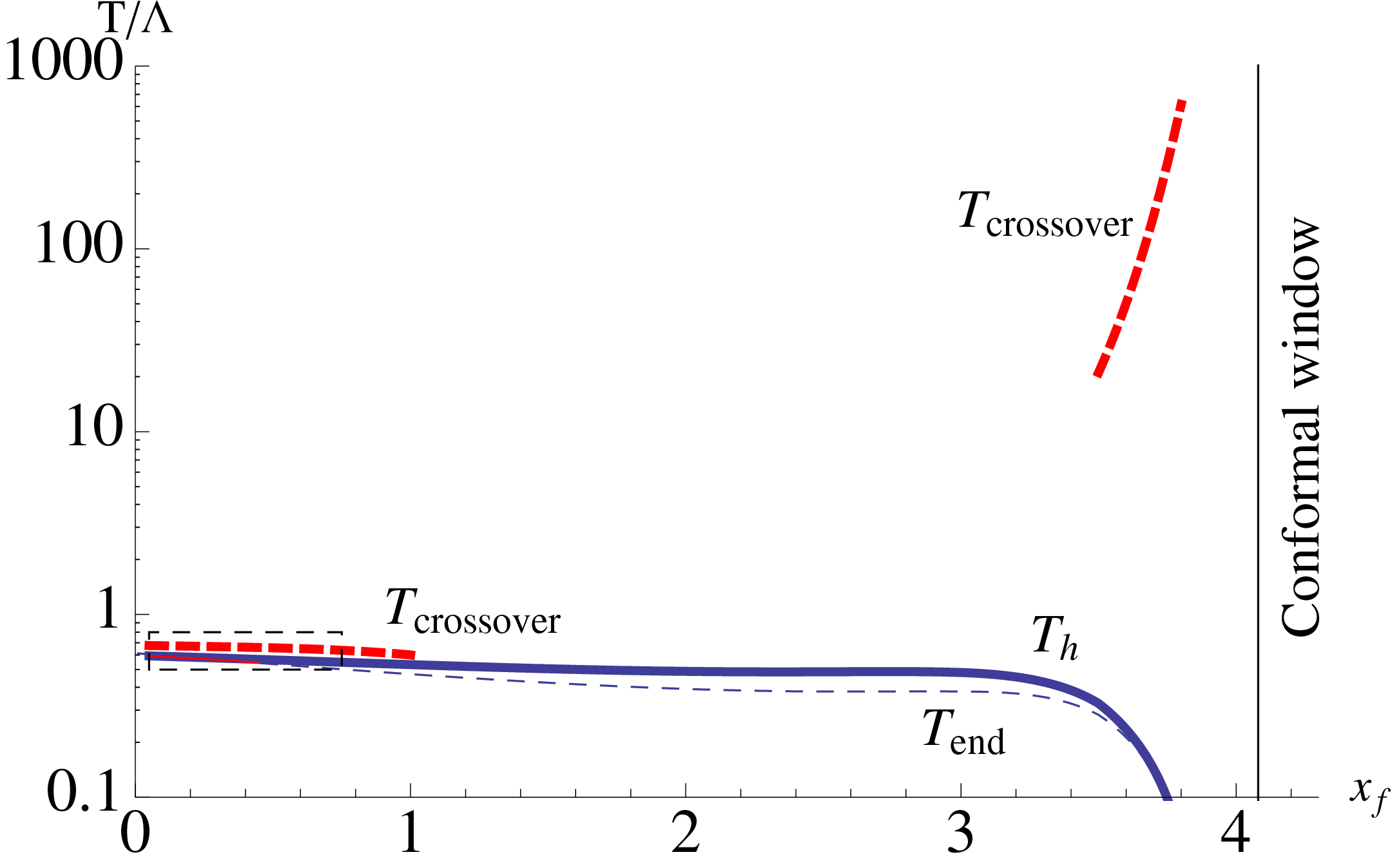}
\includegraphics[width=0.49\textwidth]{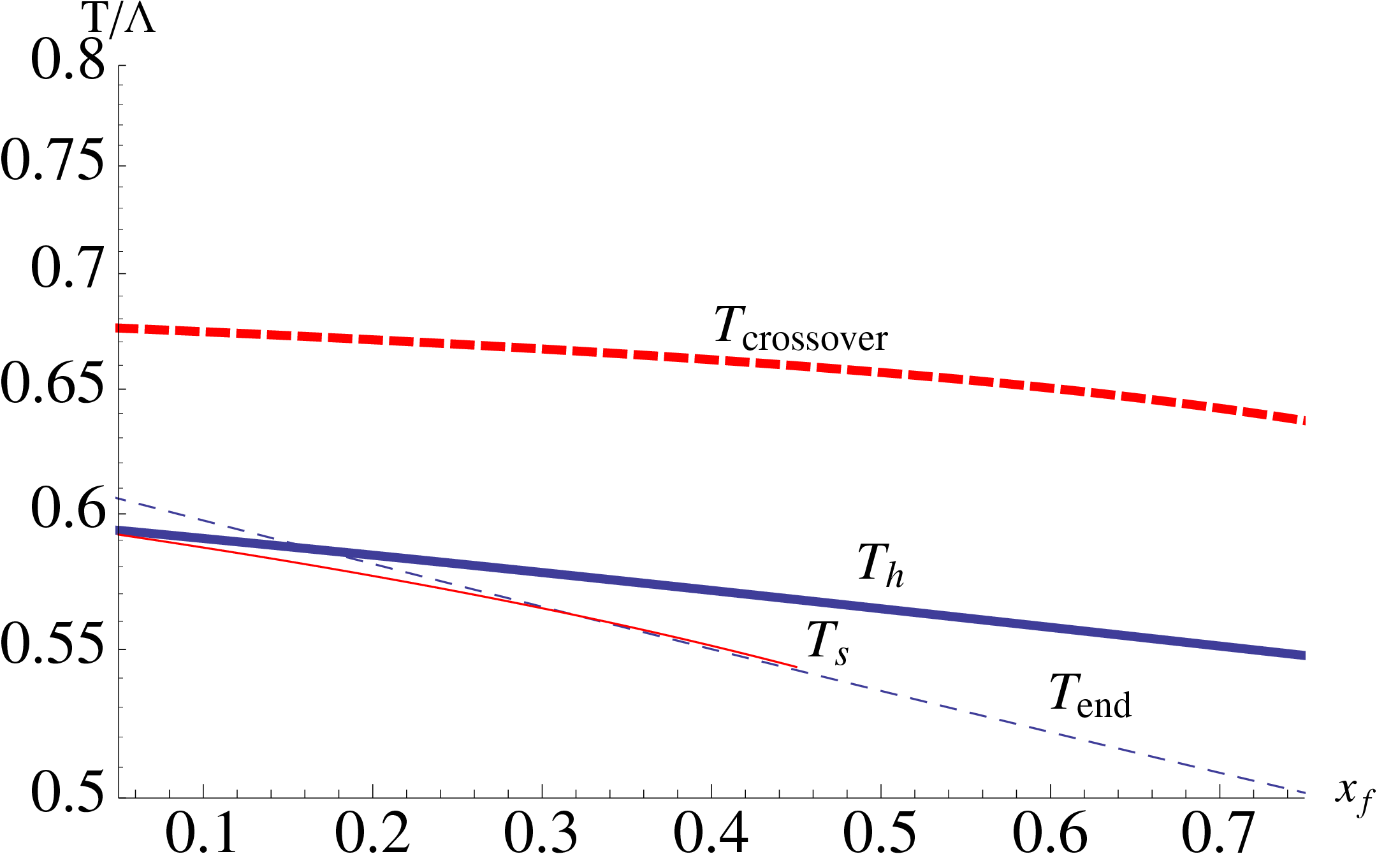}

\caption{\small \emph{Left:} The phase diagram for potential
$\mathrm{II}$ at $W_0 = 12/11$, $x_c=4.08$.
The dashed box shows the region detailed in the other plot. For discussion, see text.
 }
\label{figPhasesPotIIW012per11}
\end{figure}

The phase diagram for potential $\mathrm{II}$ at $W_0 = 12/11$ is shown in
Fig.~\ref{figPhasesPotIIW012per11}.
The main difference with respect to the previous cases is that $T_\rmi{end}<T_h$ for all values of $x_f$,
so the region with $T_\chi>T_d$ does not exist.
Notice that $T_\rmi{end}$ is close to $T_h$ for $x_f \to x_c$ as seen from Fig.~\ref{figPhasesPotIIW012per11} (left).
Because the region with small $x_c-x_f$ is numerically challenging, we do not have reliable data for $x_f \gtrsim 3.8$.
However, nontrivial structure apart from the Miransky scaling, such as rapid changes in the
ratios of the various temperatures, are not expected in this region (see discussion below in Sec.~\ref{xfto0}).
The chirally symmetric crossover transition $T_\rmi{crossover}$ is the highest
temperature stable transition where it exists. The next stable transition is everywhere $T_h$, and
as already pointed out, $T_\rmi{end}$ is in the metastable branch of the solution.
Details of further metastable structure at small $x_f$ are shown in the right hand plot.
At $x_f \sim 0.25$, the first order $T_\rmi{s}$ transition appears in the
metastable branch slightly below $T_h$, see Fig.~\ref{figTransitionsTs}.
This transition develops into the YM transition in the $x_f \rightarrow 0$ -limit.
$T_\rmi{end}$ crosses above the $T_h$ transition, but it is also in the metastable branch,
see Fig.~\protect\ref{figTransitionsUnstab} for details.

\begin{figure}[!tb]
\centering

\includegraphics[width=0.49\textwidth]{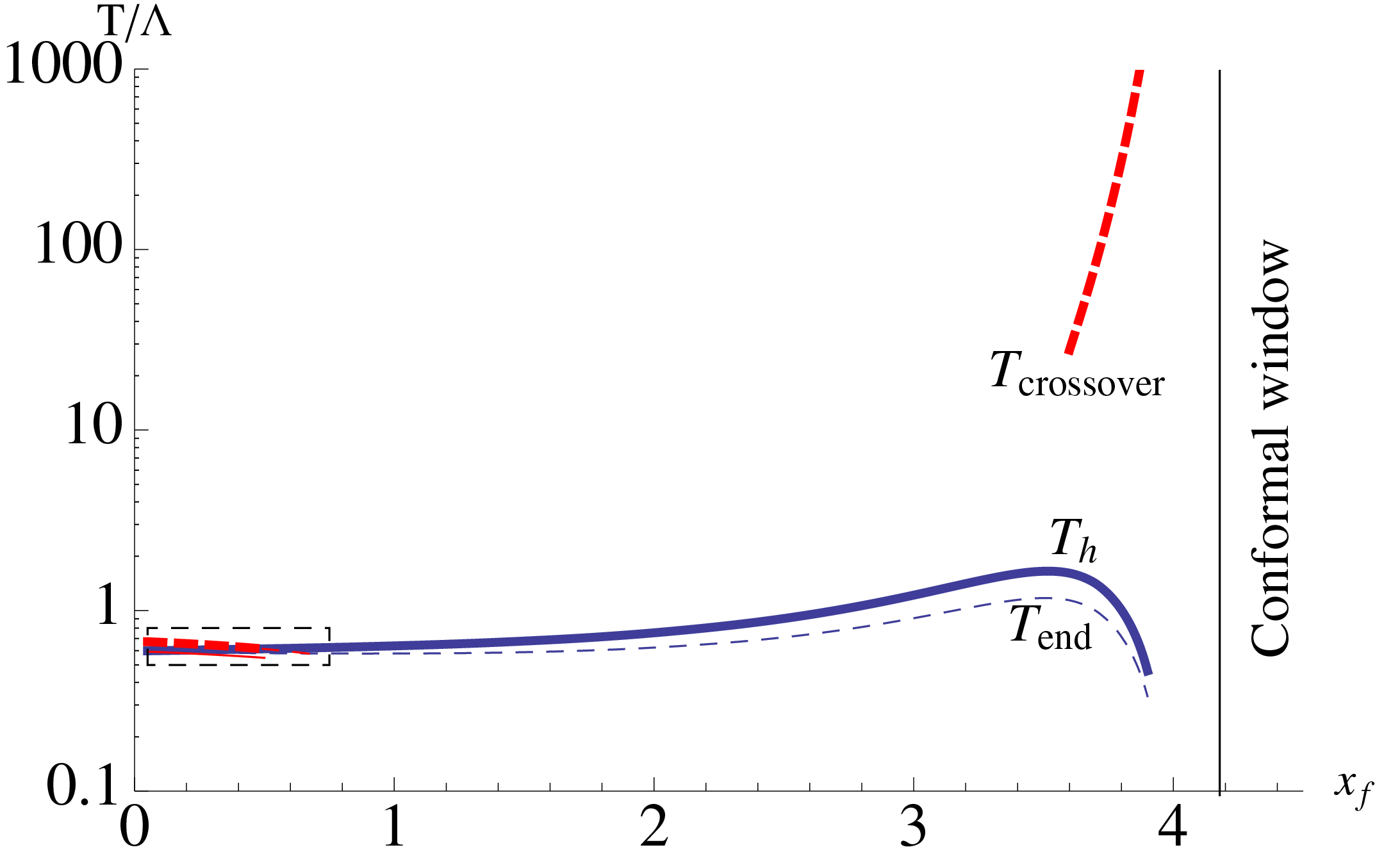}
\includegraphics[width=0.49\textwidth]{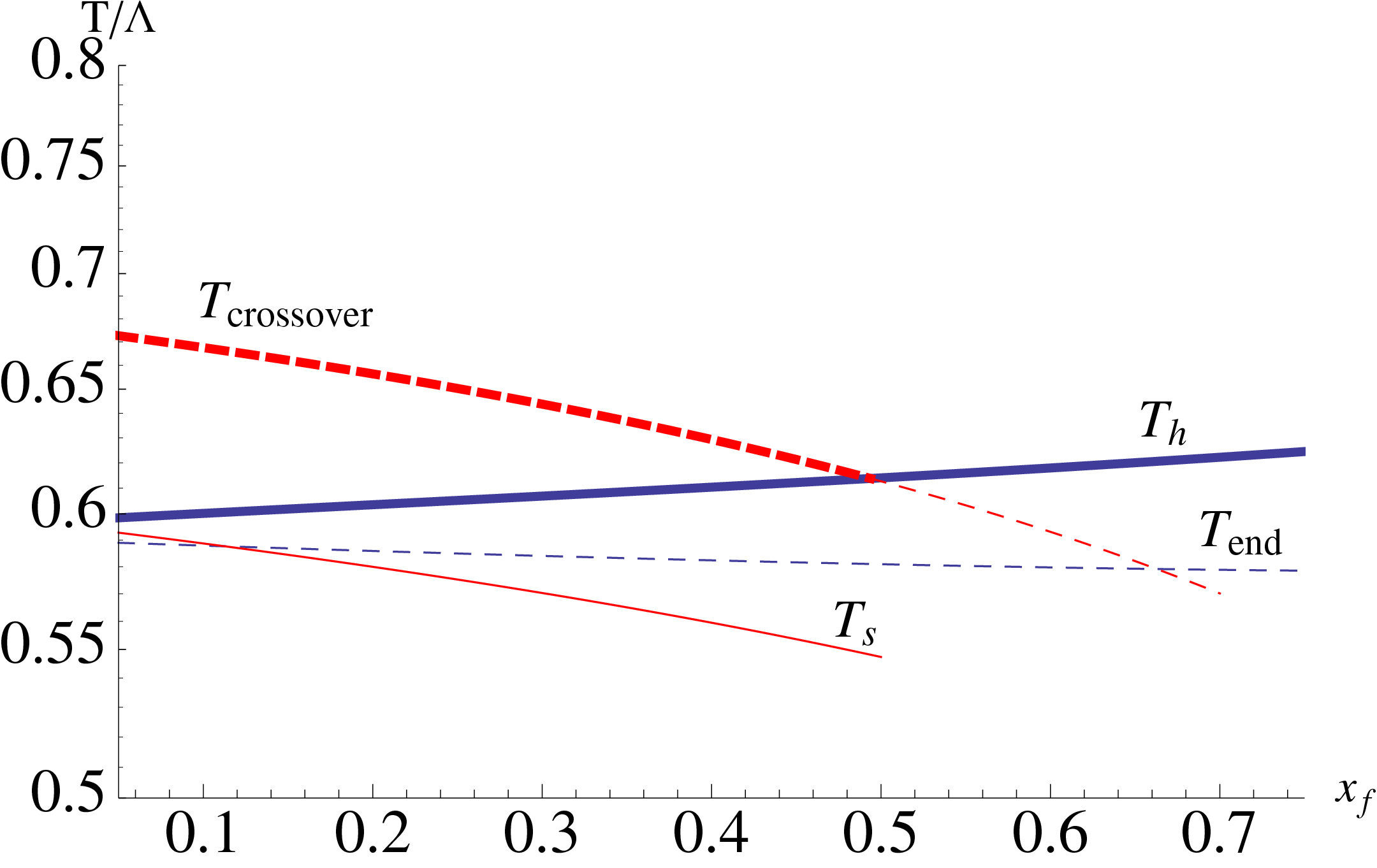}

\caption{\small \emph{Left:} The phase diagram for potential
$\mathrm{II}$ at $W_0 = 0$, $x_c=4.18$.
The dashed box shows the region detailed in the other plot. For discussion, see text. $T_\rmi{crossover}$
continues into the conformal window.
}
\label{figPhasesPotIIW0_0}
\end{figure}

The phase diagram for potential $\mathrm{II}$ at $W_0 = 0$ is shown in
Fig.~\ref{figPhasesPotIIW0_0}.
For all points shown,
$T_\rmi{end}$ is below $T_h$ and in
the metastable branch. The crossover exists  when $x_f \gtrsim 3.6$ and again
between $x_f = 0$ to $\sim 0.7$. The close-up of the small $x_f$ -region in the right hand plot
shows the crossover and the hadronisation transition $T_h$, with the
$T_\rmi{end}$ and $T_{s}$  transitions in the metastable branch. As a new feature the
crossover also becomes metastable for $0.5 \lesssim x_f \lesssim 0.7$.

Finally, let us comment on the $x_f$ dependence of the transition temperature(s).
For SB normalised $W_0$ or $W_0=24/11$ (Figs.~\ref{figTTransitionsSB} and~\ref{figTTransitions}),
$T_h$ and $T_\rmi{end}$ decrease with increasing $x_f$, in qualitative agreement with estimates
based on field theory \cite{Tcxf}.
Decreasing $W_0$ to $12/11$ (Fig.~\ref{figPhasesPotIIW012per11}), however, the $x_f$ dependence
becomes almost flat, and for $W_0=0$ (Fig.~\ref{figPhasesPotIIW0_0}) the temperatures increase
with $x_f$ up to $x_f \simeq 3.5$.
Rather similar behavior with varying $W_0$ will be found for potentials I below, where the
$x_f$-dependence is partially disturbed by the additional structure appearing at low $x_f$.

\subsection{Class-II$_*$ Potentials}

In this section, we consider the phase diagram corresponding to the
potential II$_{*}$.
Recall that the
star subscript
refers to the fact that the potential
$V(\l)=V_g(\l)-x_fV_{0f}(\l)$ has an extremum
only for
$x_*<x_f$,
while for the cases discussed earlier such extremum
exists for all values of $x_f$; see Sec. \ref{sec2pt2pt2} for detailed definitions.

\begin{figure}[!tb]
\centering

\includegraphics[width=0.4\textwidth]{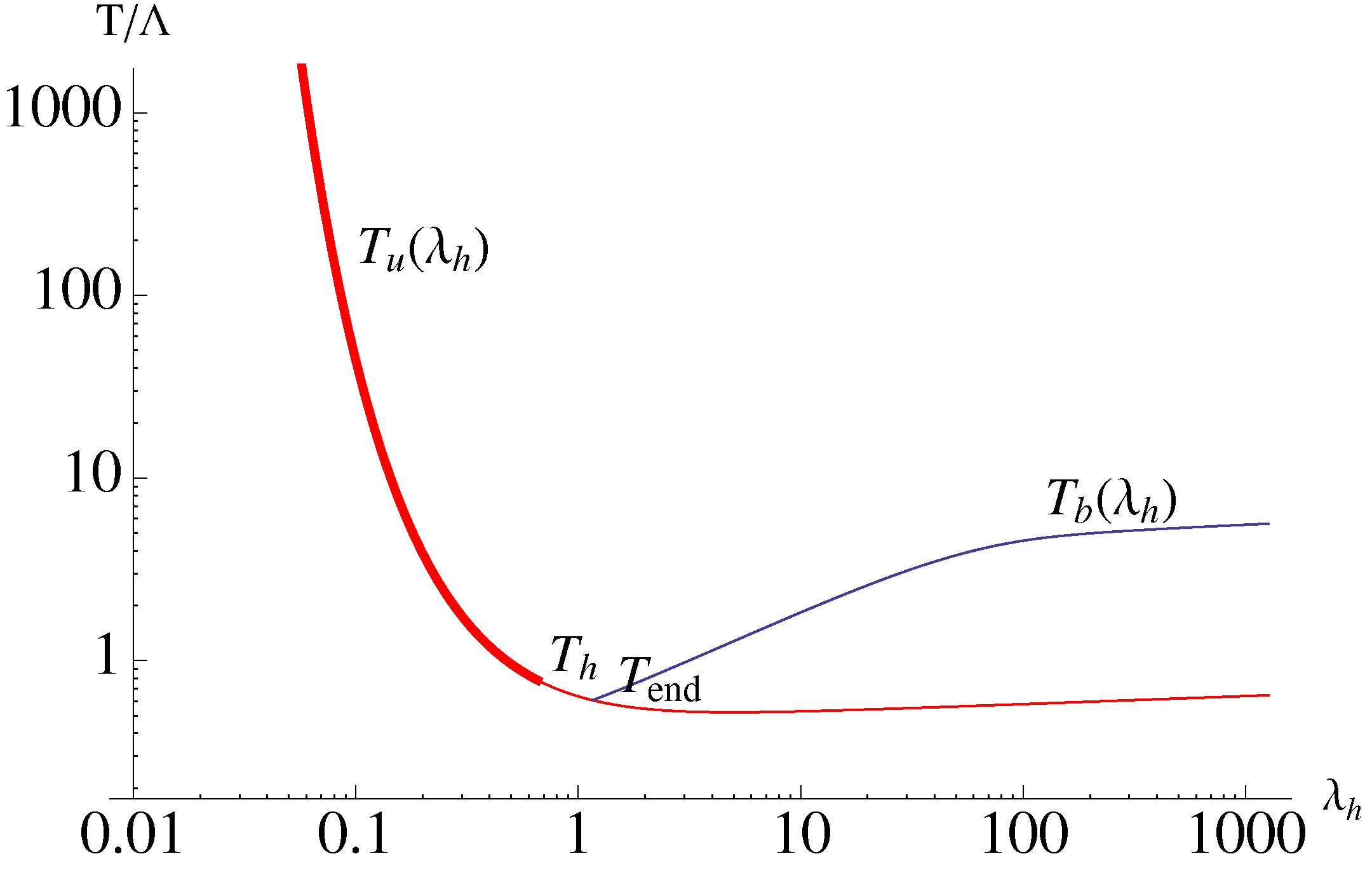}
\includegraphics[width=0.4\textwidth]{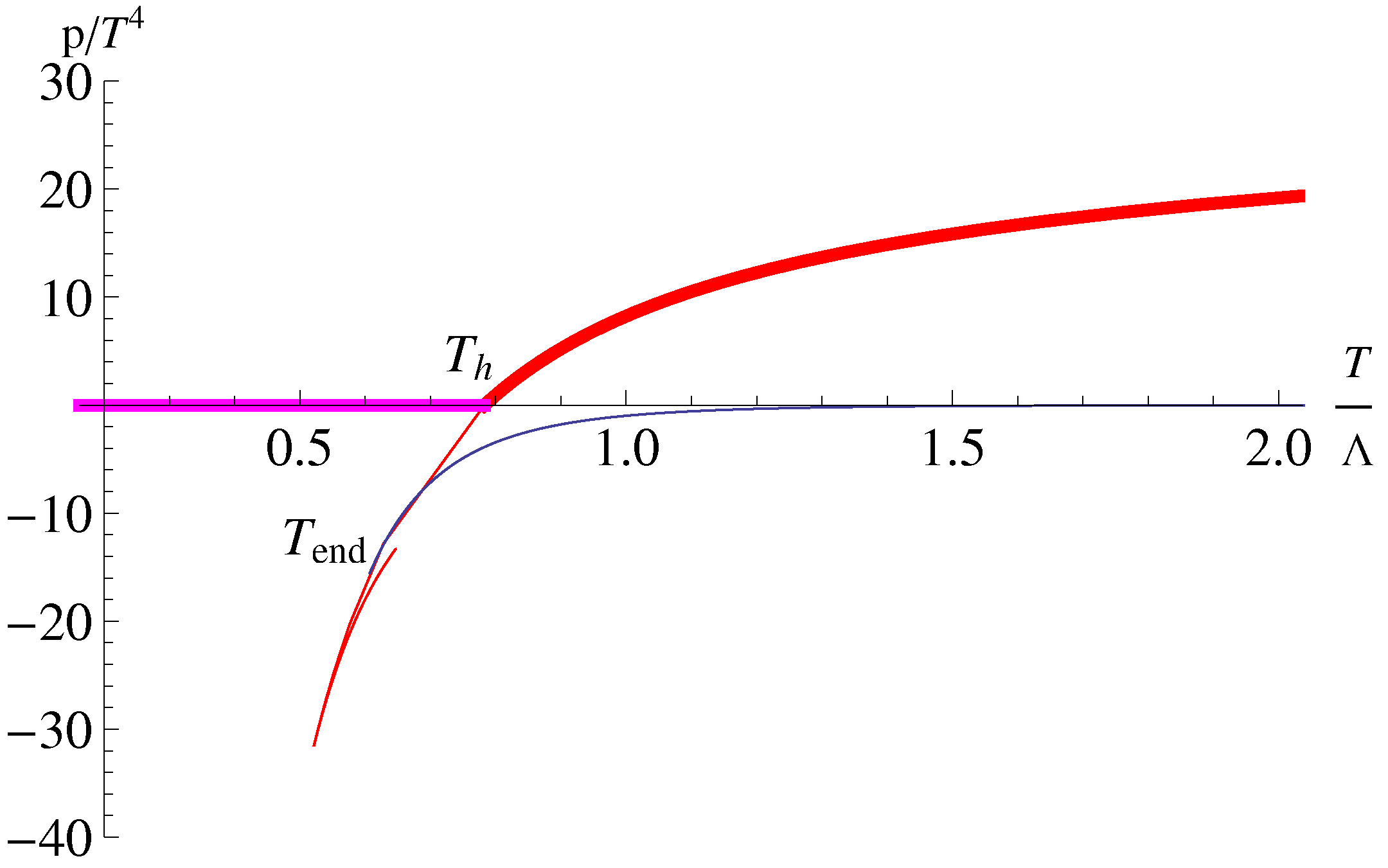}

\vspace{4mm}
\includegraphics[width=0.4\textwidth]{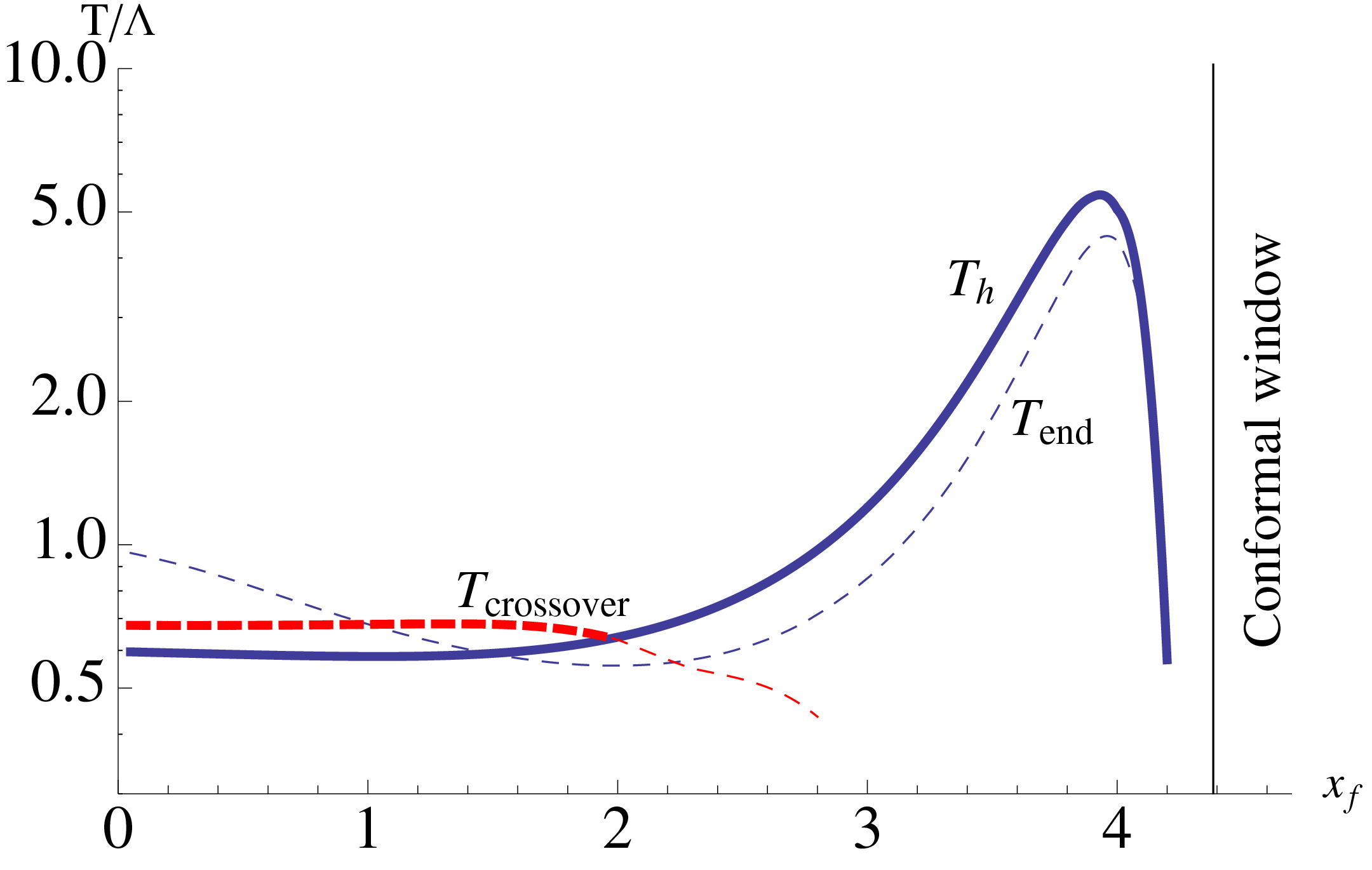}

\caption{\small Phase diagram for potential $\mathrm{II}_\ast$ with $W_0$ SB normalised (bottom).
The top panel shows $T=T(\l_h)$ and $p/T^4$ at $x_f=2.5$. $T_\rmi{crossover}$ reappears at
about $x_f=4$
at a temperature well above the range shown here, 
and continues into the conformal window.}
\label{figPhasesModPotII}
\end{figure}

The resulting $(x_f,T)$ -phase diagram is shown in Fig.~\ref{figPhasesModPotII}, the top
panel shows how the phase diagram is derived at $x_f=2.5$. Starting at large $T$ one is in the
tachyonless black hole phase (thick red curve). At $T_h\approx0.8\Lambda$ pressure goes to zero and the
ground state is the thermal gas phase with $p=0$. If one could supercool further one would at
$T_\rmi{end}$ meet the chirally broken tachyonic black hole phase. It has a higher free energy
than the stable broken phase and therefore is unstable.

The main features are that the crossover exists only for small values
of $x_f$, $x_f\lesssim 2$ where it nearly coincides with $T_h$, 
and again at larger values $x_f\gtrsim 3.5$, where it
is clearly separated from $T_h$. The second order endpoint $T_{\rmi{end}}$ remains
in the unstable phase for $x_f \le x_c$. Below the conformal window, for values
$2\lesssim x_f\lesssim 4$
both $T_h$ and $T_{\rmi{end}}$  increase.
They reach their maximum and finally start to decrease (as predicted by the Miransky scaling)
only
around
$x_f=4$, very near the boundary of the conformal window.
This suggests
that the modification of the potential has the tendency to ``squeeze'' the walking region.

\subsection{Class-I Potentials}
\begin{figure}[!tb]
\centering

\includegraphics[width=0.4\textwidth]{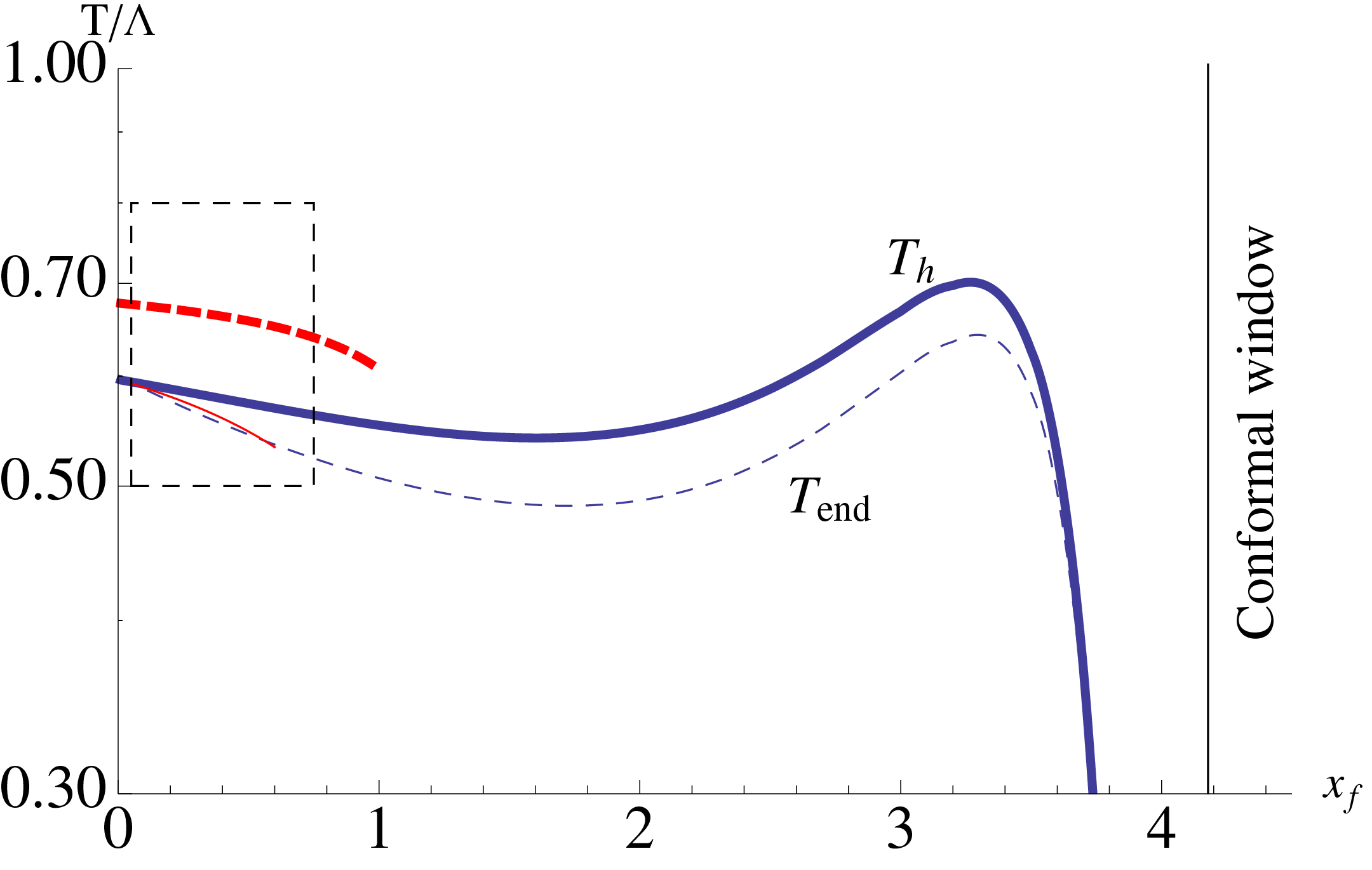}\hfill
\includegraphics[width=0.4\textwidth]{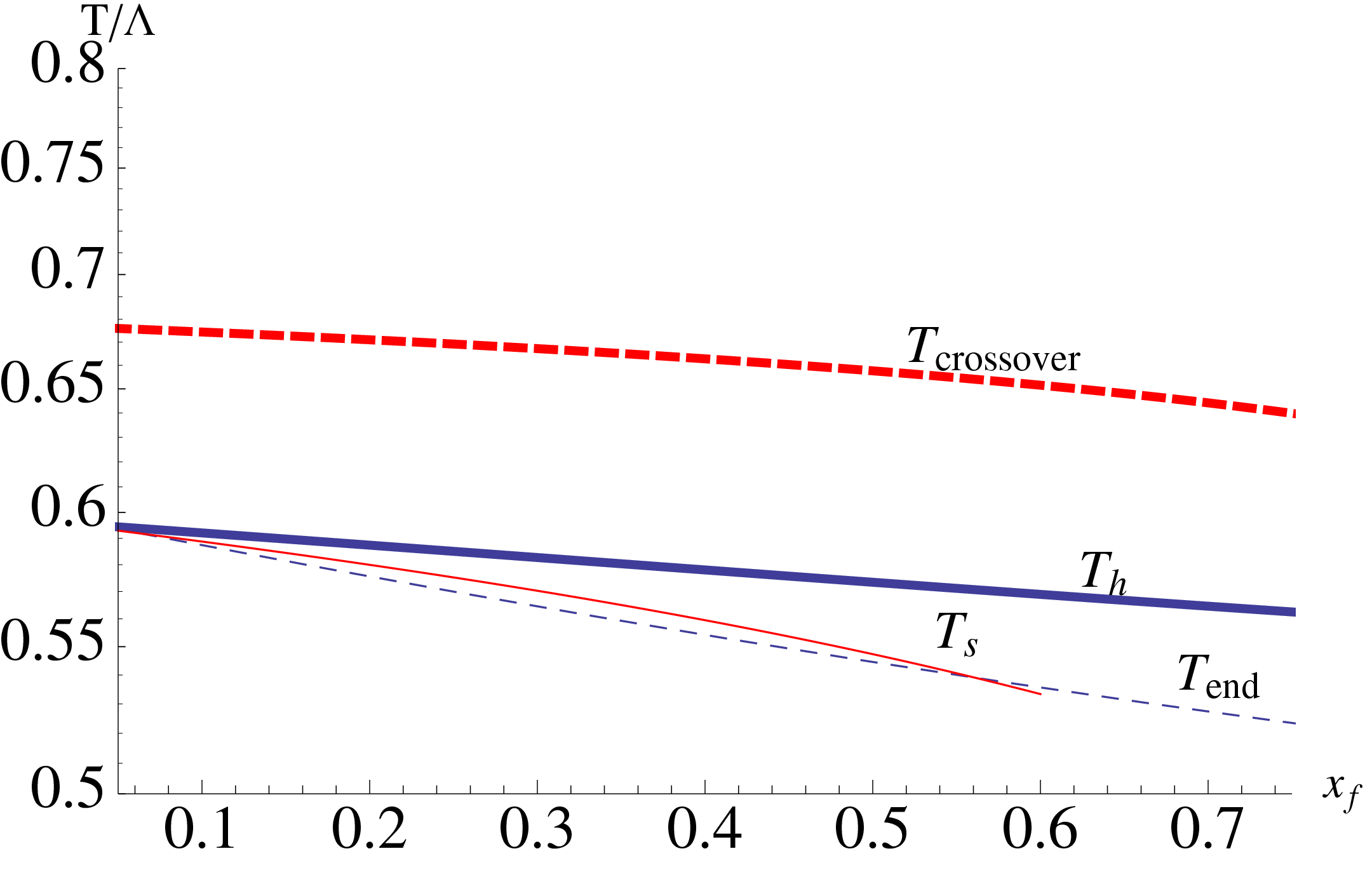}

\vspace{0.01mm}
\includegraphics[width=0.4\textwidth]{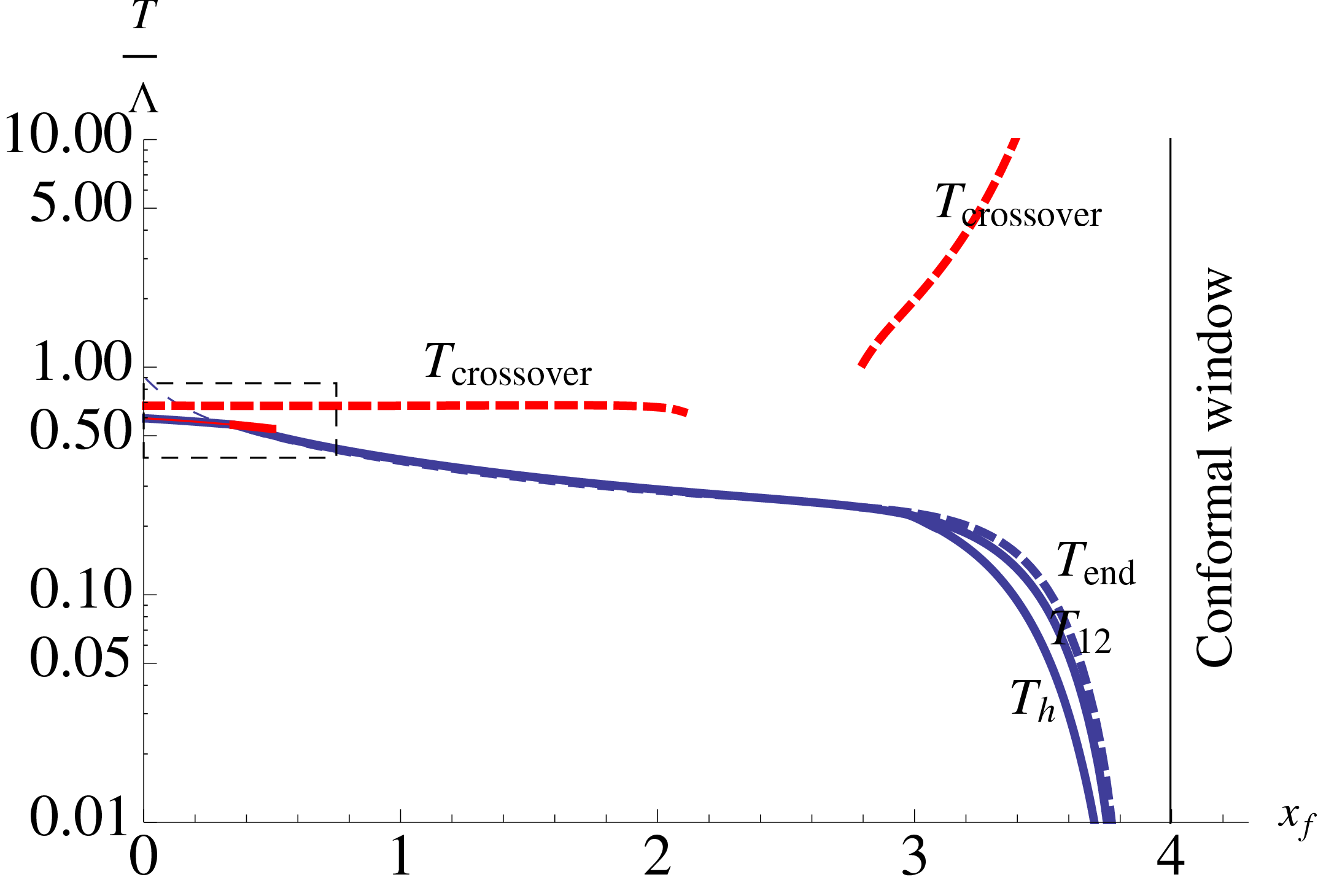}\hfill
\includegraphics[width=0.4\textwidth]{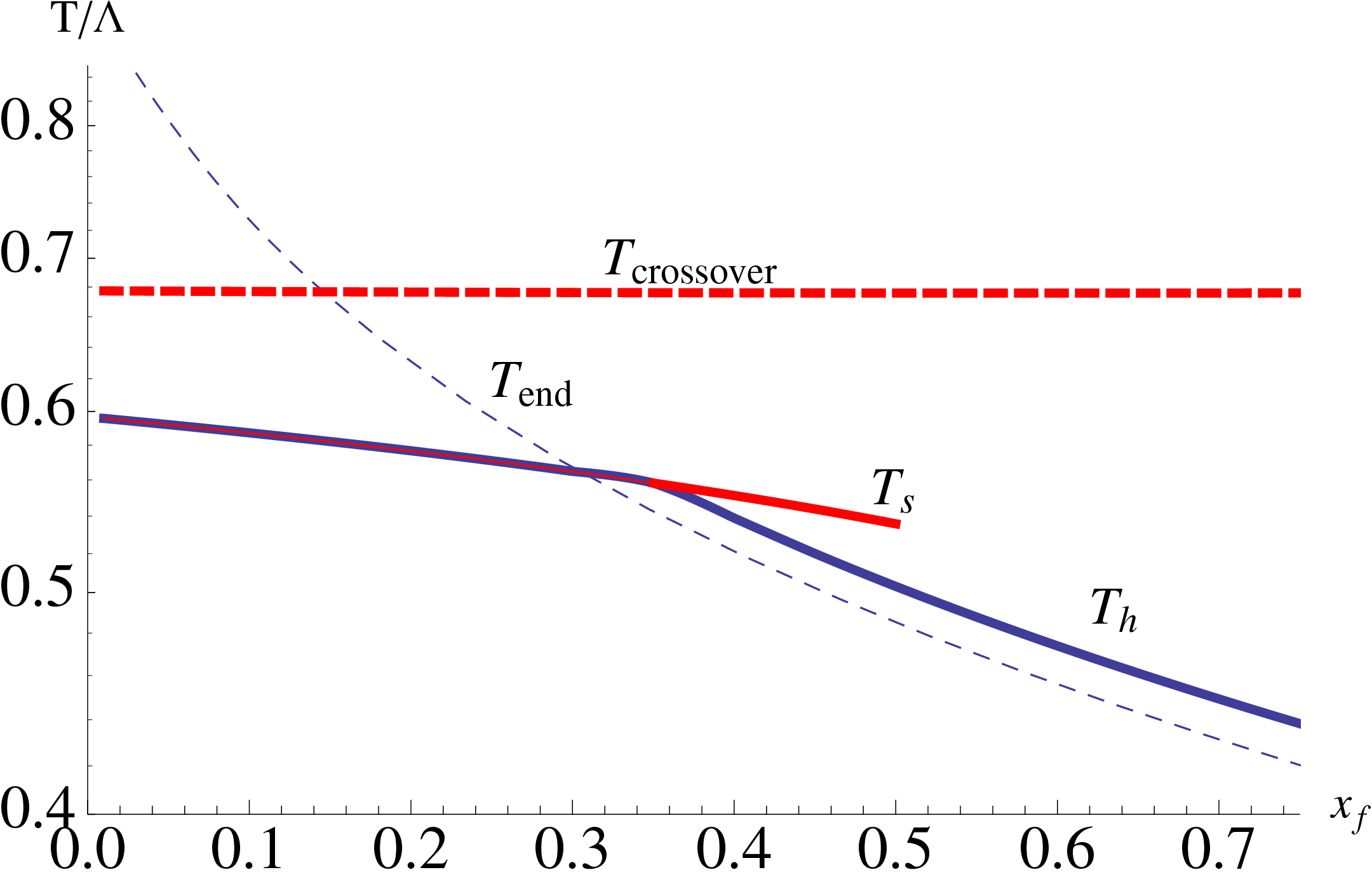}

\vspace{0.01mm}
\includegraphics[width=0.4\textwidth]{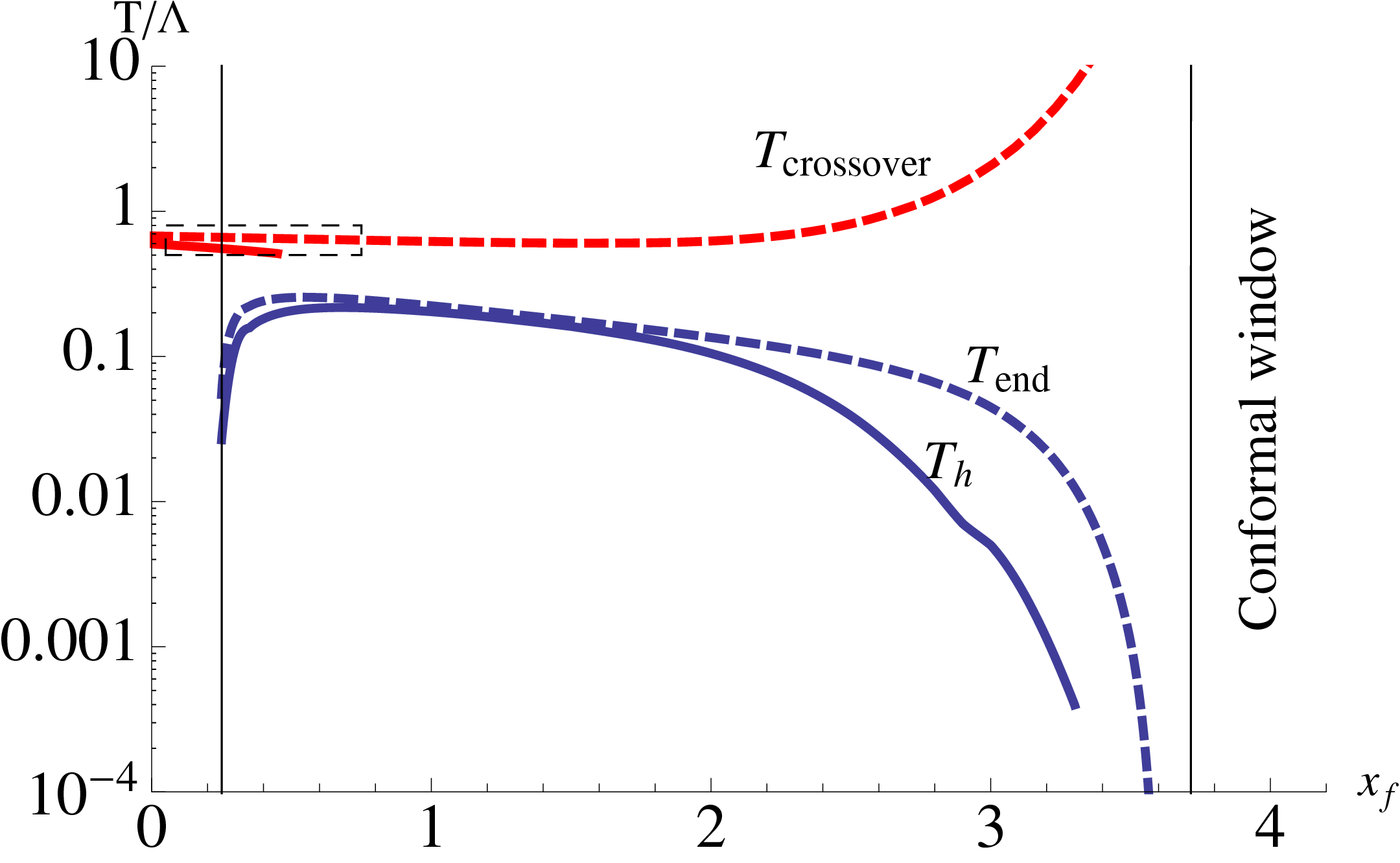}\hfill
\includegraphics[width=0.4\textwidth]{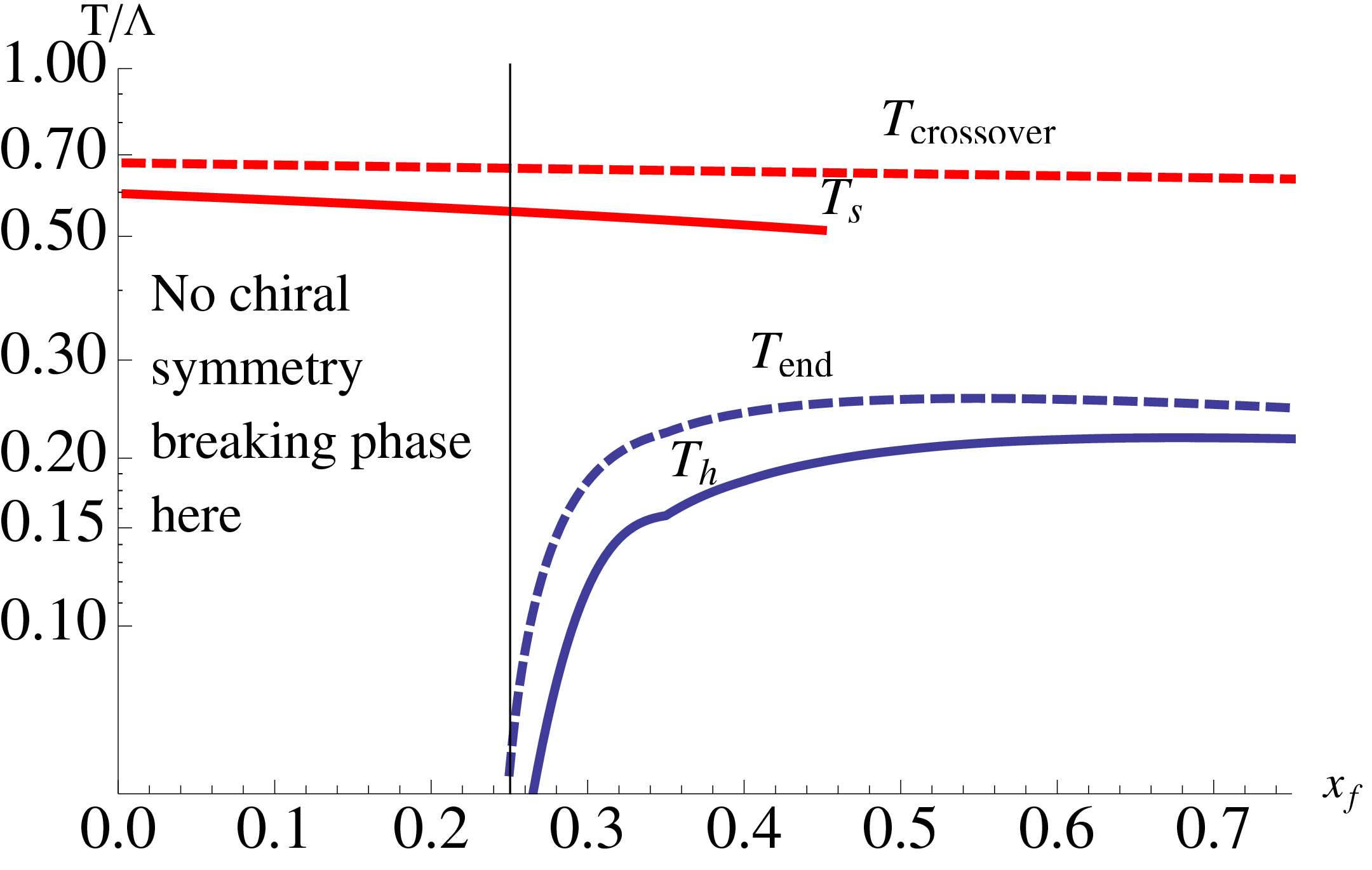}

\includegraphics[width=0.4\textwidth]{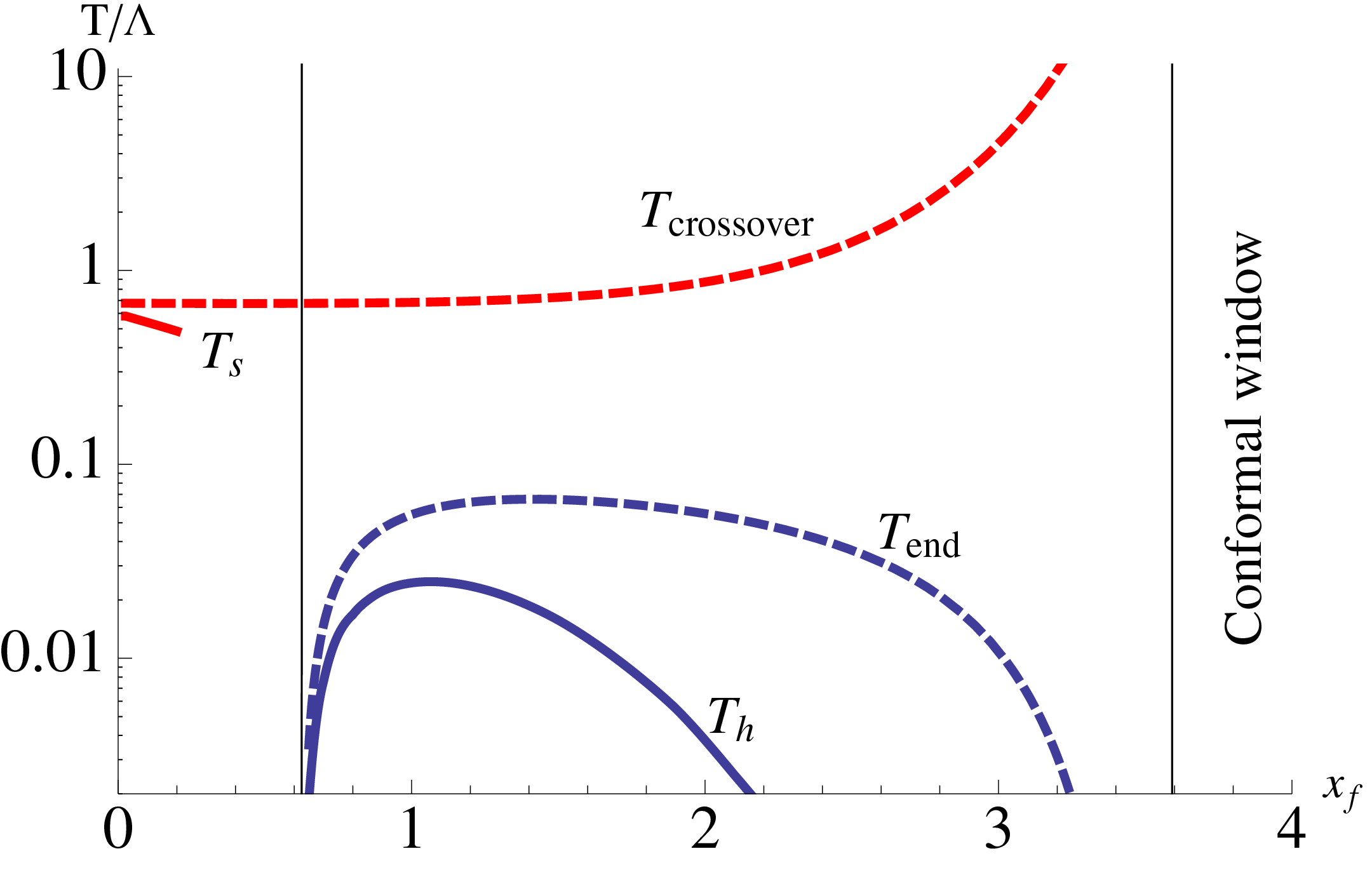}

\caption{\small Phase diagrams for potential $\mathrm{I}$. \emph{Top:} $W_0=0$,
\emph{Middle ones:} $W_0=12/11, \,24/11$, \emph{Bottom:} SB normalisation of $W_0$. A blow-up of the small
$x_f$ region is shown at right separately for three top rows. The leftmost vertical line in the bottom three figures denotes the 
value of $x_f$ below which chiral symmetry breaking solutions do not exist.
}
\label{figPhasesPotIW0}
\end{figure}

For class I potentials Fig.~\ref{figPhasesPotIW0} shows phase diagrams for
$W_0=0,\,12/11,\,24/11$ and for the SB-normalised case.
Recall that for these potentials the tachyon diverges exponentially in the IR. The choices
of $a$ and $\kappa$ are given in Eqs.~\eqref{akappa}.
We also remind that transitions
between stable phases are plotted as thick lines. Transitions plotted as thin
lines can be seen only if the system is, e.g., supercooled, so that they are
not there in the thermodynamic limit.

One can observe several characteristic features for varying $W_0$:
\bi
\item The first observation is
the striking structure near $x_f=0$ which is observed at large $W_0$, i.e., for $W_0=24/11$ or SB normalized.
The temperatures $T_h$ and $T_\rmi{end}$ drop rapidly with decreasing $x_f$ near $x_f=0$
and reach zero at a finite value of $x_f$. Below this critical value, all phases are chirally symmetric.

This behavior is related to the tachyon mass at the IR fixed point, shown in
Fig.~\ref{tachyonmassfig}. For PotI (the absolute value of) the squared tachyon mass is
below the BF bound for low values of $x_f$. Therefore it is not guaranteed that a
solution with zero quark mass and nontrivial tachyon profile 
exists (at any temperature) in this region. For large $W_0$
it actually turns out that the solution with $m_q=0$ and nontrivial tachyon profile
does not exist
for very low $x_f$,
which explains the absence of
chiral symmetry breaking.
This implies that this potential is not describing a QCD-like theory. However,
the applicability of PotI can be rescued by a simple logarithmic modification of
$\kappa(\l)$, see Section~\ref{sectlogkappa} and Fig.~\ref{figmumod}.

\item
The symmetric$\to$ symmetric transition $T_s$ becomes
a stable transition when
$W_0=24/11$ or SB normalized.
For comparison, for PotII it was always in a metastable phase.
This happens mostly in the region of very low $x_f$ where all phases are chirally
symmetric so that $T_h$ and $T_\rmi{end}$ are absent.
For $W_0=24/11$ we observe a region with $0.25 \lesssim x_f \lesssim 0.45$ where
these transitions are also
present. In this case the order of transitions is $T_s>T_\rmi{end}>T_h$,
and chiral symmetry is broken in the middle one.
For $W_0$ SB normalized we find instead a region with $0.2 \lesssim x_f \lesssim 0.5$
where only the crossover exists, so that the phase structure is similar to the conformal window.

\item At large $x_f\gta 3$, $W_0=12/11$, one observes the splitting of the 1st order line $T_h$
into two 1st order lines $T_{12}>T_h$.
The order of the transitions is
$T_\rmi{end}>T_{12}>T_h$, chiral symmetry is broken at the largest one, $T_\rmi{end}$.
The holographic action therefore gives two consecutive 1st order transitions within the
chirally-broken phase. It is an open issue what the nature of these
transitions is. It is plausible that PotI at large $W_0$ is not related to QCD-like theories.

\item The high temperature crossover
exists over a larger and larger range when $W_0$ increases
and ultimately appears at all $x_f$. This is the same tendency seen also for potentials in the II class.
\ei

\subsection{Class-I$_*$ Potentials}

\begin{figure}[!htb]
\centering

\includegraphics[width=0.45\textwidth]{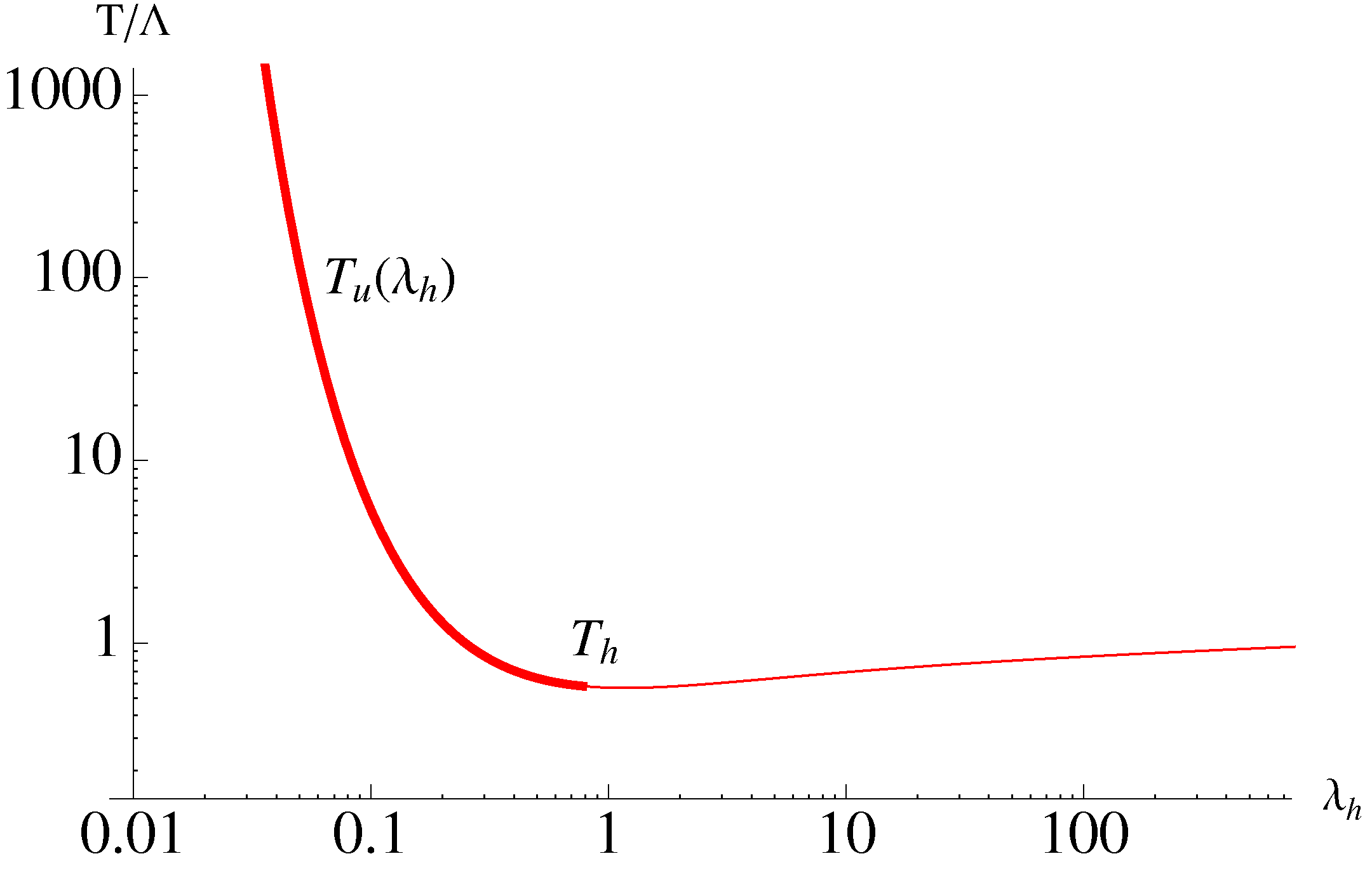}\hfill
\includegraphics[width=0.45\textwidth]{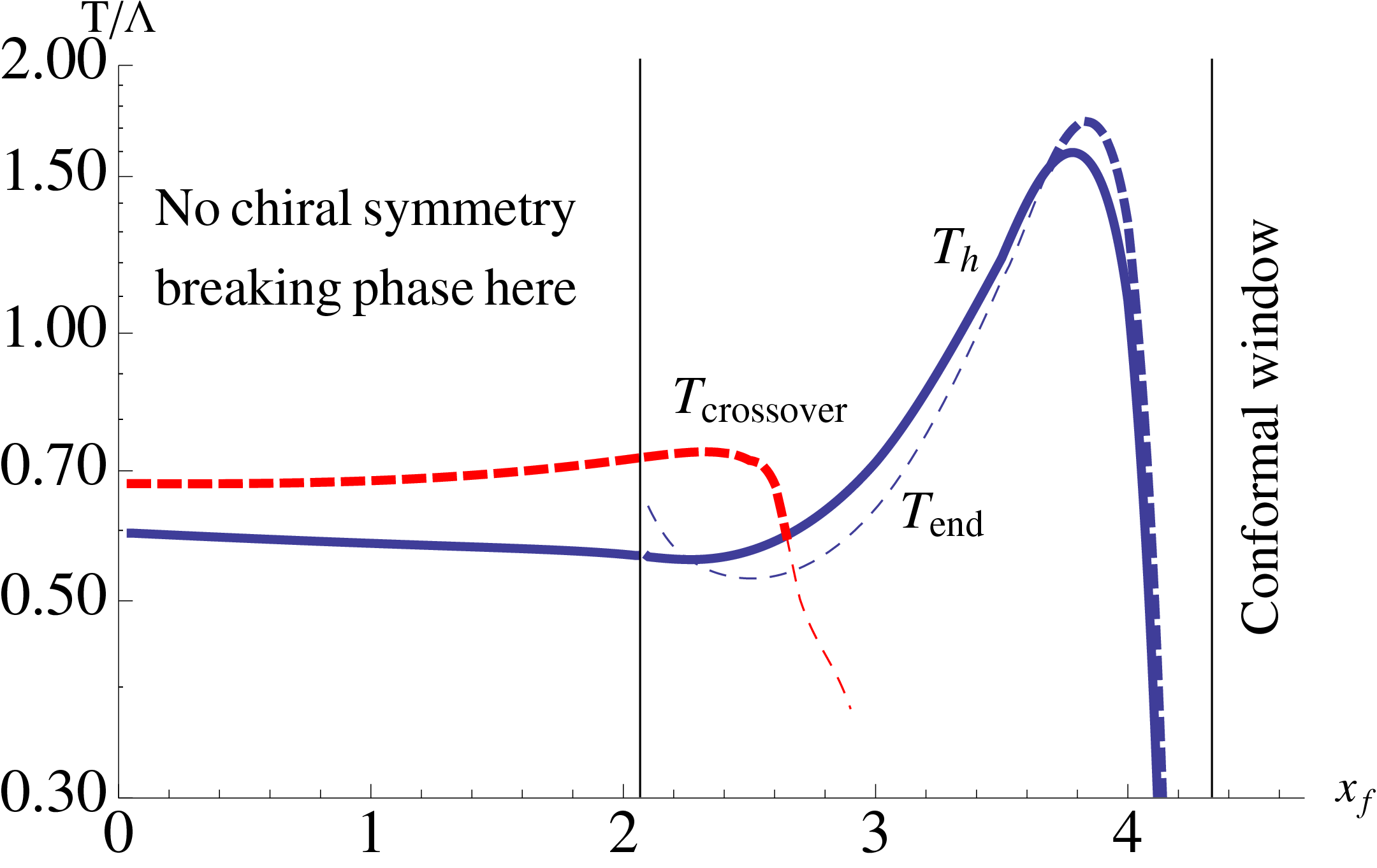}

\caption{\small Phase diagram for potential $\mathrm{I}_\ast$ with SB normalisation of $W_0$.
The left panel shows
$T=T(\l_h)$ at $x_f=1$: no tachyonic black hole!}
\label{figPhasesModPotI}
\end{figure}

Finally, we present the phase diagram corresponding to the potential I$_*$ in Fig.
\ref{figPhasesModPotI}.
The striking difference between the phase diagram of the potential I$_*$ in comparison
with potential II$_*$ considered earlier is that for small values of
$x_f\lesssim 2$
there are no solutions with broken chiral symmetry, not even
at low temperatures; all phase boundaries here are between chirally symmetric phases.
There is $T_h$, but now it describes a chirally symmetric $\to$ symmetric transition.
To illustrate this we show explicitly $T=T(\l_h)$ at $x_f=1$. It is very structureless,
and has no solutions with nonzero tachyon.
Thus the ($\l_f,T$) -diagram is qualitatively similar to the Yang-Mills case \cite{YM2}.
Only above $x_f\sim 2$ and below $x_c$ is
chiral symmetry broken at low temperatures.

Otherwise the
overall features are similar to those in the case of potential II$_*$.
For small values of $x_f$,
$x_f\lesssim 2$, the crossover
nearly coincides
with $T_h$ . The second order endpoint, $T_{\rmi{end}}$, is in the unstable branch for
small values of $x_f$, but enters into the stable branch at $x\sim 4$. Below the conformal window,
for values
$2\lesssim x_f\lesssim 4$
 both $T_h$ and
$T_{\rmi{end}}$  increase.
They start to decrease 
only
at
$x_f\sim 4$, very
near the boundary of the conformal window.

We have also studied the potentials I$_*$ for the case of fixed $W_0$ and found qualitatively
similar results for the phase structure
for $W_0=12/11,\,24/11$.
For $W_0=0$ the problematic region without chiral symmetry breaking is absent, and the
phase diagram is similar to PotII$_*$.
This implies that, like Pot I, this type of potential is probably not applicable for QCD-like theories
when $W_0$ is large.

\subsection{PotI with logarithmic correction to $\kappa(\l)$}
\label{sectlogkappa}
The function $\kappa(\l)$ in the action \nr{lagrang} represents the effects of going from the
string frame (to which the derivation of the DBI action as
the $\alpha'\to0$ limit of open strings leads) and
the Einstein frame (where the gravity dual is formulated). Extending the conformal
transformation relating these to UV by $\l\to1+\l$ one has,
in terms of the metric
functions,\footnote{Notice that we introduced additional constants in the formulas \nr{akappa}
and \nr{akappa2} in order to match with the perturbative anomalous dimensions in QCD.}
\be
\kappa(\l)={1\over(1+\l)^{4/3}}={b^2\over b_s^2}.
\ee
where $b$ and $b_s$ are the metric factors in the Einstein and string frames, respectively.

The potential \nr{VgSB}
carries the factor $\l^{4/3}$, but also the logarithmic factor
$(\ln\l)^{1/2}$, which plays a quantitatively important role: for $(\ln\l)^P$ the excitation
spectrum is $m\sim n^P$ \cite{YM1} and one wants the Regge-like spectrum, $P=1/2$. Also
numerically $\ln\l$-effects are important, see Fig.\ref{figTlah}. To study these
effects in $\kappa$ we use the parametrisation
\be
\kappa(\l)= {[1+\ln(1+\l)]^{\bar\mu}\over[1+\fra34(\fra89 b_0+1+\bar\mu)\l]^{4/3}}.
\label{kappamu}
\ee

There are constraints on this parametrisation from the UV and IR. First, to maintain the
proper mass anomalous dimension equation \nr{massanomdim} at small $\l$, $\bar\mu$ has to
appear also in the denominator as shown in \nr{kappamu}. Secondly, for $\bar\mu=0$ the
tachyon grows exponentially in $r$ according to Eq.~\nr{exptachyon}. The effect of
$\bar\mu$ on this comes from the change $b^2/\kappa=\sqrt{\ln\l}/(\ln\l)^{\bar\mu}\sim
r^{1-2\bar\mu}$ (in the IR $r\sim\sqrt{\ln\l}$, see \nr{IRresl}).
This effect propagates through the computation of the $r$ dependence
which comes out to be $\tau(r)\sim\exp(Cr^{1-2\bar\mu})$, indicating that $\bar\mu<1/2$.

\begin{figure}[!tb]
\centering

\includegraphics[width=0.49\textwidth]{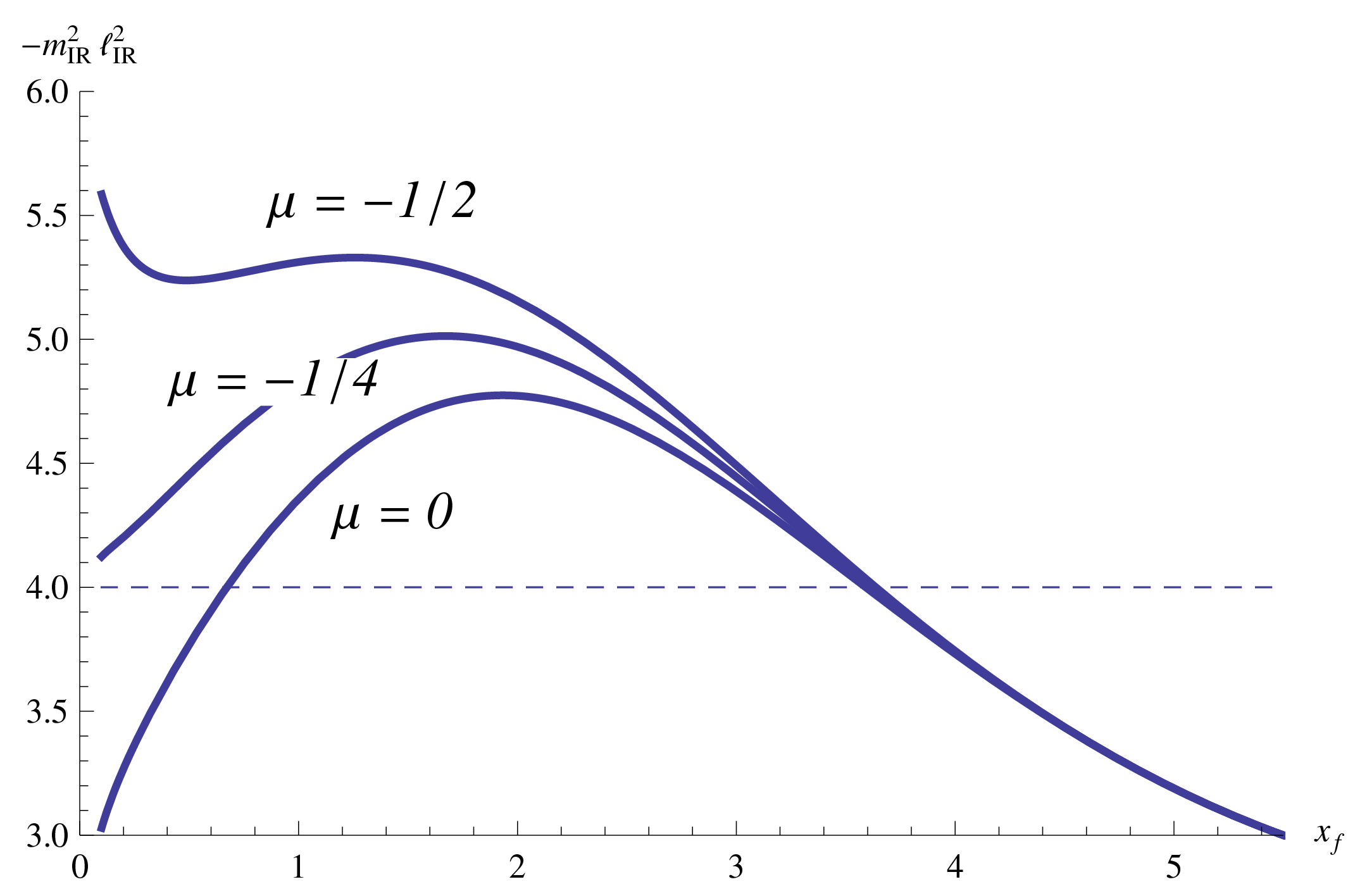}\hfill
\includegraphics[width=0.49\textwidth]{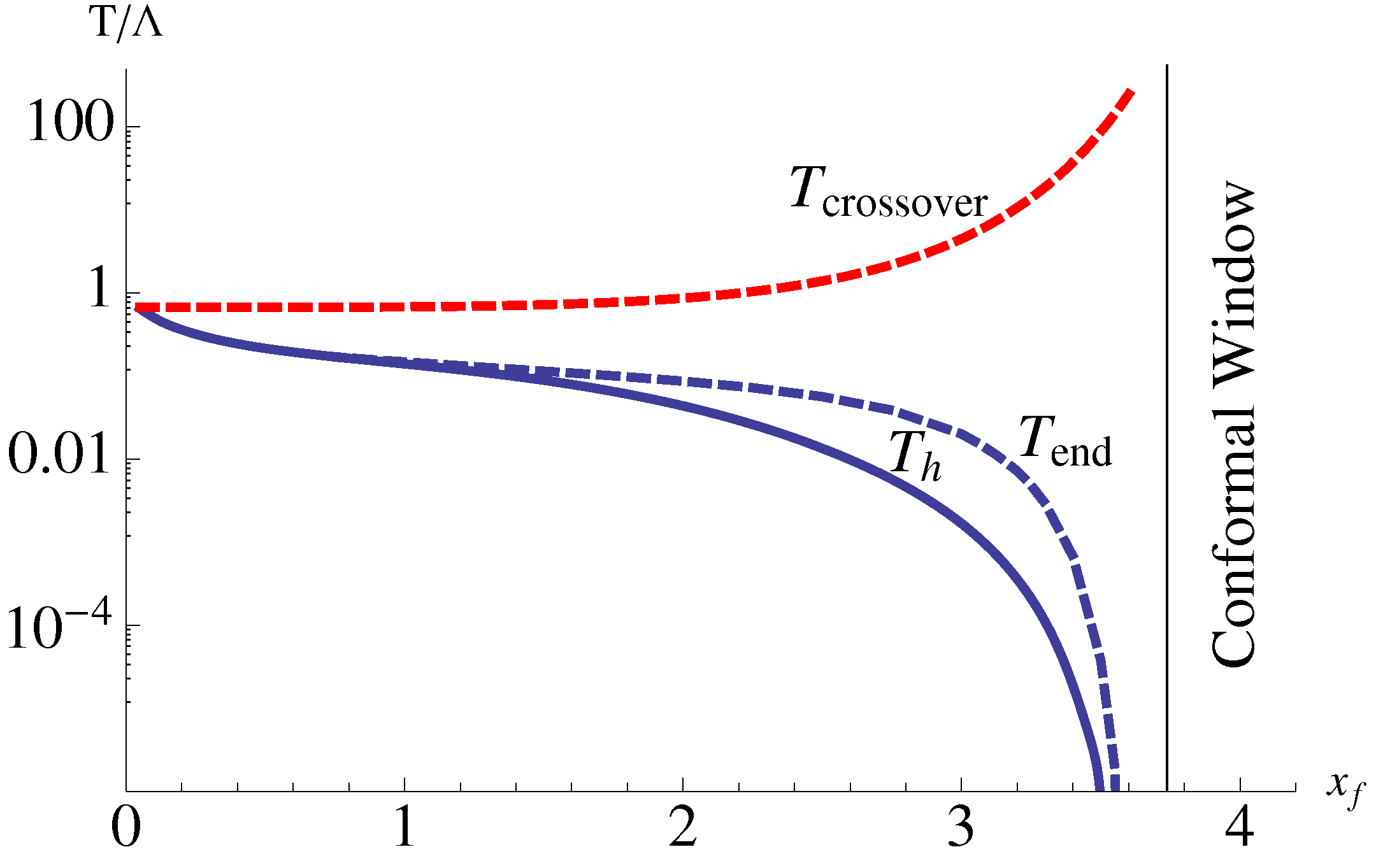}

\caption{\small Left: The tachyon mass at $\l^*$ for PotI with log-modified $\kappa(\l)$,
evaluated using Eqs. \protect\nr{emel} and \protect\nr{kappamu} for $\bar\mu=0,\,-\fra14,\,-\fra12$.
The $\bar\mu=0$ curve is the same as the red dotted curve in left panel of Fig.~\protect\ref{tachyonmassfig}.
Right: The phase diagram for PotI with SB-normalised $W_0$
with $\kappa(\l)$ given by \protect\nr{kappamu} with $\bar\mu=-\fra12$. $T_\rmi{end}$ intersects $T_h$ at
$x_f=x_\chi=0.72$. $T_\rmi{crossover}$ is the same as in Fig.~\protect\ref{figPhasesPotIW0}, bottom.
}
\label{figmumod}
\end{figure}

The most interesting effect comes from evaluating the tachyon IR mass using \nr{emel}.
The result is shown in Fig.~\ref{figmumod}, to be compared with Fig.~\ref{tachyonmassfig}.
The difficulty with PotI was that at small $x_f$ the curve in the left panel of
Fig.~\ref{tachyonmassfig} dropped below the BF bound. The reason for this is easy to
see analytically by studying the $\l^*\to\infty$ limit of \nr{emel},
which gives $-m_\rmi{IR}^2\ell_\rmi{IR}^2 \sim (-\ln x_f)^{-1/2}$ in this case.
For small $x_f$, $\l^*$ approaches infinity and obviously negative values of $\bar\mu$ increase
the tachyon mass $-m_\rmi{IR}^2\ell_\rmi{IR}^2$,
so that $-m_\rmi{IR}^2\ell_\rmi{IR}^2 \sim (-\ln x_f)^{-\bar\mu-1/2}$
For $\bar\mu<-1/2$ it even grows without
bounds as for PotII in Fig.~\ref{tachyonmassfig}. This is seen in Fig.~\ref{figmumod}.

As a consequence, the phase diagram for PotI with log-modified $\kappa$
does not suffer from the problems at small $x_f$ described earlier for PotI. The phase diagram
computed for $\bar\mu=-\fra12$ is shown in Fig.~\ref{figmumod} and, in fact, resembles
qualitatively those for PotII. This is very gratifying since PotI also leads to a
Regge-like particle spectrum \cite{mesons}. PotI with log-modified $\f(\l)$ \nr{kappamu}
thus seems to be the gravity dual leading to the simplest thermodynamics in Fig.~\ref{figxTphases}
and the expected Regge-like hadron spectrum. It is interesting that also PotII, a dual with
spectrum of type $m\sim n$, also leads to the simple thermodynamics in Fig.~\ref{figxTphases}.

\subsection{The conformal window}
\begin{figure}[!tb]
\centering

\includegraphics[width=0.49\textwidth]{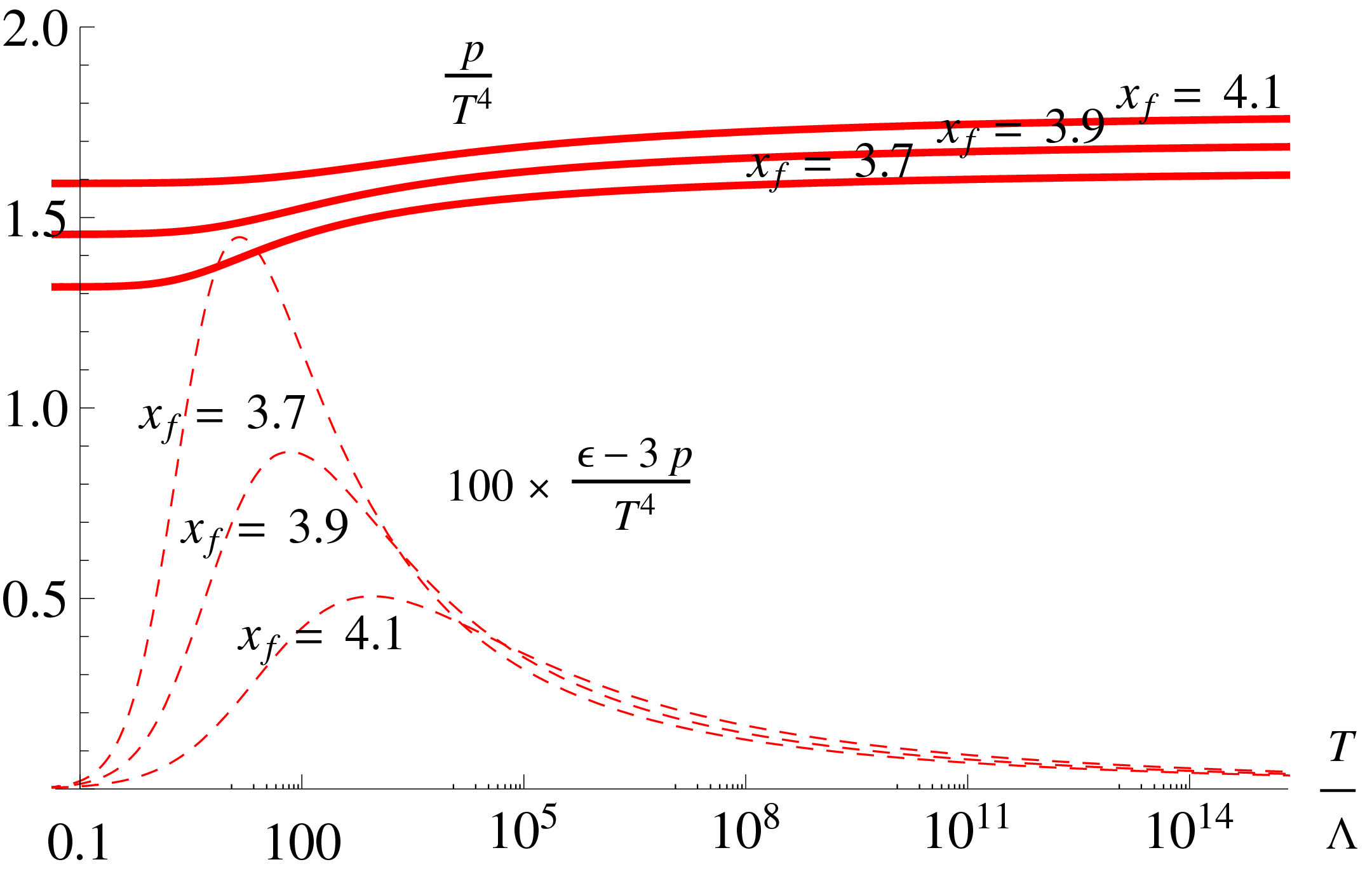}


\caption{\small Thermodynamics for some values of $x_f$ within the conformal
window, computed for PotI. Note that in the conformal window always $\tau=0$
and the functions $a(\l),\,\f(\l)$ do not affect the result.
}
\label{figconfwin}
\end{figure}

A detailed
picture of thermodynamics in the conformal window is shown in
Fig.~\ref{figconfwin}. Here $p/T^4$, i.e. the effective number of degrees of freedom,
is plotted for some values of $x_f>x_c$.
At large $T$ it is normalised so that it approaches the SB limit \nr{SBpressure}
for any $x_f$. For $T$ approaching zero, $p/T^4$ approaches
another constant, the value of which decreases when $x_f$ approaches $x_c$ from
above. For all $x_f$, the vacuum phase has zero pressure,
and at the limit $T\rightarrow 0$ there is a transition
from the black hole to the thermal gas phase.
When $x_f$ approaches the upper end of the conformal
window $11/2$, the behavior of the curves can be worked out analytically in
perturbation theory \cite{tuominen} since the coupling then is small.

For the present potential the finite temperature transition between the low and high
temperature phases inside the conformal window is a smooth crossover.
Fig.~\ref{figconfwin} also plots the interaction measure, the maximum of which
defines the critical temperature for this crossover. Note that even if
the transition here is smooth crossover, the transition can also be of 1st order
in different theories \cite{kajantieIRFP}. What determines this behavior is the
overall magnitude of the beta function. For illustration, consider the beta function of
large $N_f$ QCD, $\beta=-b_0\l^2-b_1\l^3+\dots$. The values of the coefficients
behave as $b_0\sim {\cal{O}}(1)\ll |b_1|$, while the results of \cite{kajantieIRFP}
suggest $b_0\sim{\cal{O}}(10)\sim |b_1|$ for 1st order phase transition.
For the models we have considered here, we find that the nonperturbative
beta function extracted from the gravity solution is small over the entire range
$0\le\l\le\l_\ast$ inside the conformal window.

The large temperature values appearing in Fig.~\ref{figconfwin} may appear somewhat
surprising. However, they have a simple explanation.
The region in which $p/T^4$ is nearly constant and approaching its large $T$ limit
is the perturbative region $\l$ small. The conformal window is within $0<\l<\l^*(x_c)$
and the upper limit is always small, $\ll1$, so that the conformal window is perturbative,
down to $T=0$. From Fig.~\ref{figxfscan} one sees quantitatively
how this holds even somewhat below the conformal window.
To 1-loop $\log T=\exp(1/(b_0\l))$
and clearly for $b_0\to0$ this grows fast. Somewhat more quantitatively,
the beginning of the large $T$ region can be
estimated by computing the value of $T$ for which the 2-loop correction term
in the perturbative expansion of $1/\l(\mu)$ equals the 1-loop term.
One finds that the 2-loop correction is smaller than the 1-loop term if
$T>(2\log T)^{|b_1|/b_0^2}$, $|b_1|/b_0^2=3|13x_f-34|/(2(11-2x_f)^2)$. This is
always true for $T>1=\Lambda$ if $x_f<3.6$. However, for $x_f>3.6$ this gives
a lower limit of $T$ which grows extremely fast when $x_f$ grows within the
conformal window. $T_\rmi{crossover}$ is somewhat below the solution
of this equation.
Numerical values are in qualitative agreement with Fig.~\ref{figconfwin}.

\subsection{The limits $x_f \to 0$ and $x_f \to x_c$}\label{xfto0}

\begin{figure}[!tb]
\centering

\includegraphics[width=0.49\textwidth]{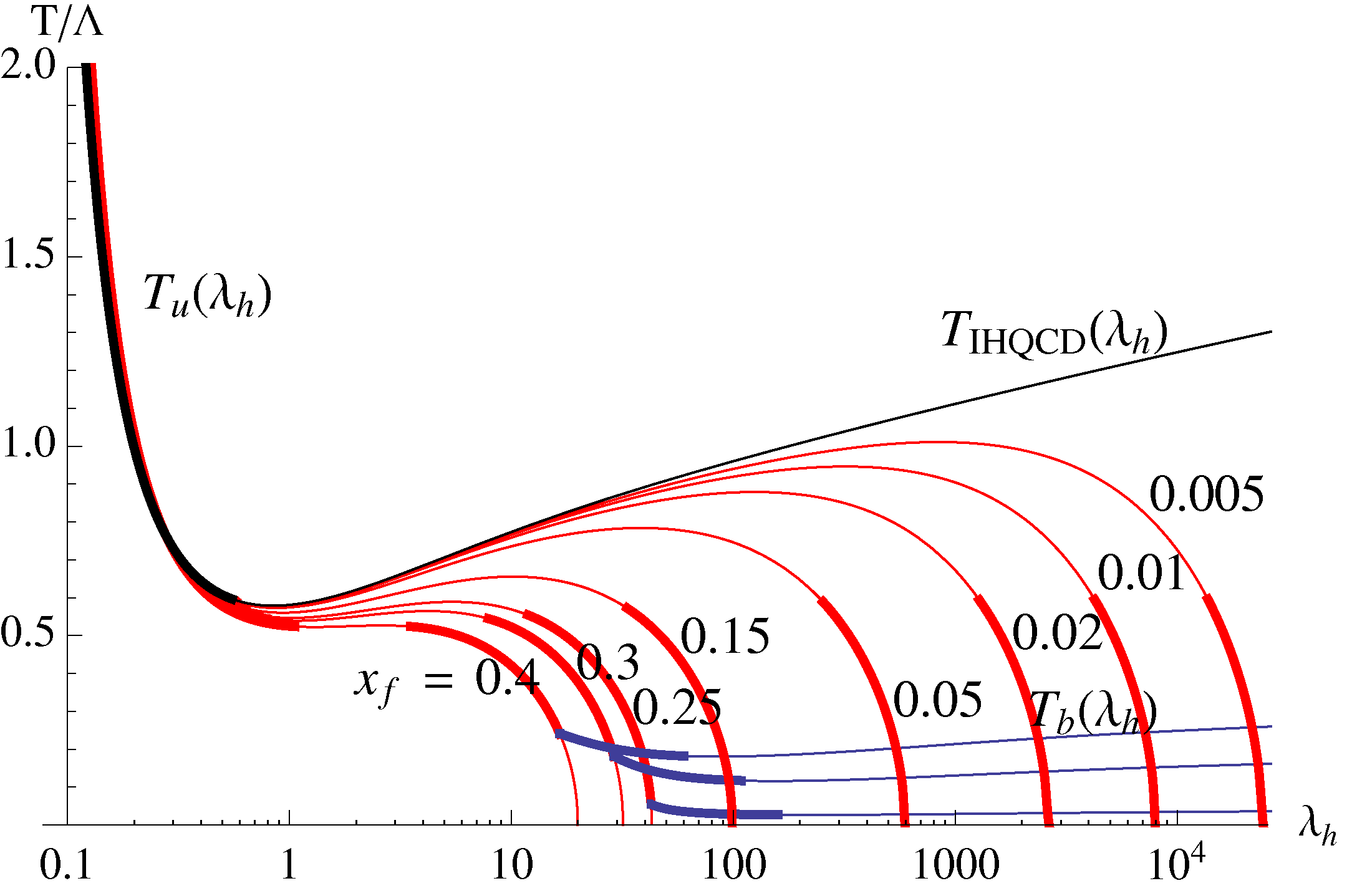}
\includegraphics[width=0.49\textwidth]{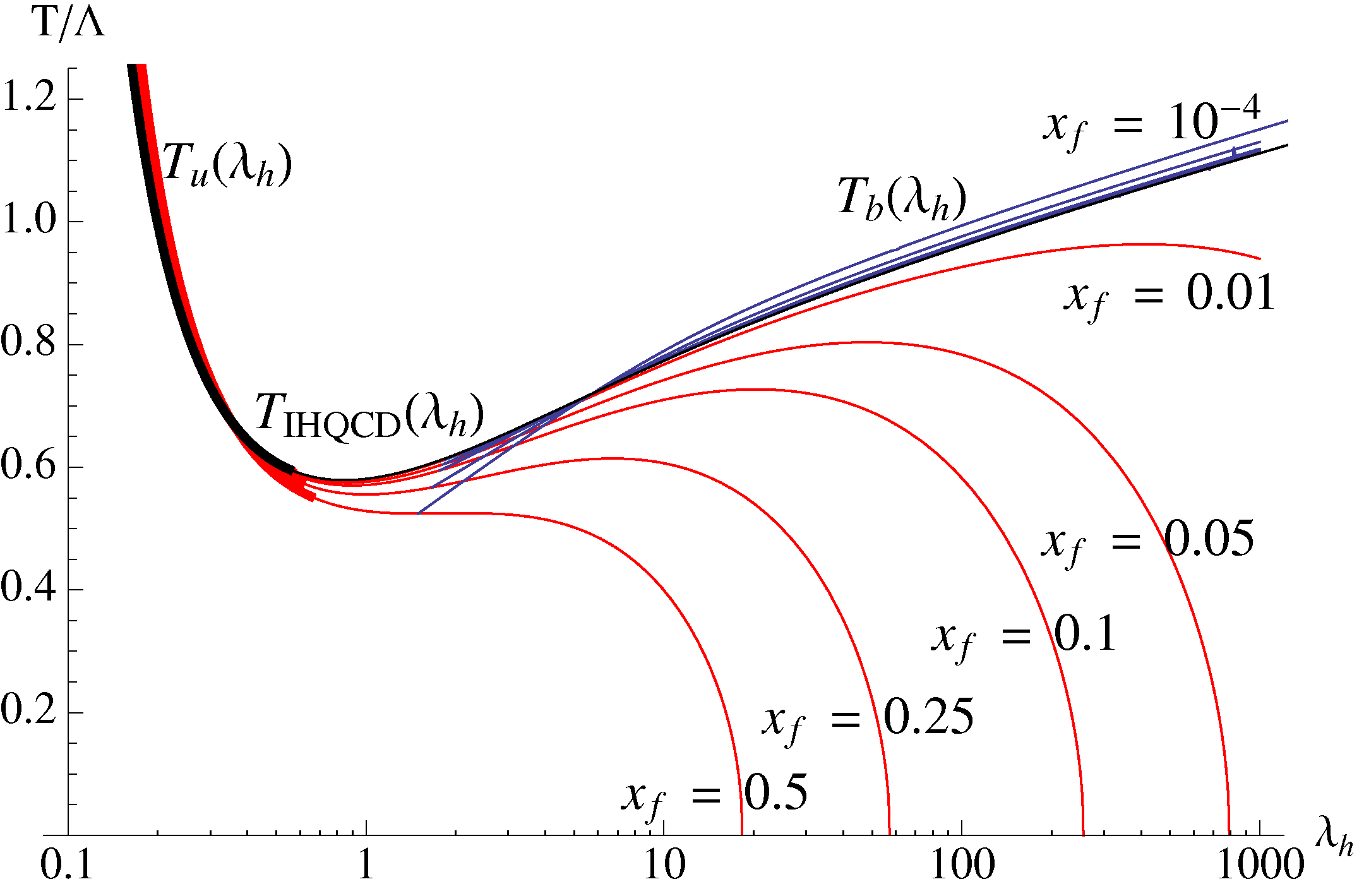}

\caption{\small $T(\lambda_h)$ for various small values of $x_f$ and for potential I,
$W_0 = 24/11$ (\emph{Left}) or for potential II, $W_0 = 12/11$ (\emph{Right}).
The black curve is the IHQCD limit. The chirally unbroken $T_u(\l_h)\equiv T(\l_h,\tau=0)$
branch asymptotes to the IHQCD curve as $x_f \rightarrow 0$, for both potentials.
The chirally broken $T_b(\l_h)\equiv T(\l_h,\tau_h(\l_h,m_q=0))$ branches behave
very differently for PotI and PotII. For PotI $T_b$ is
absent at such low $x_f$ and all phases are chirally symmetric
 (see also Fig.~\protect\ref{figPhasesPotIW0}).
For PotII the curves $T_b$ follow very closely $T_s$ and, correspondingly, $T_h\approx T_s$
(see Fig.~\protect\ref{figPhasesPotIIW012per11}).
}
\label{figPotxf0limit}
\end{figure}

The V-QCD
 models at $x_f = 0$ are equivalent to an IHQCD model with potential $V_g$.
One thus
expects that the hadronisation transition $T_h$ will approach
the 1st order deconfining transition of SU($N_c)$ YM
theory\footnote{Note that strictly speaking the limit of YM theory demands $N_f=0$ and falls outside
the Veneziano limit $N_f\to\infty$ of QCD. This may explain the nontrivial structures
observed at $x_f\to0$.} when $x_f  \rightarrow 0$.
However,
for PotI and large $W_0$ this cannot be the case, since the hadronisation transition does not
exist for very low $x_f$, as we pointed out above. Indeed we see from the phase diagrams of
Fig.~\ref{figPhasesPotIW0}, that the symmetric transition $T_s$, which only exists for $x_f\lta0.4$,
is the precursor of the YM transition in this case.

Let us then discuss in detail what happens in the $x_f \to 0$ limit in the two cases and at finite temperature.
Since thermodynamics is determined by the set of $T(\l_h)$ curves in Fig.~\ref{figTlah},
one should study how this configuration behaves when $x_f\to0$. The $T(\l_h,\tau=0)$ curve
(shown in red in Fig.~\ref{figTlah}) exists
only for $\l<\l^*$ and, according to the definition \nr{defoflastar} $\l^*\to\infty$ when
$x_f\to0$. In more detail, the limit is given by
\be
\l_*^{2/3}={8\over W_0+\fra{20568}{4619}}\,{1\over x_f}\,\sqrt{\ln\l_*},\quad
V(\l_*)={18476\over 729}\l_*^{4/3}\sqrt{\ln\l_*}.
\label{limitlastar}
\ee
where the equation on the left determines $\lambda_*$ while the equality on the right expresses
$V(\l_*)$ as a function of $\l_*$.
Eq.~(\ref{limitlastar})
is valid both for potentials I and II.
The red curves of Fig.~\ref{figTlah},
therefore, stretch to the right when $x_f\to0$. Quantitatively
what happens is shown in Fig.~\ref{figPotxf0limit} and one sees that they approach the
$T(\l_h)$ curve of IHQCD when $x_f\to0$. This is as expected since only $V_g$ remains in
the limit. It is thus obvious that
$T_s$ approaches the transition temperature of IHQCD as $x_f \to 0$ (but it may be a transition
between two metastable phases rather than a physical transition).

To find the relative behavior of $T_s$ and $T_h$ one needs the asymptotic properties
of the curves $T_b(\l_h)\equiv T(\l_h,\tau_h(\l_h,m_q=0))$
(shown in blue in Fig.~\ref{figTlah})
which only exist for $\l>\l_\rmi{end}$.
Here PotI and PotII behave in considerably different ways, as is
already seen from Figs. \ref{figPhasesPotIIW012per11} and \ref{figPhasesPotIW0}.

The crucial difference between PotI
(at large $W_0$)
and PotII comes from the fact that for PotI the
value of $\l_\rmi{end}$
(the endpoint of the blue $T_b(\l_h)$ curves in Fig.~\ref{figPotxf0limit}) 
grows rapidly when $x_f$ decreases, while for PotII
$\l_\rmi{end}$ remains almost constant. Since always $T_u(\l_\rmi{end})=
T_b(\l_\rmi{end})$ and $T_u(\l_h)$ decreases rapidly at large $\l_h$, also
the temperature $T_\rmi{end}$ becomes
small at small $x_f$ for PotI.
This drives the whole
curve $T_b(\lambda_h)$ towards zero and since  $T_h$ is determined by integration along
$T_b(\lambda_h)$ also $T_h\to0$.
Finally $\l_\rmi{end}$ ceases to exist when $x_f$ goes below a critical value $\sim 0.25$,
the temperatures $T_\rmi{end}$ and $T_h$ reach zero, 
and the
low 
temperature chiral symmetry breaking phase disappears.

For PotII $T_b(\l_h)$ follows very closely $T_\rmi{IHQCD}$ above it
(Fig.~\ref{figPotxf0limit}) and it is thus natural that $T_h\gta T_s$ and that they approach the same
limit. $T_s$ is actually metastable (Fig.~\ref{figPhasesPotIIW012per11}).

One can also illustrate the connection of the behavior of $\l_\rmi{end}$ to the BF bound of the
tachyon (Fig.~\ref{tachyonmassfig}) by analyzing the linearized tachyon equation motion as discussed
in Appendix~\ref{Applaend}.

In the limit of $x_f\to x_c$ one expects that
all dimensionful quantities sensitive to the IR vanish as
specified by Miransky scaling \nr{i4}. All our numerical
results are compatible with this, but
conclusive numerical verification would require extensive further work.
Analytic arguments supporting the scaling, similar to those presented in Sec.~10 of \cite{jk},
can also be constructed in the finite temperature case. We shall
here, however, only briefly comment on the scaling as well as the overall behavior of the
solutions as $x_f \to x_c$ from below.

We start with the case of zero temperature which is simpler.
For $x_f<x_c$ the dominant background is the one with nontrivial tachyon, and chiral symmetry is broken.
As $x_f \to x_c$ the solution comes closer and closer to the fixed point, and the near conformal region grows.
As it turns out, the pieces for $\l>\l_*$ and $\l<\l_*$ approach separately fixed solutions in this limit,
which do not talk to each
other.\footnote{More precisely, keeping fixed the scale $\Lambda_\rmi{IR}$ defined by the
IR expansions, the background approaches pointwise a ``IR'' limiting solution which flows
from the good IR singularity to the fixed point at $\l=\l_*$. This solution which approaches
$\l_*$ from the ``wrong'' side is possible due to the presence of the tachyon.
Keeping the UV scale $\Lambda_\rmi{UV}$ fixed instead, the ``UV'' limiting solution is the
one that flows from the IR fixed point at $\l=\l_*$ to the standard UV fixed point at $\l=0$
with zero tachyon.}
Thus any observable which can be expressed only in terms of either the UV or the IR solution
approaches a fixed value in the $x_f \to x_c$ limit. The ratio of the characteristic scales
of the two pieces diverge as specified by the Miransky scaling factor of Eq.~\eqref{i4}.

\begin{figure}[!tb]
\begin{center}
 \includegraphics[width=0.49\textwidth]{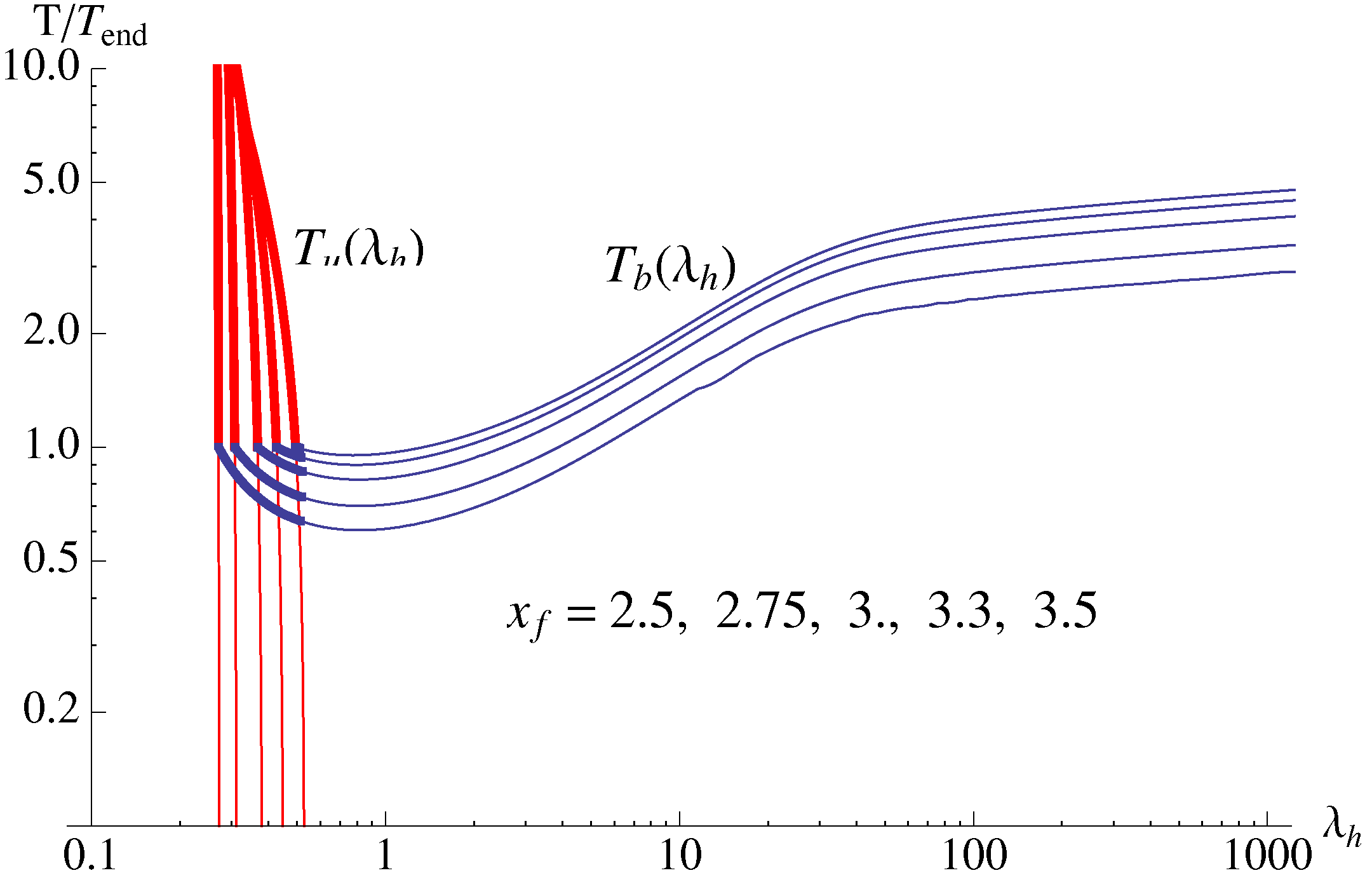}
\end{center}
\caption{\small The temperatures $T_u(\l_h),\,T_b(\l_h)$ of the two black hole branches,
unbroken and broken, scaled 
to the temperature $T_\rmi{end}$ for PotII, $W_0$ SB normalized. The values of $x_f$ from 
top to bottom are $2.5$, $2.75$, $3$, $3.3$, and $3.5$. Compare to 
Figs.~\protect\ref{figTransitionsTend} and~\protect\ref{figTTransitionsSB}.}
\label{figTendscaled}
\end{figure}

It is hard to find simple examples of such observables at zero temperature, but as it turns out,
at finite temperature there are plenty. However, the analysis of the limit is more involved,
since we have the additional parameter $\l_h$ we can be either in the UV ($<\l_*$) or in the IR ($>\l_*$) regions.
The curve which determines the main features of the thermodynamics is $T_b(\l_h)$, which
lies mostly in the IR region. Its endpoint $\l_\rmi{end}$ is however smaller than $\l_*$.
When $\l_h>\l_*$ we expect that the background solution breaks into two parts similarly as for
$T=0$ in the limit $x_f \to x_c$, and the temperature is determined solely by the IR
piece.\footnote{There is a subtlety here as comparing temperatures at different $\l_h$ requires fixing
the units of energy, which we do by calculating $\Lambda_\rmi{UV}$.
However for $\l_h>\l_*$ the UV part of the solution, and hence practically the units of energy,
become independent of $\l_h$ in the limit $x_f \to x_c$. Thus we could equally well define
the units in terms of the behavior of the IR piece of the solution near $\l=\l_*$.}
In the limit $x_f \to x_c$ we find\footnote{This is observed numerically, and can be understood
by studying the violation of the BF bound in the spirit of Appendix~\ref{Applaend}.}
that $\l_\rmi{end} \to \l_*$ from below. Therefore the whole $T_b(\l_h)$ curve is in the IR
region in the strict $x_f \to x_c$ limit, and it is plausible that it takes a
fixed shape.

This behavior is supported by the numerical study of Fig.~\ref{figTendscaled}, where we plot 
the temperatures of the two black hole branches as functions of $\l_h$ for PotII with $W_0$ SB normalized. 
The $x_f$-dependence of the curve $T_b(\l_h)$ is, up to the overall normalization, so small 
for $\l_h \gg \l_*$ that it cannot be resolved from the plot even at relatively low $x_f \lesssim 3$.
The main effect with increasing $x_f$ is that the $T_b(\l_h)$ curve is visible down to lower 
and lower $\l_h$ as $\l_*$ decreases slowly, while the shape of the curve remains fixed.
The curve $T_u(\l_h)$ approaches a vertical line when scaled to $T_\rmi{end}$, reflecting 
the fact that $\l_\rmi{end} \to \l_*$.

The values of all the critical temperatures (except $T_\rmi{crossover}$), as well as all
thermodynamics up to the transitions, are determined by $T_b(\l_h)$ as $x_f \to x_c$.
Therefore, we expect that the thermodynamics ``freezes'' in this limit, in the sense
that all ratios of the critical temperatures approach fixed values.
Moreover, the parts of the thermodynamical functions which are determined by the IR solutions,
are expected to have well defined limits.
While we have not proven these statements, they are strongly supported by the numerical 
study of Fig.~\ref{figTendscaled}. Notice however, that our data only extends up to 
$x_f=3.5$ which is still well below the critical value $x_c \simeq 3.70$. 
Therefore we cannot exclude the possibility that something drastic happens for $x_f$ even closer to $x_c$.

\begin{figure}[!tb]
\begin{center}

\includegraphics[width=0.49\textwidth]{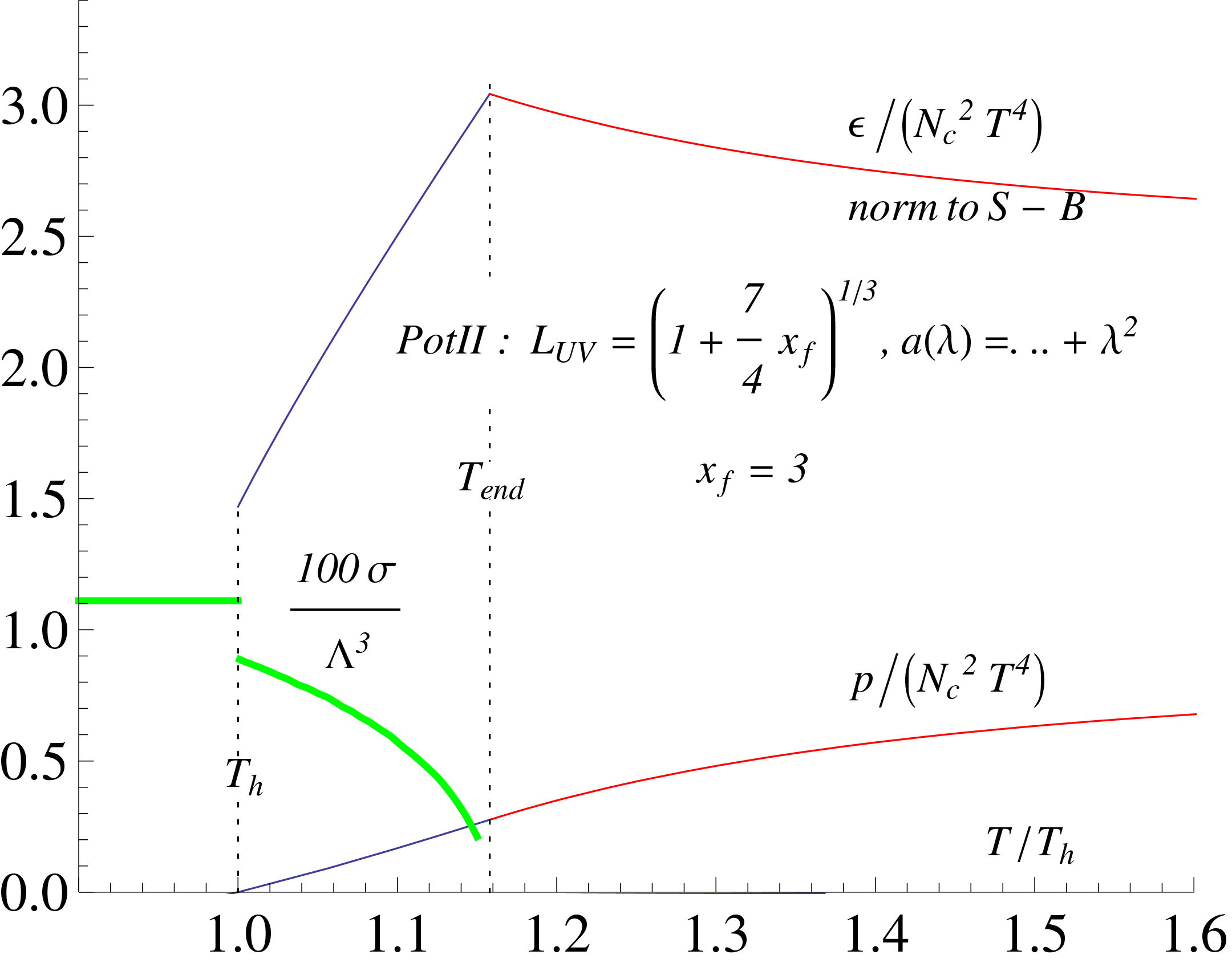}\hfill
\includegraphics[width=0.49\textwidth]{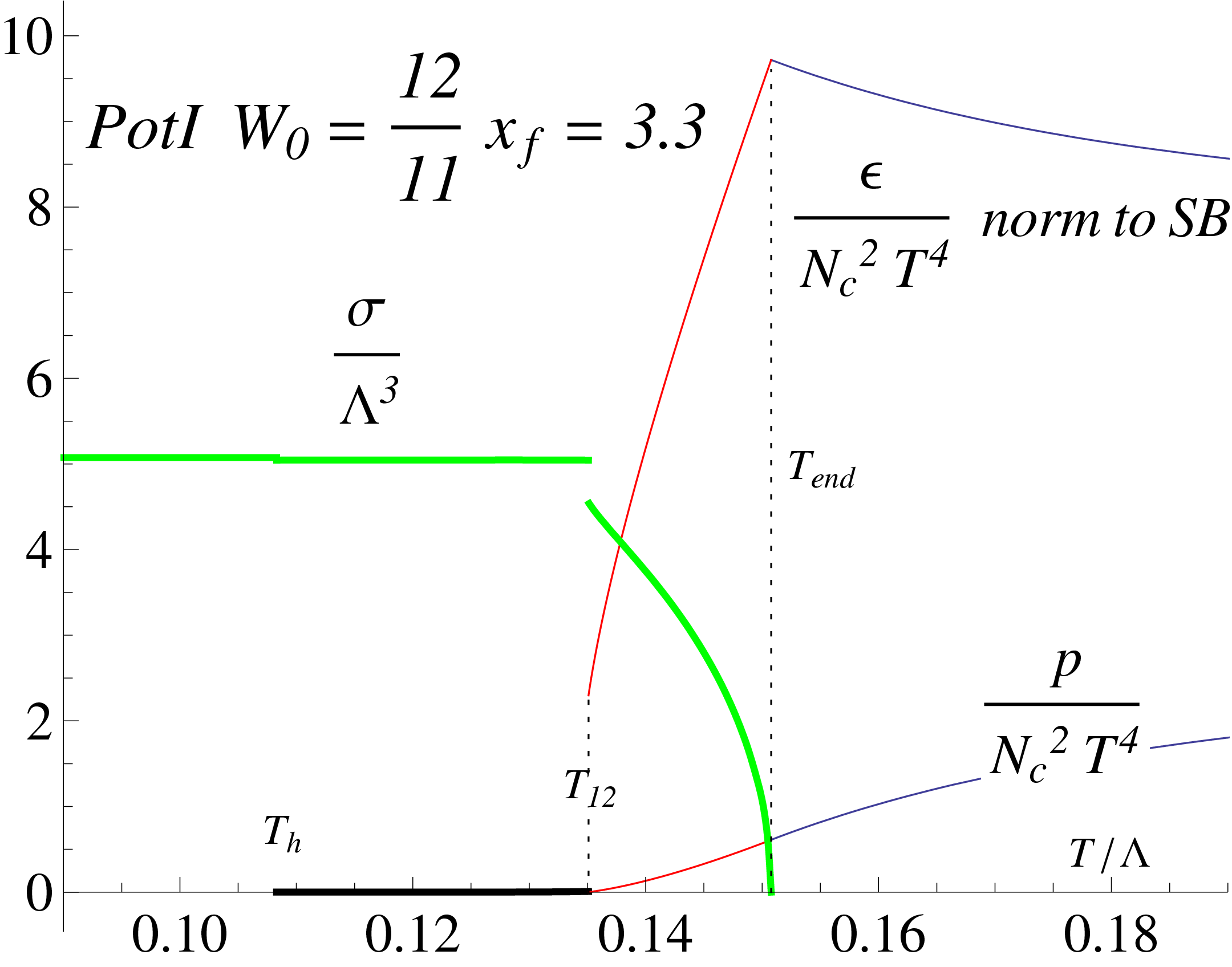}
\end{center}

\caption{\small Examples of equation of state with $\bar q q$ condensate. \emph{Left:}
Type II SB-normalised potentials at $x_f=3$ (compare
Figs.~\protect\ref{figTransitionsTend} and \protect\ref{figTTransitionsSB}). Note the scaling of
$\sigma$ by a factor 100.
\emph{Right:} Type I potentials with $W_0=12/11$ and $x_f=3.3$ (compare Fig.~\protect\ref{figPhasesPotIW0},
middle panel).
}
\label{thermosigma}
\end{figure}

\begin{figure}[!tb]
\begin{center}

\includegraphics[width=0.49\textwidth]{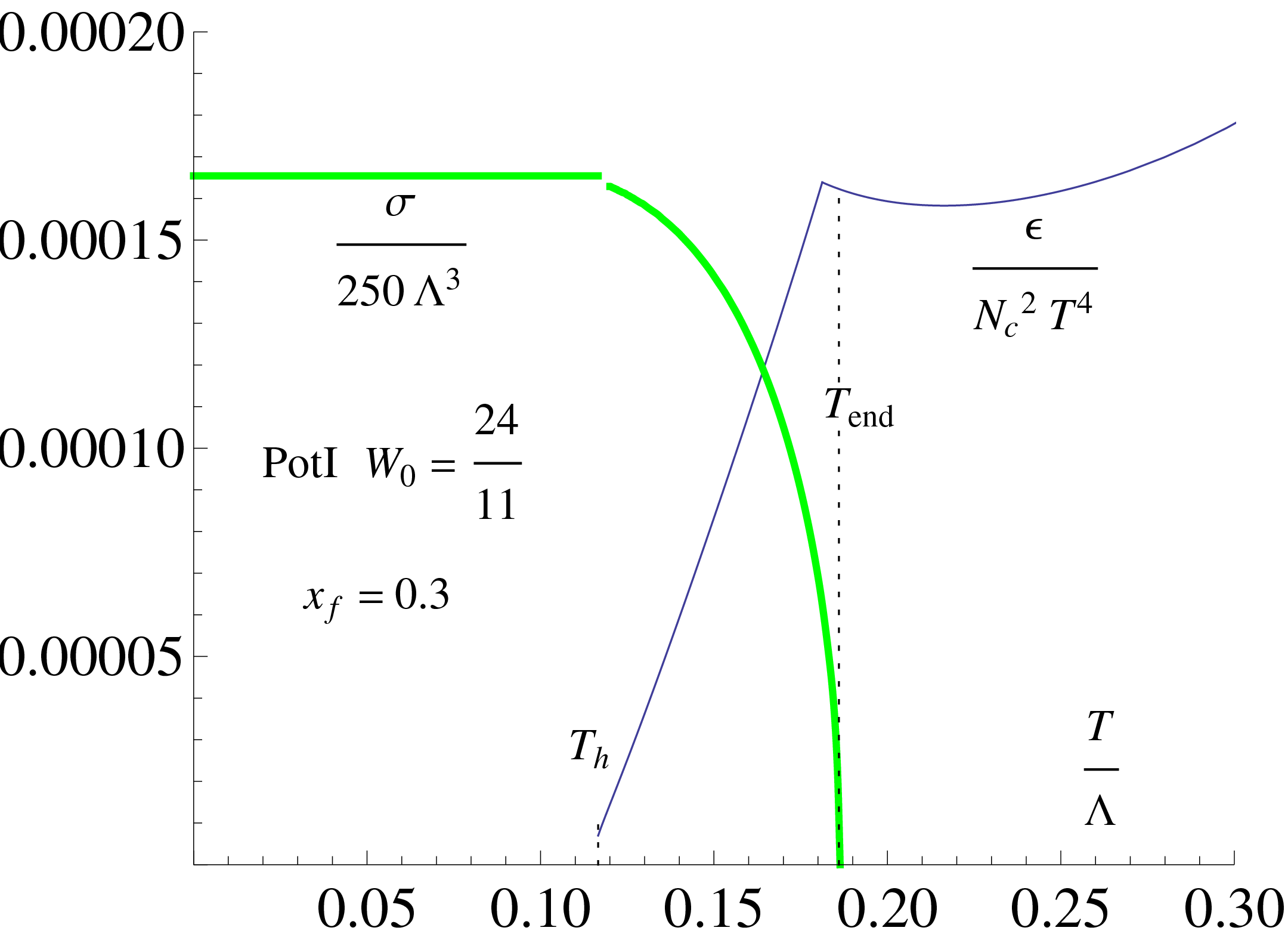}\hfill
\includegraphics[width=0.49\textwidth]{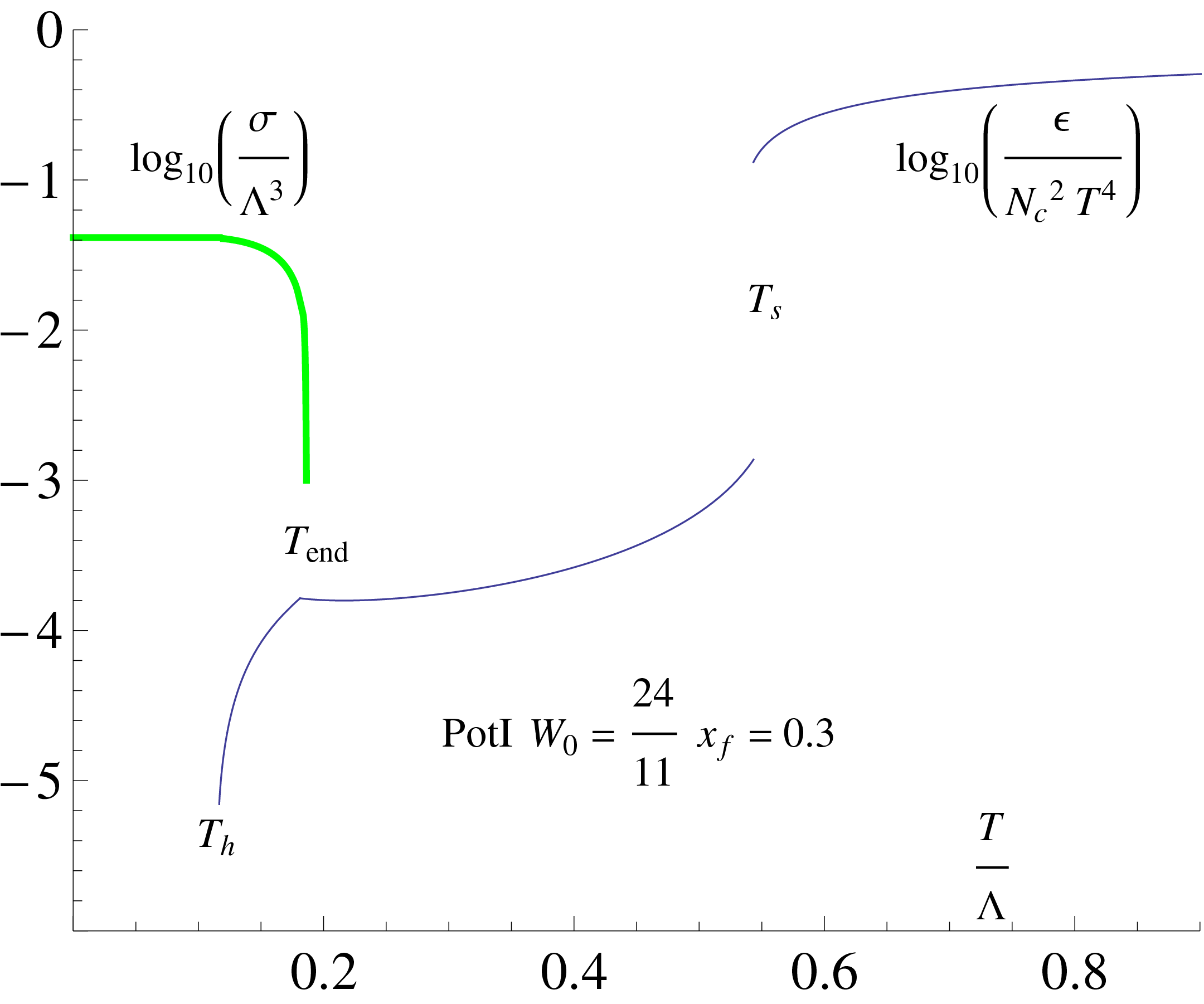}
\end{center}

\caption{\small The condensate for the Pot I, $W_0=\fra{24}{11}$ phase diagram in the right hand column of
Fig.~\protect\ref{figPhasesPotIW0} at $x_f=0.3$ in linear (left) and log (right) scale.
Note the scaling of $\sigma$ by a factor 1/250 in the left panel. The right panel shows all
the three transitions in Fig.~\protect\ref{figPhasesPotIW0}.
}
\label{thermosigmaW024per11}
\end{figure}

\section{The chiral condensate}
In principle, it should be straightforward to extract the chiral condensate $\propto \sigma$ from
the tachyon solution in the UV
\be
\tau(r) \sim \sigma r^3(-\ln(\Lambda r))^{3/(2b_0)},\quad r\to0,
\label{sigt}
\ee
as the quark mass is set to zero. However, in this model the task is actually very demanding
due to the logarithmic corrections (i.e., the running of the condensate) and the fact that
the numerical solutions have a tiny residual quark mass due to limited numerical precision.
These issues and their resolution are discussed in Appendix~\ref{AppCond}.

Examples of the correlation of $\sigma$ with that of the free energy are shown
Fig.~\ref{thermosigma}.
One expects that $\sigma$ jumps in a 1st order transition. The first case is
thermodynamics for SB-normalised type II potentials at $x_f=3$, also studied in
Figs.~\ref{figTransitionsTend} and \ref{figTTransitionsSB}.
Cooling from large $T$ chiral symmetry first breaks at the 2nd order transition $T_\rmi{end}$ and the
condensate starts from zero and increases with further cooling.
At $T_h$ the system experiences a 1st order transition and
$\sigma$ jumps the amount shown in the figure. Below that $\sigma$ remains constant in the present
models, which does not describe the thermodynamics of the low $T$ phase. In that the degrees of
freedom are $N_f^2$ massless Goldstone bosons.

A second example is thermodynamics with condensate of Type I potentials with $W_0=12/11$ and $x_f=3.3$.
This case is special in that in it the 1st order line $T_h$ splits in two 1st order transitions
at $T_h,\,T_{12}$ if $x_f\gta2.8$, as shown explicitly in Fig.~\ref{figPhasesPotIW0},
middle panel. Again the highest temperature transition is a 2nd order one at $T_\rmi{end}$,
at which the condensate starts to grow when the system is cooled.
The condensate grows up to the value $\sigma=4.537\Lambda^3$.
Then there is a 1st order
transition at $T_{12}$ with a jump in $\sigma$ and latent heat:
\be
{\Delta\sigma\over\Lambda^3}=0.508,\quad {\Delta\epsilon\over T_{12}^4}=2.29
\ee
and finally a very weak transition at $T_h$ with the value
\be
{\Delta\epsilon\over T_h^4}=1.03\cdot 10^{-7}.
\ee
It is clear that the value of $\sigma$ also jumps at the latter transition, but the size of the
jump is so small that we could not extract it reliably from the numerics.

As a third example, consider the case Pot I, $W_0=\fra{24}{11}$ at $x_f=0.3$, which is very special in
that $T_h$ is very small and it is $T_s$ which dominates, as is seen in the right hand column
of Fig.~\ref{figPhasesPotIW0}. The magnitudes vary so much that all the transitions can be
presented only on log scale, see Fig.~\ref{thermosigmaW024per11}.

\begin{figure}[!tb]
\begin{center}

 \includegraphics[width=0.49\textwidth]{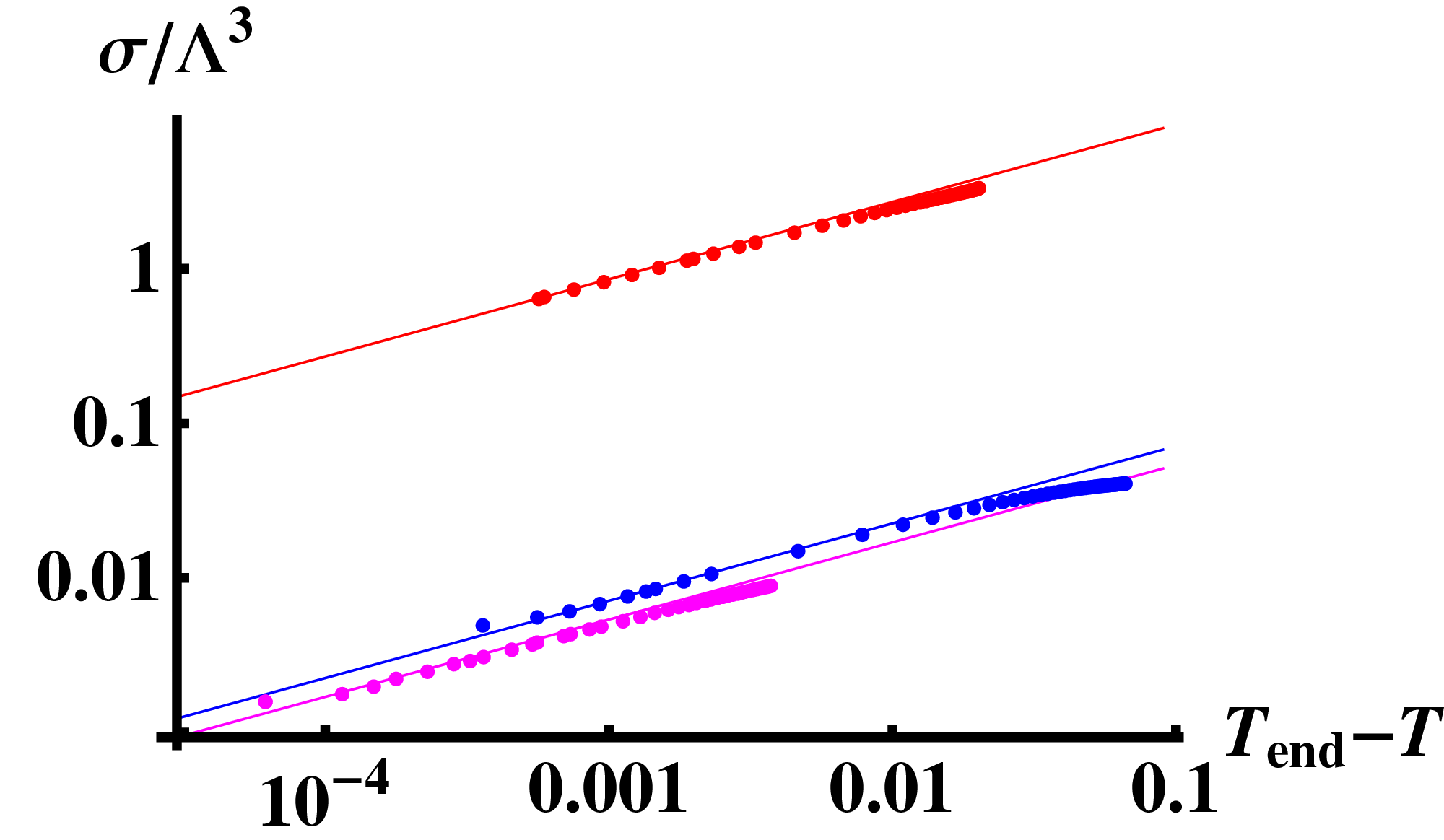}
\end{center}
\caption{\small The condensate as a function of $T_\rmi{end}-T$ in log-log scale. The dots
are our data, extracted from the background solutions, and the lines are fits having the form
$\sigma = C \sqrt{T_\rmi{end}-T}$. The choices of potentials are, from top to bottom: PotI
with $x_f=3.5$ and $W_0 =12/11$ (red); PotI with $x_f=0.3$ and $W_0=24/11$ (blue); and PotII
with $x_f=3$ and SB normalized $W_0$ (magenta). }
\label{sigmascaling}
\end{figure}

At the 2nd order transitions the condensate goes to zero continuously as the temperature approaches
$T_\rmi{end}$ from below. The curves in Figs.~\ref{thermosigma} and~\ref{thermosigmaW024per11} seem to
be compatible with the standard expectation $\sigma \sim \sqrt{T_\rmi{end}-T}$. We study this
more precisely in Fig.~\ref{sigmascaling} where we plot our data for the condensate for $T$
close to $T_\rmi{end}$ in the log-log scale for various choices of the potentials. The data
are compared to the lines $\sigma = C \sqrt{T_\rmi{end}-T}$, with the coefficients $C$ chosen
such that the lines overlap with the tails of the data at small $T_\rmi{end}-T$. It is
convincing that $1/2$ is indeed the correct exponent.

\section{Nonzero quark mass and thermodynamics}\la{secnonzero}
We have so far mainly discussed the case of vanishing quark mass, for which
chiral symmetry may hold.
Effects of nonzero quark mass have been mentioned in subsection
\ref{subsubsect_nonzeromq}
and described in Figs.~\ref{masspic} and \ref{figTlah}. They follow from the
fact that tachyonfree
black hole solutions do not exist, as a reflection of the disappearance of
chiral
symmetry. However, numerically the tachyonic small-$m_q$ solutions are very
close to the
zero mass tachyonfree solutions in the UV at small $\l$, as is seen
by comparing left and right panels of Fig.~\ref{figTlah}. Thus chiral symmetry
is always broken
even in the high-$T$ phase, but quantitative effects are small at large $T$.

\begin{figure}[!tb]
\begin{center}

\includegraphics[width=0.49\textwidth]{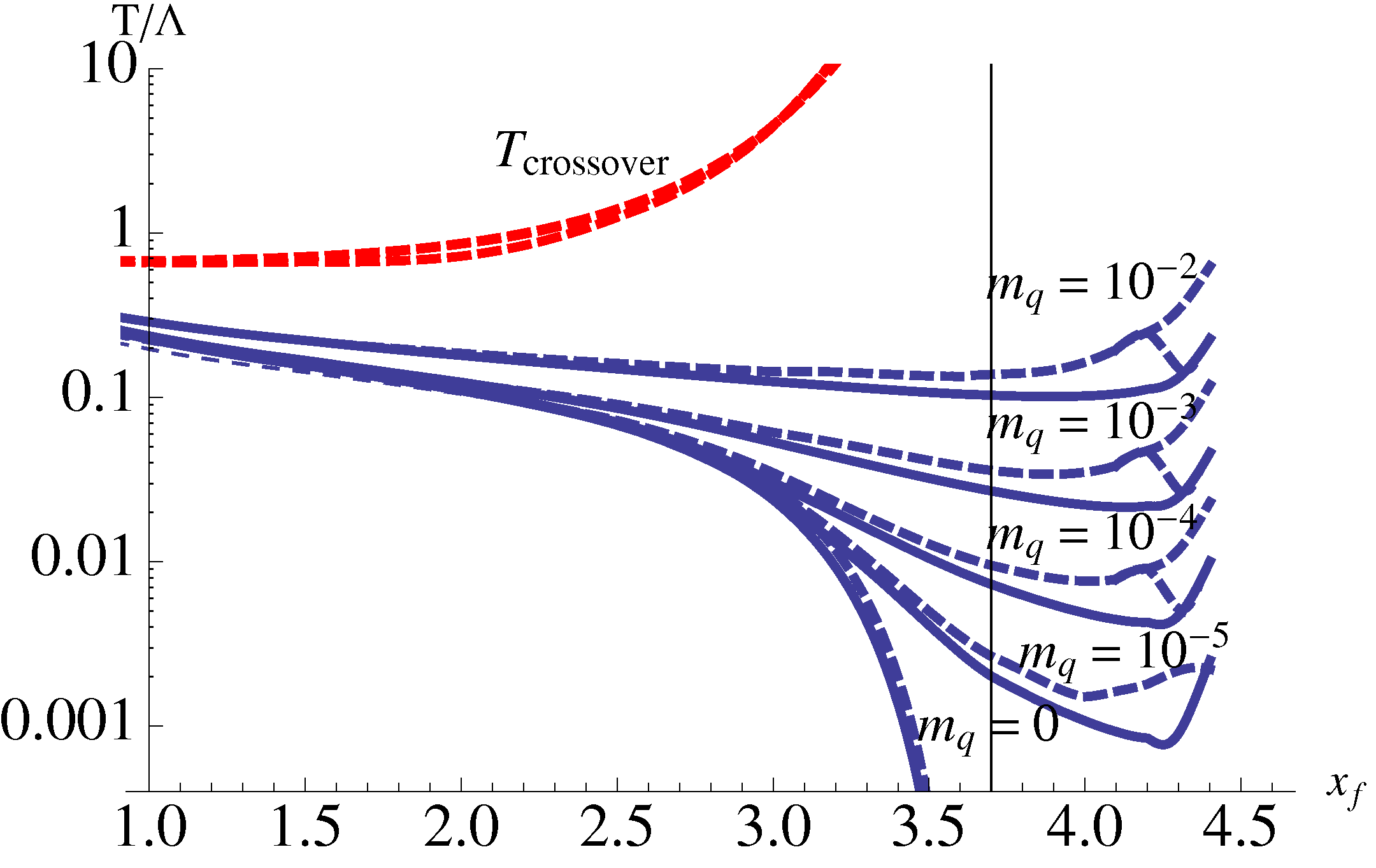}
\end{center}
\caption{\small The behavior of $T_h(x_f)$ (in units of $1/\CL_\rmi{UV}$, over the $x_f$
range in the figure this unit changes by about 30\%)
for $m_q=0$ and for small values of $m_q$. For $m_q=0$ the dashed line shows the true
2nd order chiral symmetry breaking transition. For nonzero $m_q$ the line shows the
position of one maximum of the interaction measure, a second one gives the usual
$T_\rmi{crossover}$ (which is almost independent of $m_q$). }
\label{noconfwin}
\end{figure}

The effects of small nonzero $m_q$ are shown quantitatively in Fig.~\ref{noconfwin}
and can be summarised as follows:
\bi
\item
The main effect follows from the fact that for nonzero $m_q$ even the high $T$ phase
is chirally broken.
The curve marked $m_q=0$ is the same $T_h(x_f)$ as that in Fig.~\ref{figTTransitionsSB}.
For this case the
phase at $T>T_h$ is chirally symmetric and chiral symmetry is broken when $T$ decreases
below $T_h$. For nonzero $m_q$ also the phase at $T>T_h$ is chirally
broken and the effective order parameter of the transition is the jump in entropy
or energy density. There is also a jump in the condensate, but the condensate is
nonzero also for $T>T_h(x_f)$.
\item
For $x_f$ clearly below
$x_c$ the effects of small $m_q$ on the phase diagram are small.
Particularly interesting is the pattern of approach towards
$m_q=0$. The smaller $m_q$, the higher is the value of $x_f$ where the curves
start deviating significantly.
\item
The 2nd order transition at $T_\rmi{end}$ becomes a continuous one. This is
obvious from Fig.~\ref{figTlah}, there is no discontinuity near $T_\rmi{end}$.
However, a remnant of the genuine transition is a maximum of interaction
measure, also plotted in Fig.~\ref{noconfwin}.
\item
At large $x_f$ the conformal window and Miransky scaling disappear.
For $m_q=0$ the transition
temperature $T_h$ vanishes when $x_c-x_f\to0$ as dictated by Miransky scaling.
The smallest nonzero mass destroys this effect and $T_h$ curves upwards towards
larger values.
\item
The effect of nonzero mass could also be seen by plotting the beta function for
values
of $x_f$ within the conformal window. For $m_q=0$ only the tachyonless solutions
matter and
they extend only up to $\l^*$ in Fig.~\ref{figTlah}. The beta function
$\beta(\l)$ only
exists for $\l<\l^*$ and $\beta(\l^*)=0$ at the IR fixed point $\l^*$. For
$m_q>0$ the beta
function comes close to $\l^*$ but continues past it to larger values of $\l$.
\ei

\section{Outlook}

In this paper, we have used bottom-up holography to study the thermodynamics of
models that are in the universality class of  QCD with massless quarks in the Veneziano limit,
(large $N_c$ and $N_f$ but fixed $x_f=N_f/N_c$).

The temperature dependence of the pressure $p(T)$ and the condensate
$\sigma(T)=\langle\bar q q\rangle(T)$ was computed
at various $x_f$ up to the loss of asymptotic freedom at $x_f=11/2$ with a conformal window appearing
at $x_c<x_f<11/2$, $x_c\approx4$. The singularities of these quantities define the phase
diagram of the system.

One expects that the system has two phases, a low temperature phase
with broken chiral symmetry and a chirally symmetric high $T$ phase. The simplest alternative
is that these are separated by a 1st order transition at some $T=T(x_f)$, see Fig.~\ref{figxTphases}.
In holography, the high $T$ phase is a metric with a black hole and a new feature appears: this
phase can be either chirally symmetric (no tachyon) or chirally broken (nonzero tachyon).
The same doubling applies to the low $T$ thermal gas phase. The
phase structure can be correspondingly more complicated. For given gluonic and fermionic
potentials
of the V-QCD action
the thermodynamics is fixed and calculable.

A typical prediction of the model is that there indeed is a 1st order line for $x_f$ from $0$ up
to some value $\le x_c$. In holography this is a transition at some $T=T_h(x_f)$ between a
non-tachyonic black hole metric and a thermal gas metric with a tachyon. 
In field theory language one would say that
at this temperature there is deconfinement and chiral symmetry breaking with coinciding
deconfinement and chiral temperatures, $T_d=T_\chi$. However, a new interesting feature is that
at larger $x_f$ this 1st order line can split in two: first at higher $T$
chiral symmetry is broken in a 2nd order transition, then at a lower $T$ there is a 1st order
deconfinement transition (see, e.g., Fig.~\ref{figTTransitionsSB}), $T_\chi>T_d$.

A particular feature of the phase diagram is that with $x_f$ approaching $x_c$ all transition
temperatures, as all mass scales, decrease as specified by Miransky scaling. Associated
with this approach is quasiconformality and walking. At very large $T$ there is always a
weak coupling region into which one enters at $T_\rmi{crossover}$.
This increases when $x_f$ grows, behaves regularly at
$x_c$ and continues further into the conformal window increasing faster and faster
(see again Fig.~\ref{figTTransitionsSB}).

Detailed predictions of the model depend on the gluonic and fermionic scalar potentials.
There are several physical constraints in deriving them, but they are not uniquely
determined. One crucial constraint is missing: in contrast to the application of this
model to hot SU($N_c$) theory \cite{YM2}, there is no 4d lattice data in the large $N_c,\,N_f$
limit. Thus one is genuinely making predictions and we find that also phase diagrams deviating
rather drastically from the above baseline prediction follow from holography. Particularly
striking examples are shown in Fig.~\ref{figPhasesPotIW0}. In these, the chirally symmetric
phase can extend all the way down to zero temperature.

Although we do not expect this to happen in QCD, it may happen in other large N-theories.
The last diagram of  Fig.~\ref{figPhasesPotIW0} has a structure that is very reminiscent of
the phase diagram of high-T$_c$ cuprates, although here we are at zero charge density.
In particular the intermediate dome-like structure with chiral symmetry breaking
corresponds to the superconducting dome \cite{dome} in the cuprate diagram.

One may also note that the models predict a very rich structure of metastable states.

There are many directions in which the model could be further sharpened and developed:
\bi
\item
All the computations are numerically very demanding and much further work would be useful.
This holds, in particular, for the approach to the conformal window, $x_f\to x_c$, and for
computations of the condensate.
\item
Effects of non-zero quark mass on the thermodynamics should be studied
beyond the discussion in Section \ref{secnonzero}.

\item
The extension to $x_{f1}$ flavors of mass $m_1$, etc., is possible but requires a
non-abelian version of the Sen action.
This is in principle possible to deal with, although we have much less information from
string theory on the details of such an action.

\item
It would be very interesting to accumulate lattice Monte Carlo data in the Veneziano limit.
For pure SU($N_c$) theory already $N_c=3$ is very close to $N_c=\infty$ \cite{teper1,panero}.
One might thus see that, e.g., $N_f=4,8,12$ at $N_c=3$ would already give useful
information.

\item
A chemical potential for baryons should be included. This necessitates the inclusion of a bulk
U(1) baryon vector field $A_\mu$. It would be particularly interesting to know the fate
of the $T_\chi>T_d$ phase at large $x_f$ and $\mu$.

\ei

\section{Acknowledgments}

We thank Ken Intriligator, Costas Kounnas, Vasilis Niarchos, Marco Panero, Gabriele Veneziano, for discussions.
This work was in part supported by grants  FP7-REGPOT-2008-1-CreteHEPCosmo-228644,
 PERG07-GA-2010-268246, the EU program ``Thales'' ESF/NSRF 2007-2013, and by the European Science
Foundation ``Holograv" (Holographic methods for strongly coupled systems) network.
It has also been co-financed by the European Union (European Social Fund, ESF) and
Greek national funds through the
 Operational Program ``Education and Lifelong Learning'' of the National Strategic
 Reference Framework (NSRF) under
 ``Funding of proposals that have received a positive evaluation in the 3rd and 4th Call of ERC Grant Schemes''.
 TA thanks the Vaisala foundation for financial support.

\appendix
\renewcommand{\theequation}{\thesection.\arabic{equation}}
\addcontentsline{toc}{section}{APPENDIX\label{app}}
\section*{APPENDIX}

\section{UV expansions} \label{AppUV}

The expansions near the standard UV boundary can be computed in a straightforward manner. Let us fix
the location of the boundary at $r=0$.  The blackening factor $f$ is very close to unity in the UV,
\be
 f(r) = 1 + \mathcal{O}(r^4) \ .
\ee
Therefore the leading finite temperature and zero temperature expansions of the various fields coincide.
Moreover, as the tachyon vanishes at least linearly in $r$ for $r \to 0$, it can be set to zero when
solving for the leading behavior of the coupling $\l$ and the warp factor $A$. Hence the expansions
take the form familiar from earlier works \cite{YM1,YM2}. We reproduce here the leading expansions of $\l$
and $A$ as well as the expansions of the non-normalizable and normalizable tachyon expansions
both in terms of $r$ and $A$.

\subsection{Fields $\l $ and $A$}

Setting the tachyon to zero, the equations of motion for $\l$ and $A$ involve the effective potential
\be \label{Videfapp}
 V_{\rm eff}(\l)=V_g(\l)-x_f V_f(\l,0)=\frac{12}{\CL_\rmi{UV}^2}\left[1 + V_1 \l +V_2
\l^2+\cdots \right] \ .
\ee
Then the (leading) UV expansions of $A$ and $\l$ can be found by substituting suitable Ans\"atze
in the equations of motion. The result reads
\bea \label{UVexpsapp}
A(r) &=& -\ln\frac{r}{\CL_\rmi{UV}} + \frac{4}{9 \ln(r \Lambda)}  \\
&&+ \frac{
  \frac{1}{162} \left[95  - \frac{64 V_2}{V_1^2}\right] +
   \frac{1}{81} \ln\left[-\ln(r \Lambda)\right] \left[-23 + \frac{64
V_2}{V_1^2}\right]}{
  \ln(r \Lambda)^2} +{\cal O}\left(\frac{1}{\ln(r\Lambda)^3}\right) \nn \\
  V_1 \l(r)&=&-\frac{8}{9 \ln(r \Lambda)} + \frac{
   \ln\left[-\ln(r \Lambda)\right] \left[\frac{46}{81} - \frac{128 V_2}{81
V_1^2}\right]}{\ln(r \Lambda)^2}+{\cal
O}\left(\frac{1}{\ln(r\Lambda)^3}\right) \ .
\eea
Two combinations of the series coefficients of the effective potential appear here. As the potential
is matched with perturbative QCD, they become
\bea
 V_1 &=&  \frac{8}{9} b_0  = \frac{88-16x_f}{27}  \\
 \frac{V_2}{V_1^2} &=& \frac{23}{64}+\frac{9 b_1}{16 b_0^2} =\frac{1}{64}
 \left(23+\frac{54 (34-13 x_f)}{(11-2 x_f)^2}\right)
\eea
where $b_i$ are the coefficients of the perturbative QCD beta function.
Notice that these coefficients are indeed the same for all potentials used in our study
and in particular independent of the choice of $W_0$.

Let us also present the expansions in terms of $A$, as we use it as a coordinate in all
numerical calculations. The result after the conversion reads
\bea 
 \ln r(A) &=&-A+\ln (\CL_\rmi{UV} )-\frac{4}{9 A} \\\nn
&&-\frac{ 72  \ln (\CL_\rmi{UV}
   \Lambda )-95 +\frac{64 V_2}{V_1^2} + \left(46 -\frac{128V_2}{V_1^2}\right)
   \ln A}{162 A^2} + \CO\left(A^{-3}\right) \\
  V_1 \l(A) &=& \frac{8}{9A}+\frac{ \left(46 -\frac{128 V_2 }{V_1^2}\right)\ln A + 72 \ln
   (\CL_\rmi{UV} \Lambda )}{81 A^2} + \CO\left(A^{-3}\right) \ .
\eea

\subsection{The tachyon}

As the tachyon is decoupled near the UV boundary, its UV behavior can be studied by
inserting the expansions calculated above for $\l$ and $A$ into the tachyon EoM.
We also develop the potentials as series in the UV:
\bea \label{Vhexps}
 V_{\rm eff}(\l) &=& V_g(\l)-x_f V_f(\l,0)=\frac{12}{\CL_\rmi{UV}^2}\left[1 + V_1 \l +V_2
\l^2+\cdots \right] \\\nn
 x V_f(\l) &=& W_0 + W_1 \l +W_2 \l^2+\cdots  \\
 \frac{\f(\l)}{a(\l)} &=& \frac{2 \CL_\rmi{UV}^2}{3}\left[1 + \f_1 \l +\f_2 \l^2+\cdots
\right] \ .
\eea
Here the leading coefficient of $\f/a$
was already fixed in order to have the correct UV mass of the tachyon
\cite{ckp}.
It is enough to study the linear terms in the tachyon EoM, which become
\bea
 &&\tau''(r) + \Bigg[-3+ {\cal O}\left(\frac{1}{\ln(r\Lambda)^2}\right)\Bigg]\frac{\tau'(r)}{r}\\ \nn
&&+ \Bigg[3+\frac{8 (\f_1\!+V_1\!)}{3 V_1 \ln(r\Lambda)}
  + {\cal O}\left(\frac{1}{\ln(r\Lambda)^2}\right)
\Bigg] \frac{\tau(r)}{r^2} = 0 \ .
\eea

The general solution for $r \to 0$ reads
\bea \label{TUVres}
 \frac{1}{\CL_\rmi{UV}}\tau(r) &=& m_q r
(-\ln(r\Lambda))^{\frac{4}{3}+\frac{4 \f_1}{3
V_1}} \left[1+ {\cal
O}\left(\frac{1}{\ln(r\Lambda)}\right)\right] \\ \nn
&&+\sigma r^3
(-\ln(r\Lambda))^{-\frac{4}{3}-\frac{4 \f_1}{3
V_1}} \left[1+ {\cal
O}\left(\frac{1}{\ln(r\Lambda)}\right)\right] \ .
\eea
Here matching with the perturbative anomalous dimension of the quark mass in QCD gives
\be
 \frac{4}{3}+\frac{4 \f_1}{3
V_1} = -\frac{\gamma_0}{b_0} = -\frac{9}{22-4 x_f}
\ee
where $\gamma_0$ is the leading coefficient of the anomalous dimension of the quark mass in QCD.

The result can be again written in terms of $A$, and it becomes
\bea \label{TUVresA}
 \tau(A) &=& m_q\, \CL_\rmi{UV}^2\, e^{-A} \,
(\ln A)^{-\frac{\gamma_0}{b_0}} \left[1+ {\cal
O}\left(A^{-1}\right)\right] \\ \nn
&&+\, \sigma \, \CL_\rmi{UV}^4 \, e^{-3 A}\, (\ln A)^{\frac{\gamma_0}{b_0}} \left[1+ {\cal
O}\left(A^{-1}\right)\right] \ .
\eea

\subsection{Finite temperature}
The basic relations
\be
f(r)=1-{\int_0^r\,dr/b^3(r)\over \int_0^{r_h}\,dr/b^3(r)},\quad
{1\over4\pi T}=b_h^3\int_0^{r_h}{dr\over b^3(r)},
\ee
can be evaluated in the UV by inserting from \nr{UVexpsapp}
\be
b=e^A={\CL_\rmi{UV}\over r}\biggl[1+{4\over9\ln(\Lambda r)}+
{4\over9\ln^2(\Lambda r)}
\biggl({b_1\over b_0^2}\ln(-\ln(\Lambda r))+\fr{11}9-{b_1\over 2b_0^2}\biggr)\biggr].
\ee
Terms of the order of $\ln^2(\ln)/\ln^3$ are neglected; for these, see \cite{KKVV}. One finds
\footnote{Ref.~\cite{YM1}, second paper, equation (D.3), has a different constant in the expansion
of $T$. We have checked the 1/3 here also numerically.}
\be
\int_0^r\,{dr\over b^3(r)}={r^4\over4\CL_\rmi{UV}^3}
\biggl[1-{4\over3\ln(\Lambda r)}-{4\over3\ln^2(\Lambda r)}
\biggl({b_1\over b_0^2}\ln(-\ln(\Lambda r))+\fr{7}{12}-{b_1\over 2b_0^2}\biggr)\biggr],
\ee
\be
T={1\over\pi r_h}\biggl(1+{1\over3\ln^2(\Lambda r_h)}\biggr).
\ee
For the quantity $b_h/T$ needed for the latent heat one has
\be
{b_h\over T}=\pi\CL_\rmi{UV}\biggl[1+{4\over9\ln(\Lambda r_h)}+
{4\over9\ln^2(\Lambda r_h)}
\biggl({b_1\over b_0^2}\ln(-\ln(\Lambda r_h))+\fr{17}{36}-{b_1\over 2b_0^2}\biggr)\biggr].
\label{bhoverTApp}
\ee

\section{IR expansions} \label{AppIR}

\subsection{Zero temperature}

Here we first discuss the expansions near the ``good'' IR singularity at zero temperature.
It is the particular solution which can be lifted to finite temperature. In the IR,
the tachyon potential in the DBI action is expected to be exponentially suppressed.
Therefore the tachyon is again decoupled, and the IR behavior of $\l$ and $A$ can be
solved separately from that of the tachyon. Moreover, the IR expansions of $\l$ and
$A$ are exactly the same as in IHQCD. We will anyhow repeat the discussion for the
particular asymptotics of $V_g$ that matches well with the IR properties of QCD
\cite{YM1},
which covers all potentials in this article.

\subsubsection{$A$ and $\l$}

Let us assume that the potential $V_g$ has the asymptotic behavior
\be
 V_g(\l) = v_0 \l^{4/3}\sqrt{\ln \l}\left[1 + {v_1\over\ln\l} +
 {v_2\over \ln^2\l} + \cdots \right] \ .
\ee
Then the asymptotic solution reads
\bea \label{IRresA}
 A &=& -\frac{r^2}{R^2}+\frac{1}{2}\ln\frac{r}{R}-\ln R-\frac{1}{2}\ln v_0+
 \frac{5}{4}\ln 2+\frac{3}{4}\ln 3 + \frac{23}{24}+\frac{4 v_1}{3}\nn \\
&&+\frac{R^2 \left(-173+512 v_1^2+1024 v_2\right)}{3456 r^2} + \CO\left(r^{-4}\right)\\
&=& -\frac{r^2}{R^2}+\frac{1}{4}\ln\frac{3r^2}{2R^2} + A_0 + \frac{23}{24}+\frac{4 v_1}{3} \nn \\
 \label{IRresA2}
&&+\frac{R^2 \left(-173+512 v_1^2+1024 v_2\right)}{3456 r^2} + \CO\left(r^{-4}\right)\\
\ln \l &=& \frac{3}{2}\frac{ r^2}{R^2}-\frac{23}{16}
-2 v_1-\frac{R^2
\left(151+512 v_1^2+1024 v_2\right)}{2304 r^2}+ \CO\left(r^{-4}\right)
\label{IRresl}
\eea
where
\be
e^{A_0}={\sqrt{24}\over R\sqrt{v_0}}.
\ee
The IR scale $R=1/\Lambda_\rmi{IR}$ is an integration constant
here.\footnote{It is not independent of the scale $\Lambda=\Lambda_\rmi{UV}$ of the UV
expansions, i.e., the complete solution from UV to the IR will have fixed
$\Lambda_\rmi{UV}/\Lambda_\rmi{IR}$.}
Recall that $r$ does not appear explicitly in the equations of motion, and therefore there is also an
integration constant related to $r$: we have the freedom of shifting any solution by $r \to r + \delta r$.
The solution having
the simple $r$-dependence of Eqs.~\eqref{IRresA} and~\eqref{IRresA2} corresponds to a special choice of $\delta r$.
It will have its UV boundary at an arbitrary value of $r$ (rather than at $r=0$). If $\delta r$ is
fixed instead by requiring the UV boundary to lie at $r=0$, a corresponding shift must be added to the asymptotic formulas.
For our choice of $V_g$,
\bea
 v_0 &=& \frac{92 \left(b^{\rmi{YM}}_0\right)^2-144 b^{\rmi{YM}}_1}{27 \CL_0^2} = \frac{18476}{243}\\
 v_1 &=& \frac{1}{2} \ ; \qquad v_2 = - \frac{1}{8}
\eea
if we set $\CL_0=1$.

Using $A$ as the coordinate, the result reads
\bea
 \frac{r^2}{R^2} &=& -A+\frac{1}{4}\ln(-\fra32 A)+ A_0+
 \frac{23}{24}+\frac{4 v_1}{3} \nn\\
 &&-\frac{655+1152 v_1+512 v_1^2+1024 v_2}{3456 A} 
-\frac{\ln(-\fra32 A)+4 A_0}{16 A} + \CO\left(A^{-2}\right)\\
\ln\l &=& -\frac{3}{2}A +\frac{3 }{8}\ln(-\fra32 A)+\frac{3}{2}A_0 
-\frac{7+16 v_1+3 \ln(-\fra32 A)+12A_0}{32 A}+ \CO\left(A^{-2}\right)
\label{loglaA}
\eea

Various other combinations may be useful. In thermodynamics one needs $b=e^A$ in terms of $\l$;
from \nr{loglaA} one can invert:
\be
b=\bigl(\fra32\bigr)^{3/4}{4\over R\sqrt{v_0}} {1\over\lambda^{2/3}}\bigl(\fra23\ln\l)^{1/4}
\left[1+\mathcal{O}\left(\frac{1}{\ln \l}\right)\right]
\label{blaIR}
\ee
For the pair of functions $q(A)=e^Ar'(A),\,\l(A)$ used in numerics one can derive, for $A\to-\infty$,
\ba
q(A)&=&-{R\over2}e^A(-A)^{-1/2}\biggl[1+{1\over8A}\biggl(\ln(-\fra32 A)+4A_0+\fra92\biggr)+
\CO\left(A^{-2}\right)
\biggr]\\
\l(A)&=& e^{-\fra32(A-A_0)}(-\fra32 A)^{3/8}
\biggl[1-{3\over32A}\biggl(\ln(-\fra32 A)+4A_0+5\biggr)+
\CO\left(A^{-2}\right)
\biggr].
\ea

\subsubsection{The tachyon}

The IR expansion of the tachyon depends on the large-$\l$ asymptotics of the potentials
$V_f$, $a$, and $\f$. Recall that the tachyon
potential $V_f(\l,\tau)$ needs to vanish
in the IR \cite{ckp} in
order to have correct kind of flavor anomalies. All power-law asymptotics for the
potentials were analyzed in \cite{jk}, and two different acceptable cases were
chosen as examples.
They are:
\begin{itemize}
 \item[I] Asymptotics with
       \be
         a(\l) \sim \l^0 \ ; \qquad \f(\l) \sim \l^{4/3} \ ; \qquad V_{f0}(\l) \sim \l^{\hat \tau}
       \ee
       where $\hat \tau<10/3$. This case includes the potentials I
       and I$_*$
       of this article
       (for which $\hat \tau=2$). The tachyon diverges exponentially for $r \to \infty$ ($A \to -\infty$),
       \be
         \tau \sim e^{C_I \frac{r}{R}} \sim  e^{C_I\sqrt{-A}}
       \ee
       where the coefficient reads for potentials I
       \be
         C_I = \frac{81\ 3^{5/6} (115-16 x_f)^{4/3} (11-x_f)}{812944\ 2^{1/6}} \ .
       \ee
 \item[II] Asymptotics with
        \be
         a(\l) \sim \l^{\hat \sigma} \ ; \qquad \f(\l) \sim \l^{4/3} \ ; \qquad V_{f0}(\l) \sim \l^{\hat \tau}
       \ee
       where $\hat \sigma>0$ and $\hat \tau$ can take any value. This case includes the
       potentials II
       and II$_*$
       of this article (for which $\hat \sigma = 2/3$ and $\hat \tau=2$).
       The tachyon diverges for $r \to \infty$ ($A \to -\infty$) as
       \be
         \tau \sim
C_{II} \sqrt{\frac{r}{R}} \sim C_{II} (-A)^{1/4}
       \ee
       where the coefficient reads for potentials II
       \be
         C_{II} = \frac{27\ 2^{3/4} 3^{1/4} }{\sqrt{4619}} \ .
       \ee

\end{itemize}

\subsection{Finite temperature}

We will work out the finite temperature IR expansions in $A$-coordinates.
Instead of writing down the explicit expansions as above, it is more convenient to state the
relations between the coefficients of the series expansions. We start by defining the series
\ba
f&=&\epsilon f'_h + \CO(\e^2)\  , \quad f'(0)=f'_h+\epsilon f''_h + \CO(\e^2),\label{horexpf}\\
q&=&q_h+\epsilon q'_h + \CO(\e^2),\\
\lambda&=&\lambda_h+\epsilon\lambda'_h + \CO(\e^2),\\
\tau&=&\tau_h+\epsilon\tau'_h+\fra12\epsilon^2\tau''_h + \CO(\e^3),
\label{horexptau}
\ea
where $\e=A-A_h$ is the distance from the horizon, which lies at $A=A_h$, and all coefficients
are to be evaluated at the horizon. The key input here is $f(A_h)=0$. Inserting
to the equations of motion one can solve for six of the nine coefficients listed above:
\ba
q_h&=&-{\sqrt{3f'_h}\over\sqrt{V_g-V_{f}}},\\
f''_h&=&-4f'_h+{q_h^4\over f'_h}\biggl[\fr1{16}\lambda_h^2
\bigl(\partial_\lambda V_g-\partial_\lambda V_f\bigr)^2+{(\partial_\tau V_f)^2\over 6V_f\kappa_h}\biggr],
\\
q'_h&=&{q_h^5\over (f'_h)^2}\biggl[\fr1{16}\lambda_h^2
\bigl(\partial_\lambda V_g-\partial_\lambda V_f\bigr)^2+{(\partial_\tau V_f)^2\over 6V_f\kappa_h}\biggr]
=q_h\biggl(4+{f''_h\over f'_h}\biggr),
\\
\lambda'_h&=&-{3\lambda_h^2q_h^2\over 8f'_h}\bigl(\partial_\lambda V_g-\partial_\lambda V_f\bigr),
\\
\tau'_h&=&{q_h^2\partial_\tau \ln V_f\over f'_h\kappa_h},
\\
\tau''_h&=&{9\partial_\tau V_f(A+B+C)+D\over 12\f_h^2\,V_f^3(V_f-V_g)^3},
\ea
with the abbreviations
\ba
A&=&6\l_h^2\f'_hV_f^3(\partial_\lambda V_g-\partial_\lambda V_f),\nn\\
B&=&V_f^2[8\partial_\tau^2V_f-3\l_h^2
(\partial_\lambda V_g-\partial_\lambda V_f)(\f_h(\partial_\lambda V_g-3\partial_\lambda V_f)
+2\f'_hV_g)],\nn\\
C&=&-2V_f\,[6\partial_\tau V_f+V_g(4\partial_\tau^2 V_f+3\l_h^2\f_h\partial_\lambda V_f
(\partial_\lambda V_g-\partial_\lambda V_f))],
\nn\\
D&=&27\l_h^2\f_h\partial_\tau\partial_\l V_f\,V_f^2(V_g-V_f)(\partial_\lambda V_g-\partial_\lambda V_f).
\ea
Here $V_g\equiv V_g(\lambda_h),\,V_f\equiv V_f(\lambda_h,\tau_h)$, $\f'_h=d\f(\l_h)/d\l,\,\f_h=\f(\l_h)$.
The so far unspecified three coefficients $\lambda_h$, $\tau_h$, and
$f'_h$ remain as free parameters. However $f'_h$ will be fixed by requiring the standard
normalization of the blackening factor $f \to 1$ in the UV. Therefore the physically
relevant parameters are $\lambda_h$ and $\tau_h$, which can be mapped to the temperature
and the quark mass after the full solution has been found.

\section{The quark mass and the Efimov solutions}\label{AppEfimov}
As detailed in \cite{jk}, the existence of the Efimov vacua is tightly linked to the tachyon
mass at the IR fixed point, plotted in Fig.~\ref{tachyonmassfig}. In particular, the existence
of the full Efimov tower of vacua with arbitrary number of tachyon nodes is guaranteed if the
tachyon mass violates the BF bound.
The same holds at finite temperature: one can always tune $\l_h$ and $\tau_h$ such that the
solution comes arbitrarily close to the fixed point.
When the BF bound is violated, the tachyon solution is oscillatory in the vicinity of the
fixed point. Thus, when approaching the fixed point the tachyon will achieve arbitrary many nodes,
which signals the presence of the full Efimov tower.
In this case the dependence of the quark mass on $\l_h$ and $\tau_h$ is the ``standard'' one, i.e.,
qualitatively as in Fig.~\ref{mass0pic}. 

There are, however, some cases where either the fixed point is absent, which is the case
for potentials I$_*$ and II$_*$ at low $x_f$, or the BF bound is not violated, which is the
case, interestingly, for potentials I at very low $x_f$  (as well as in the conformal window for all potentials).
In such cases the picture can be different from Fig.~\ref{mass0pic}. We shall not give a
detailed description of all possible cases here, but rather discuss some of the main features and give examples.

The curve $\tau_{hc}$ (which actually starts at $\l_*$) exists if and only if there is a fixed point.
If there is no fixed point, the solutions are expected to reach
the standard UV boundary for all values of $\l_h$ and $\tau_h$. For the curves
$\tau_{h0},\, \tau_{h1},\ldots$ the situation is more complicated.
At least few of these curves may still exist even if there is no fixed point or if the
BF bound is satisfied at the fixed point.
Their existence at asymptotically large $\l_h$ is linked to the existence of Efimov solutions
at zero temperature: taking $\l_h \to \infty$ with $\tau_h$ fixed along the curves, the finite
temperature Efimov configurations converge towards their zero temperature counterparts.
In particular, we expect that the chiral symmetry is broken at zero temperature if and only if
$\tau_{h0}$ exists at asymptotically large $\l_h$.
We have found numerically that the curves are always absent in the conformal window, $x_f\ge x_c$,
so that chiral symmetry is intact. This turns out to be the case also for potentials I at large
$W_0$ and low $x_f$, but only in a part of the region where BF bound is satisfied at the fixed point.
See also the phase diagrams in Fig.~\ref{figPhasesPotIW0} of Sec.~\ref{secphases} which show that
chiral symmetry is intact at low $x_f$.
For potentials I$_*$ and at low $x_f$, where no fixed point exists,
the curves are also absent, and chiral symmetry is unbroken. In this case the $m_q=0$ thermodynamics
is determined by the $\tau=0$ solution and
is qualitatively similar to the Yang-Mills one (see also Fig.~\ref{figPhasesModPotI}).
For potentials II$_*$ however, at least the leading solution $\tau_{h0}$ can always be found and
chiral symmetry is thus broken at low temperatures (see Fig.~\ref{figPhasesModPotII}).

\section{Computation of $\l_\rmi{end}$}\label{Applaend}

One can also illustrate the connection of the behavior of $\l_\rmi{end}$ to the BF bound of the
tachyon (Fig.~\ref{tachyonmassfig}), assuming that we have chosen a set of potentials and value
of $x_f$ such that the IR fixed point exists. First we recall that $\l_\rmi{end}$ can be  defined
as the endpoint of the $\tau_{h0}(\l_h)$ curve which gives the (non-node) solution with
nontrivial tachyon and zero quark mass (Fig.~\ref{mass0pic}). In particular, as $\l_h$ approaches
$\l_\rmi{end}$ from above, $\tau_h$ tends to zero, and we expect that the whole tachyon solution
from the boundary to the horizon becomes small, and the tachyon decouples from the other fields.
Therefore, in order to define $\l_\rmi{end}$ it is enough to study the behavior of the tachyon based
on the linearized tachyon EoM, evaluated on a fixed background, obtained by setting the tachyon
to zero.

The linearized tachyon equation has the form
\be \label{taulineom}
 \tau''(r) + F_1 \tau'(r) + F_2 \tau(r) = 0
\ee
where
\bea \label{taucoeffs}
 F_1 &=& 3 A'(r) + \frac{f'(r)}{f(r)} + \l'(r) \frac{\d \ln \kappa(\l)}{\d \l} + \l'(r)
 \frac{\d \ln V_{f0}(\l)}{\d \l}, \\\nn
 F_2 &=& \frac{2 e^{2 A} a(\l)}{f(r)\kappa(\l)} \ .
\eea
Here $A(r)$, $\l(r)$, and $f(r)$ are the solutions of the EoMs for $\tau\equiv 0$, which
are the same for potentials I and II. The drastic difference between the
potentials, as suggested by Fig.~\ref{tachyonmassfig}, thus arises only through the
appearances of $a$ and $\kappa$ in the coefficients~\eqref{taucoeffs}. The regular
tachyon solution, which is finite in the IR, obeys
\be
 \frac{\tau'(r_h)}{\tau(r_h)} = -\lim_{r \to r_h}\frac{F_2}{F_1}
\ee
since the double-derivative term in~\eqref{taulineom} is negligible near the horizon.

Nodes of the regular solution to the linear tachyon equation can then be used to determine $\l_\rmi{end}$.
For small
$\l_h$ perturbative analysis applies and it is not difficult to see that the solution is monotonic, without nodes.
When $\l_h$ increases the equation becomes nontrivial and has to be studied numerically.
Usually we observe, that beyond a critical value of $\l_h$ a tachyon node appears in the UV.
The leading tachyon behavior in the UV is controlled by the quark mass, which has to vanish at the critical value.
We thus identify the critical value as $\l_\rmi{end}$, which was defined as the endpoint of the curve where $m_q=0$.
Thus the regular solution to the linearized EoM has no nodes for
$\l_h < \l_\rmi{end}$ and one or more nodes for $\l_h>\l_\rmi{end}$.
It can also happen that $\l_\rmi{end}$ does not exist, and the tachyon nodes are absent
for all $\l_h$.

Since $\l_h$ can take values from zero to $\l_*$,
we can construct backgrounds
which get arbitrarily close to the IR fixed point at $\l=\l_*$. If the BF bound  for the tachyon
is violated at the fixed point,
the tachyon must 
have nodes
as $\l_h \to \l_*$.
We can conclude that $\l_\rmi{end}$, and thus also the curve $\tau_{h0}$, exist in this case. 
This makes sense, since when the BF bound is violated, chiral
symmetry breaking
takes place
also at zero temperature,
which means the the curve $\tau_{h0}$ exist also at asymptotically large $\l_h$ as discussed in Appendix~\ref{AppEfimov}.

We can also say something about $\l_\rmi{end}$ in the probe limit $x_f \to 0$. For PotII it
seems that it approaches a fixed value
as seen from Fig.~\ref{figPotxf0limit} (right).
This value can be found by solving the linearized
tachyon EoM with a background evaluated at $x_f=0$ (i.e., the IHQCD solution), and by
checking if a special value of $\l_h$ (identified as $\l_\rmi{end}$) can be found where
nodes emerge in the tachyon solution. Notice that $\l_*$ goes to $\infty$ in the probe
limit so that $\l_h$ can take any value. Existence of
the limiting value
of $\l_\rmi{end}$ as $x_f\to 0$
thus requires
that the tachyon has nodes in the limit $\l_h \to \infty$ after first taking the probe limit.
Since zero temperature solutions are obtained for $\l_h \to \infty$,
it is plausible that the behavior of  Fig.~\ref{figPotxf0limit} (right) is seen
if and only if the probe limit system admits tachyon solutions with nodes (in other words,
chiral symmetry breaking) at zero temperature.
Recall that for PotI, for which the different behavior  of  Fig.~\ref{figPotxf0limit} (left)
is found, chiral symmetry is unbroken at low $x_f$.

\section{Computation of the condensate} \label{AppCond}
In principle, the condensate for an $m_q=0$ system could be computed from the UV expansion
\be
\tau(r)/\CL_\rmi{UV}=\sigma r^3(-\ln(\Lambda r))^{3/(2b_0)},\quad r\to0,
\label{defsig}
\ee
with
\be
A-\ln\left(\Lambda \CL_\rmi{UV}\right)={1\over b_0\lambda(A)}+{b_1\over b_0^2}\ln(b_0\lambda(A))
=-\ln\left( \Lambda r \right),
\quad A\to\infty,
\label{Ator}
\ee
where we dropped corrections of $\mathcal{O}(A^{-1})$.
Using this one can define
\be
\ln\tilde \sigma(A) = \ln\tau(A)-\ln\CL_\rmi{UV}+{3\over b_0\lambda(A)}+{3b_1\over b_0^2}\ln(b_0\lambda(A))
+{3\over 2b_0}\ln(b_0\lambda(A)),
\label{logofsigma}
\ee
which approaches $\ln \sigma$ for $A \to \infty$.

However,
our solution for the tachyon, which is obtained numerically by shooting from the IR, will have
a linear term $\tau \sim m_q r$ with a tiny quark mass (typically $m_q \sim 10^{-7}$), because
the IR boundary conditions cannot be fine tuned beyond the numerical accuracy of the code.
The linear term will dominate over the cubic one of Eq.~\eqref{Ator} in the deep UV.
In order to calculate the condensate, we need to separate the linear and cubic terms
from the numerically computed $\tau(A)$, and use the cubic solution in Eq.~\eqref{logofsigma}.
For $\tilde \sigma(A)$ to be a good approximation to the condensate $\sigma$, we need to have $A \sim {\rm hundreds}$.
Direct separation of the linear $m_q$ term in this region requires
numerical accuracy on the level of $e^{-{\rm hundreds}}$, which is practically impossible to achieve.

To
illustrate
the difficulty and its resolution, consider a concrete case. Let us take Potential II,
SB normalised, $\CL_\rmi{UV}=(1+\fra74 x_f)^{1/3}$, $x_f=3$. This system, when cooled,
has a 2nd order transition at $T_\rmi{end}=1.158T_h$, above a 1st order transition at $T_h$.
This is concretely seen in Fig.~\ref{figTransitionsTend}.
Since chiral symmetry is broken at $T_\rmi{end}$ we expect that $\sigma(T)$ starts growing from zero
at $T_\rmi{end}$ and grows when the system is cooled towards $T_h$.
As an example, we evaluate the condensate
when $T$ has been cooled to $T=0.95T_\rmi{end}=1.1T_h$.

\begin{figure}[!tb]
\begin{center}

\includegraphics[width=0.49\textwidth]{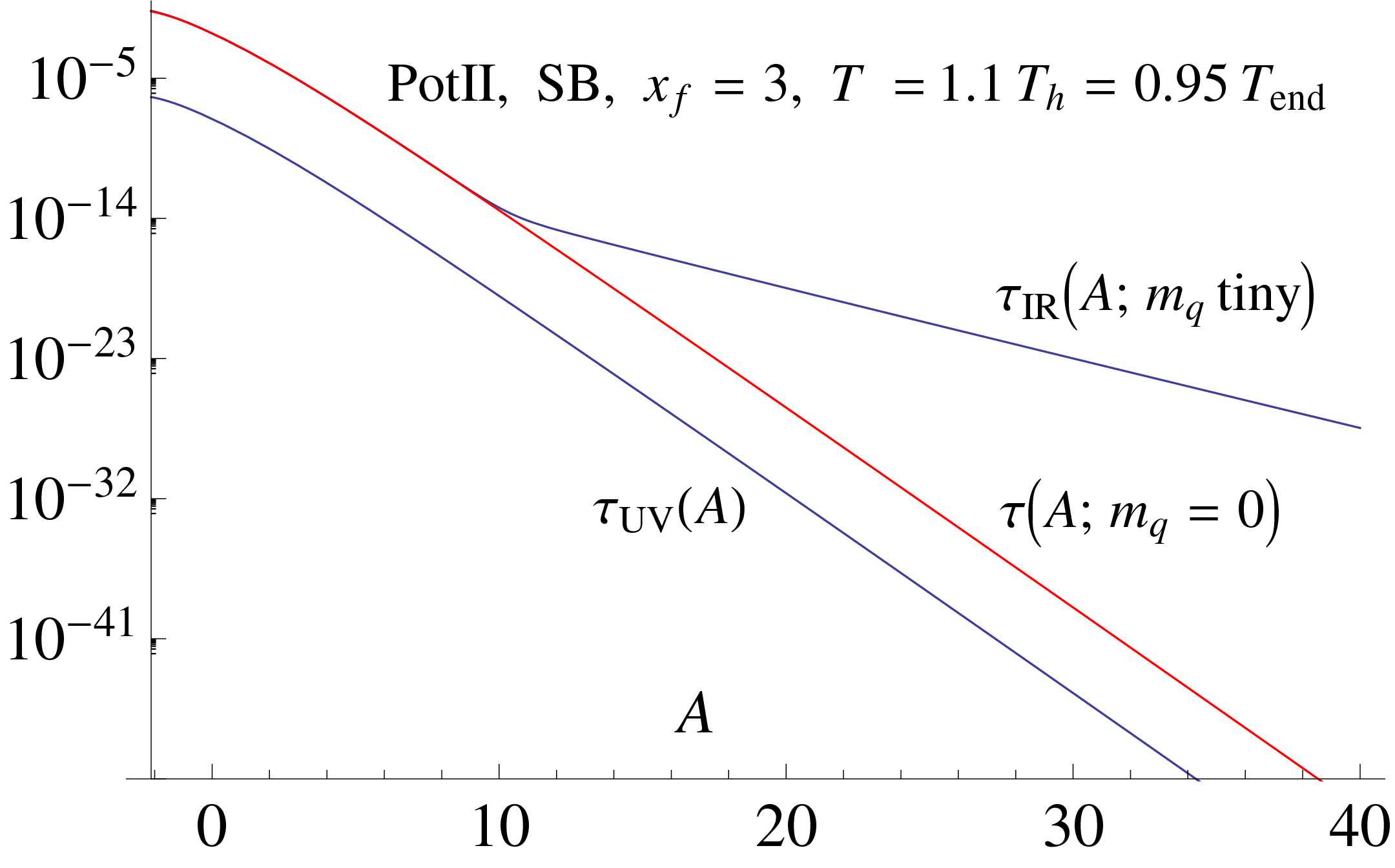}\hfill
\end{center}

\caption{\small The tachyon computed for $T=0.95T_\rmi{end}=1.1T_h$. The curve
$\tau_\rmi{IR}(A,m_q={\rm tiny})$ is obtained by integrating Einstein's equations from the
horizon and tuning $m_q=0$ as accurately as possible. The curve $\tau_\rmi{UV}(A)$
is obtained by integrating the tachyon equation of motion from the UV at
$A=400$
using
the bulk field $\lambda(A)$ from Einstein's equation and imposing $\tau\sim r^3$ in the UV.
The normalisation can be fixed by matching to $\tau_\rmi{IR}$ in the $A=2...10$ range
and a reliable $\tau(A,m_q=0)$ for the true $m_q=0$
tachyon is obtained. }
\label{sigtaufig}
\end{figure}

Numerical solution of Einstein's equations required knowing the values of $\lambda_h,\,\tau_h$
leading to a certain $T$ with $m_q=0$. For this potential and $T$ they were $\lambda_h=0.4017564,
\tau_h=\tau_{h0}(\lambda_h,m_q=0)=0.217984$. The computed $\tau(A)$ is shown in Fig.~\ref{sigtaufig}.
For $A$ up to about 10 one discerns the required $r^3\sim e^{-3A}$ behavior, but beyond that
$r\sim e^{-A}$ sets in and extends up to the end point of the computation at $A=400$. It is
impossible to shoot from the horizon and get $m_q=0$ more accurately; note that the tachyon
has already decreased to $10^{-14}$ from $0.22$ at the horizon.

To impose $\tau(r)\sim r^3$ one must shoot from the boundary, $r=0,\,A=\infty$. In this limit the
evolution of $\tau$ decouples from the other bulk fields, of which only $\lambda(A)$ is relevant since
$f\approx1$. We can thus integrate the tachyon equation from some large
$A(=400)$
using the
$\lambda(A)$ from Einstein's equations
and imposing as the initial condition $\tau(A)=e^{-3A},\,\tau'(A)=-3e^{-3A}$ with
small enough
normalisation.\footnote{It is not important to have precisely correct UV boundary conditions,
since corrections to the $\tau \sim r^3$ solution will decay fast as the system is solved
toward the IR. One should only make sure that the tachyon is much less than one in the whole
region of interest $(A \gtrsim 0)$ in order to suppress nonlinear effects, or alternatively
use explicitly linearized differential equation for the tachyon.} The result is plotted as
the curve $\tau_\rmi{UV}(A)$ in
Fig.~\ref{sigtaufig}. One observes that in the range $A=2...10$ the curve behaves accurately
as a constant$\times\tau_\rmi{IR}(A)$ and the normalisation can thus be determined. In this
way the true $\tau(A;m_q=0)$ plotted in Fig.~\ref{sigtaufig} is obtained.

Now that the accurate $\tau(A)$ is known, $\ln\tilde \sigma(A)$ can be plotted using
Eq.~\nr{logofsigma}, see Fig.~\ref{sigtaufig}. For the extrapolation it is even more
convenient to plot as a function of $\lambda(A)$, see also Fig.~\ref{sigtaufig}. One
obtains a nice linear behavior with the asymptotic value $\ln\sigma=-5.1558,\,
\sigma(T=0.95T_\rmi{end})=0.005766$.

If one used the original $\tau(A,m_q=0)$ at the largest value of $A$, $A=10$,
where the $r^3$ behavior was obtained, one
would have
$\tilde \sigma=0.0106$.
This is too
large by a factor $1.82$, not very far off,
but actually slightly larger than the expected 10\% error from neglecting the
$\mathcal{O}(A^{-1})$ corrections
in Eqs.~\eqref{Ator} and~\eqref{logofsigma}
at this value of $A$.
If we tried using the $\tau(A,m_q=0)$ solution directly, reliable extraction of
$\sigma$ would thus require much higher numerical precision, as already mentioned above.

\begin{figure}[!tb]
\begin{center}

\includegraphics[width=0.49\textwidth]{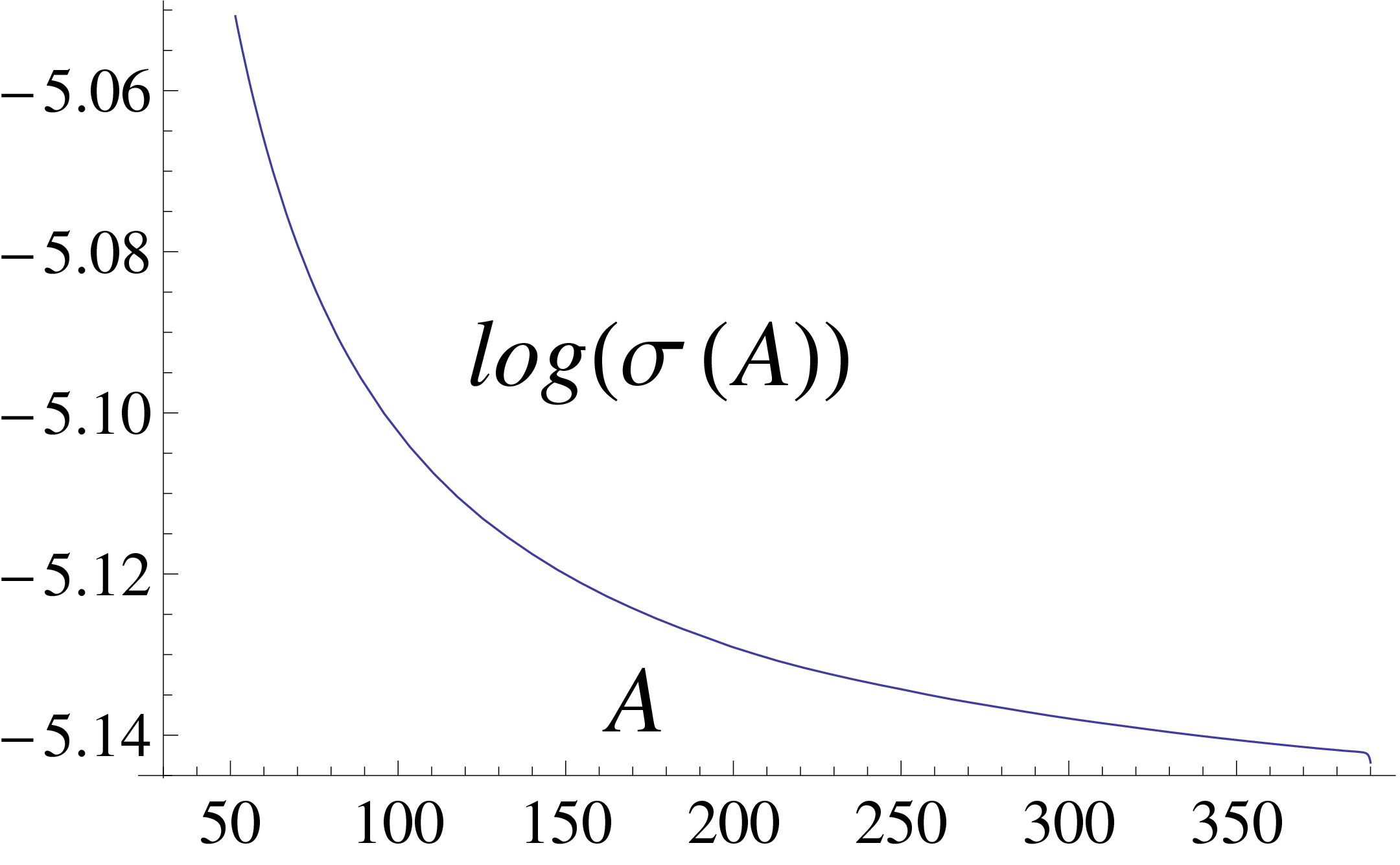}\hfill
\includegraphics[width=0.49\textwidth]{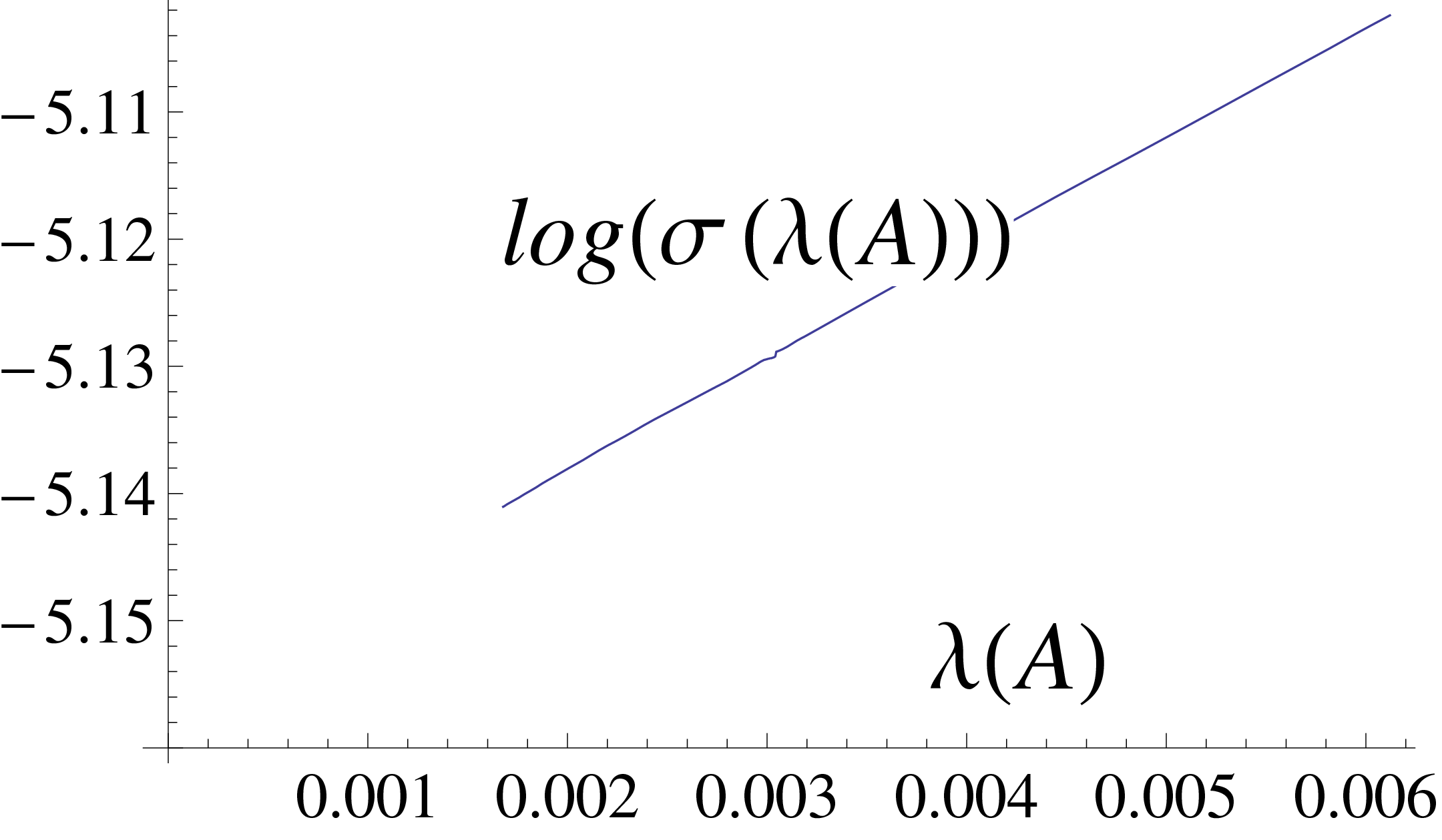}
\end{center}

\caption{\small Extrapolation of the value of $\ln\sigma$ to $r=0$, $A=\infty$, $\lambda=0$, for the
potential and temperature as in Fig.~\protect\ref{sigtaufig}
}
\label{logsigfig}
\end{figure}

After applying the procedure discussed above, the dominant error in the value of $\sigma$
arises actually from the matching of the two tachyon solutions that were obtained by
shooting from the UV and from the IR. The solutions are not exactly proportional for
$0\lesssim A \lesssim 10$ due to nonlinearities in the tachyon EoM and coupling to other fields.
The error can be reduced by introducing a further subtraction trick that effectively
reduces the value of $m_q$ of the solution that was obtained by shooting form the IR,
so that the matching can be done for slightly higher values of $A$ where the coupling
effects are considerably reduced.

We follow \cite{jk} and construct two solutions $\tau_{1,2}$ with small but different values
$m_{q1,q2}$. Optimal choice is to take $|m_q|$ as small as possible and choose one solution
with a positive value and another with a negative one. Then we construct
\be \label{taudeltadef}
 \tau_\rmi{IR}(A) = \frac{1}{1-\frac{m_{q1}}{m_{q2}}}\left(\tau_1(A) - \frac{m_{q1}}{m_{q2}}\tau_2(A)\right)
\ee
where the ratio $m_{q1}/m_{q2}$ can be accurately determined as the ratio of the solutions
$\tau_1/\tau_2$ at large $A$ (say $A=400$).\footnote{There is a small technicality
involved in this procedure as the two solutions will in general have different values
of $\Lambda$. Changing $\Lambda$ is equivalent with shifts of $A$ in $A$-coordinates
(see Eq.~\eqref{Ator}), so we can fix the issue by shifting, say, the solution
$\tau_2$ by a small $\Delta A$, obtained by requiring that the corresponding
solutions $\lambda_{1,2}$ for the coupling match at large $A$. Notice also that
the resulting $\tau_\rmi{IR}$ is only useful in the UV region $A \gtrsim 0$.}
The point is that the constructed $\tau_\rmi{IR}$ has its $m_q$ several orders
of magnitude closer to zero than either of the solutions $\tau_{1,2}$. Moreover,
the residual dependence of $\sigma$ on $m_q$ is drastically reduced: the linear
corrections cancel in~\eqref{taudeltadef} (see \cite{jk}). The improved value of
$\sigma$ can now be found by matching $\tau_\rmi{IR}$ with the solution $\tau_\rmi{UV}$, which was
obtained by shooting from the UV, as discussed above.

\end{document}